\documentclass[prx,aps,superscriptaddress,twocolumn,notitlepage,floatfix,10pt]{revtex4-2}
\usepackage{amsmath,mathtools,amsthm,amssymb,pifont}
\usepackage{mathrsfs}
\usepackage[utf8]{inputenc}
\usepackage[american]{babel}
\usepackage{graphicx,xcolor,bbold}
\usepackage{braket}
\usepackage{MnSymbol}
\usepackage{color}
\usepackage[colorlinks,
citecolor=red,
linkcolor=red,
urlcolor=red]{hyperref}
\usepackage{orcidlink}

\newcommand{\Com}{\mathrm{Com}}
\newcommand{\End}{\mathrm{End}}
\newcommand{\Span}{\mathrm{span}}
\newcommand{\GT}{\mathsf{GT}}
\newcommand{\Twirl}{\Phi}
\newcommand{\CM}{\mathfrak{M^\mathcal{C}\!\!\!}}
\newcommand{\MG}{\mathfrak{M}}
\newcommand{\Tr}{\mathrm{Tr}}
\newcommand{\sgn}{\mathrm{sgn}}
\newcommand{\Pf}{\mathrm{Pf}}
\newcommand{\CorrMat}{\mathscr{M}}
\newcommand{\Spec}{\mathrm{Spec}}
\newcommand{\Spin}{\mathrm{Spin}}

\usepackage{physics}

\begin{document}
\title{Theory of the Matchgate Commutant}

\author{Piotr Sierant~\orcidlink{0000-0001-9219-7274}}
\email{piotr.sierant@bsc.es}
\affiliation{Barcelona Supercomputing Center Plaça Eusebi G\"uell, 1-3 08034, Barcelona, Spain}

\author{Xhek Turkeshi~\orcidlink{0000-0003-1093-3771}}
\email{turkeshi@thp.uni-koeln.de}
\affiliation{Institut f\"ur Theoretische Physik, Universit\"at zu K\"oln, Z\"ulpicher Strasse 77, 50937 K\"oln, Germany}

\author{Poetri Sonya Tarabunga~\orcidlink{0000-0001-8079-9040}}
\email{poetri.tarabunga@tum.de}
\affiliation{Technical University of Munich, TUM School of Natural Sciences, Physics Department, 85748 Garching, Germany}
\affiliation{Munich Center for Quantum Science and Technology (MCQST), Schellingstraße 4, 80799 M{\"u}nchen, Germany}

\begin{abstract}
In quantum information theory and statistical physics, symmetries of multiple copies, or replicas, of a system play a pivotal role. 
For unitary ensembles, these symmetries are encoded in the replicated commutant: the algebra of operators commuting with the ensemble across $k$ replicas.
Determining the commutant is straightforward for the full unitary group, but remains a major obstacle for structured, computationally relevant circuit families.
We solve this problem for matchgate circuits, which prepare fermionic Gaussian states on $n$ qubits. 
Using a Majorana fermion representation, we show that operators coupling different system copies generate the orthogonal Lie algebra $\mathfrak{so}(k)$, endowing the space of invariants with rich and tractable structure.
This underlying symmetry decomposes the matchgate commutant into irreducible sectors, which we completely resolve via a Gelfand--Tsetlin construction. We provide an explicit orthonormal basis of the matchgate commutant for all $k$ and $n$, together with a formula for its dimension that grows polynomially in $n$.
Furthermore, we characterize the commutant of the Clifford--matchgate subgroup,  showing that restricting to signed permutations of Majorana modes yields a commutant that qualitatively diverges from the matchgate case for $k \geq 4$ replicas.
Ultimately, our orthonormal basis turns algebraic classification into a working toolbox.
Using it, we derive closed-form expressions for matchgate twirling channels and a fermionic analogue of Weingarten calculus, the projector encoding all moments of the Gaussian state orbit, state and unitary frame potentials, the average nonstabilizerness of fermionic Gaussian states, a systematic hierarchy of non-Gaussianity measures, and a fermionic de Finetti theorem.
\end{abstract}

\maketitle

\section{Introduction}
Symmetries form the bedrock of physics, orchestrating everything from the classification of fundamental particles in high-energy physics to the phenomenological description of phases within the Landau paradigm~\cite{weinberg1995quantum,haag1996local,streater2000pct,landau1951statistical}. 
Traditionally, the profound connection between continuous symmetries and conserved quantities is formalized by Noether's theorem~\cite{Noether1918}, often expressed through the commutation relations between a system's Hamiltonian and the generators of the underlying symmetry group~\cite{vonNeumann1955mathematical,SakuraiNapolitano2020}. 
In the modern era of quantum information, this continuous time evolution is increasingly complemented by discrete descriptions using quantum circuits and ensembles of unitary operations~\cite{Collins2006,harrow2013church,Collins2015,fisher2023random}. 
In this discrete setting, characterizing a system naturally shifts toward analyzing the symmetries across multiple copies, \textit{or replicas}, of the system. 
The conserved quantities for a $k$-replica system are algebraically encapsulated by the $k$th \textit{commutant}~\cite{goodman2009symmetry,Zhang2014,roberts2017chaos,gross2021schurweylduality}.

The commutant construction and replica symmetries have far-reaching applications across theoretical physics. 
In condensed matter, replica symmetries are famously essential for untangling the complex energy landscapes of spin glasses and glassy physics~\cite{Parisi79, parisi1983order,mezard1984nature,kirkpatrick1987dynamics,franz1997phase,mezard1999thermodynamics,Sachdev_1993}.
Similarly, in quantum information, these symmetries lie at the heart of quantum error correction~\cite{steane1996error,calderbank1996good,Gottesman1998,roy2009unitary,gullans2021quantum,kroll2026errorcorrectionbrickworkclifford}, aiding in entanglement entropy calculations~\cite{Calabrese04,amico2008entanglement,calabrese2009entanglement} and the statistical mechanical mapping of error-correcting codes~\cite{dennis2002topological,bombin2012topological,ygfz-crvp,trebst2024,nishimori2023,choi2020quantum,jian2020measurement,bao2020theory,gullans2020dynamical,chubb2021statistical,behrends2025statistical}. 
More broadly, characterizing unitary ensembles via their $k$-th moments, or equivalently, their commutants~\cite{Moudgalya23symm, Bao21enriched}, provides a powerful symmetry-based framework to study higher-order~\cite{PhysRevLett.123.230606,PhysRevE.99.042139,Pappalardi2024microcanonical,vallini2025longtimefreenesskicked,PhysRevB.111.054303,PhysRevLett.134.140404,10.21468/SciPostPhys.18.4.136,wang2025eigenstatethermalizationhypothesiscorrelations,10.21468/SciPostPhys.19.2.050,dowling2025freeindependenceunitarydesign,vallini2025refinementseigenstatethermalizationhypothesis,mishra2026theoryoutoftimeorderedtransport} and deep thermalization~\cite{Ippoliti22deep,ho2022exact,Choi2023,cotler2023emergent,claeys2022emergent,varikuti2025deep,chang2025deep,loio2026quantumstatedesignsmagic,lucas2023generalized,bejan2025matchgate}, measure quantum complexity via unitary designs~\cite{Brown10,Harrow09,Brandao2016,hunterjones2019unitary,haferkamp2022random,grevink2025glueshortdepthdesignsunitary,Schuster2025,lami2025anticoncentration,sauliere2025universality,lami2025quantumstatedesignemergent,cotler2017black,dalzell2022random,mele2024introductiontohaar,fava2024designsfreeprobability}, and rigorously benchmark computational quantum advantage~\cite{boixo2018characterizing,arute2019quantum,wu2021strong,zhong2020quantum,madsen2022quantum,morvan2023phase,bouland2019complexity}. 
Furthermore, the algebraic structure of these commutants and unitary moments is of practical importance for efficiently probing physical properties and verifying quantum states in experimental platforms via shadow tomography~\cite{huang2020predicting, aaronson2018shadow, elben2020cross, akhtar2023classical, hu2025dual}.

Despite their broad utility, mapping the commutant structure for highly structured, sub-universal quantum systems remains a formidable mathematical challenge.
For generic, Haar-random unitaries, Schur--Weyl duality elegantly reduces the commutant problem to well-understood permutation algebras~\cite{weyl1946classical,fulton1991representation,brouwer1996diagrammatic,Collins2006}.
However, this classification breaks down for restricted ensembles motivated by quantum computation~\cite{Preskill2018quantumcomputingin,hangleiter2023computational}.
A prime example is the Clifford group, central to quantum error correction and fault-tolerant computing~\cite{gottesman1998heisenbergrep,nielsen00,aaronson2004improvedsimulationof,Bravyi16}, whose commutant has only now been understood through a series of recent breakthroughs~\cite{gross2021schurweylduality,zhu2016clifford,nezami2020multipartite, montealegre2021rank, Montealegre_Mora_2025,magni2025anticoncentrationcliffordcircuitsbeyond,Bittel25commutant}.
Another key ensemble is that of matchgate unitaries~\cite{Valiant01,Terhal02,Jozsa08,knill2001fermionic,jozsa2010matchgate,brod2014computational,oszmaniec2017universal,kivlichan2018quantum}, a distinguished class of fermionic quantum circuits that is efficiently simulatable~\cite{Bravyi05flo,Dias24classical,ReardonSmith24improved,somma2002simulating,brod2016efficient}, yet with a rich structure that captures nontrivial scrambling and design-like behavior within symmetry-constrained sectors~\cite{Braccia2025flo,garciamartin2024architectures,hackl2021bosonic,bianchi2022volume,gluza2019equilibration}. 
Matchgates are closely connected to free-fermion physics~\cite{RevModPhys.88.035001,Bravyi05flo}, while the non-Gaussian resources beyond them underpin a theory of fermionic quantum advantage~\cite{Oszmaniec22,hebenstreit2019computational,Sierant25faf,Coffman25magic,Lyu24NGE,Lumia24}. 
Yet, unlike the Clifford case, the replica-symmetry structure of matchgate ensembles remains elusive and poorly understood. 
Existing results cover only low replica numbers, $k\leq 3$~\cite{Wan23,Oszmaniec22}, while for higher orders, $k\geq 4$, the replica symmetry, i.e., commutant structure,  is fundamentally incomplete.

In this work, we develop a complete and constructive theory of the $k$-replica commutants for the matchgate unitaries and for their discrete Clifford--matchgate subgroup, for arbitrary numbers of fermionic modes $n$ and arbitrary replica order $k$. 
Our goal is twofold. On the conceptual side, we seek a symmetry-based classification of replica invariants for structured fermionic circuits, in the same spirit in which Schur--Weyl duality organizes moments of the full unitary ensemble. 
On the practical side, we aim to turn this classification into a working toolbox for exact moment calculations, twirling maps, frame potentials, and probes of many-body complexity.

The detailed roadmap of this work is given in Sec.~\ref{sec:notation}.
The starting point is to formulate the commutant problem directly in the Majorana language. 
On $k$ replicas, operators can be expanded in tensor products of Majorana strings, and the invariance condition under matchgate evolution becomes a constrained orthogonal-invariance problem for the corresponding coefficient tensors. This immediately reveals a natural family of invariant building blocks: inter-replica Majorana \textit{pairing} operators. At low replica number, these pairing structures already account for the full commutant. At higher replica order, however, the structure becomes qualitatively richer: several independent invariant operators can live in the same Majorana sector, so counting pairings is no longer enough to resolve a basis.

Our central observation is that these inter-replica pairings are not merely combinatorial objects, but generate a \textit{replica symmetry algebra}. More precisely, the \textit{bridge} operators built from bilinears of Majorana modes across different replicas furnish a representation of the Lie algebra $\mathfrak{so}(k)$. This provides a \textit{fundamental organizing principle for the matchgate commutant}: the invariant operator space decomposes into irreducible replica-symmetry sectors, and the multiplicities that first appear at higher replica number can be resolved systematically by passing to a Gelfand--Tsetlin construction. In this way, we obtain an orthonormal basis of the matchgate commutant that is explicit and adapted to the full replica symmetry.

This representation-theoretic structure has two important consequences. First, it yields a complete characterization of the matchgate commutant and leads to closed formulas for its dimension, extending the previously known low-replica results to arbitrary replica order. Second, because the resulting basis is orthonormal, it turns the abstract classification into a practical calculational framework. 
Crucially, this circumvents the severe bottlenecks of non-orthogonal bases, where growth of matchgate commutant subspaces spanned by non-orthogonal operators with system size $n$ otherwise prevents exact twirling and higher-moment calculations.

We also analyze the corresponding problem for the Clifford--matchgate group. Although this discrete subgroup acts on Majorana operators only by signed permutations, its commutant admits an independent explicit construction with a transparent combinatorial labeling. Comparing the two structures makes precise how the continuous orthogonal symmetry of matchgates differs from its discrete Clifford restriction at the level of replica invariants. In particular, this comparison quantifies the separation between the two ensembles from the perspective of higher moments and design-like properties.

Once this complete characterization is in place, a range of applications follows naturally. The commutant bases provide explicit formulas for matchgate and Clifford--matchgate twirling channels, yielding a fermionic analogue of Weingarten calculus for these structured ensembles. They enable exact calculations of state and unitary frame potentials, and therefore provide sharp diagnostics for how closely finite-depth fermionic circuits approximate ideal moment statistics. 
At the same time, this framework also identifies Gaussian-invariant observables of replicated states, supplying a systematic algebraic foundation for measures of fermionic non-Gaussianity, including the recently introduced fermionic antiflatness and more general probes associated with replica symmetry sectors. Finally, the twirl interpretation gives a natural route to a fermionic analogue of de Finetti-type statements, in which Gaussian-symmetric states are approximated by tensor product of Gaussian states.

\section{Overview and Roadmap}
\label{sec:notation}
We consider a system of $n$ qubits with Hilbert space $\mathcal{H}_n \simeq (\mathbb{C}^2)^{\otimes n}$, of dimension $d=2^n$. 
The set of Hermitian operators on a vector space $\mathcal{V}$ is denoted by $\mathcal{B}(\mathcal{V})$.  
Via the Jordan--Wigner transformation~\cite{Jordan1928}, the qubit system is equivalently described as $n$ fermionic modes with $2n$ Majorana operators $\gamma_\mu$, $\mu\in[2n]:=\{1,\dots,2n\}$, satisfying the canonical anticommutation relations $\{\gamma_\mu,\gamma_\nu\}=2\delta_{\mu\nu}\mathbb{1}$. The explicit construction and notation for Majorana strings, replica sectors, and the Hilbert--Schmidt inner product are collected in Sec.~\ref{sec:matchgate_commutant_pairing_tensors}.

Let $\mathrm{G}\subset \mathcal{U}(\mathcal{H}_n)$ be a subgroup of the unitary group. Its $k$-replica commutant is
\begin{equation}
\Com_k(\mathrm{G}):=
\left\{
X\in\mathcal{B}(\mathcal{H}_n^{\otimes k})
\,\middle|\,
[X,U^{\otimes k}]=0,\, \forall\,U\in \mathrm{G}
\right\}.
\end{equation}
Thus, $\Com_k(\mathrm{G})$ is the space of operators on $k$ replicas that are invariant under the diagonal action of the ensemble. Moreover, $\Com_k(\mathrm{G})$ is the image of the twirling map
\begin{equation}
\Twirl_{\mathrm{G}}^{(k)}(O):=\int_{\mathrm{G}} dU \,U^{\otimes k}O(U^\dagger)^{\otimes k},
\end{equation}
where $dU$ denotes Haar measure for continuous groups and the uniform discrete measure for finite groups. In fact, the map $\Twirl_{\mathrm{G}}^{(k)}$ is the Hilbert--Schmidt projector onto $\Com_k(\mathrm{G})$.

In this work, we study two fermionic ensembles. 
Our primary focus is the matchgate group $\MG_n$~\cite{Valiant01,Terhal02,Jozsa08}, which coincides, up to phases, with the group of fermionic Gaussian unitaries
\begin{equation}
U_Q=\exp\!\left(\frac{1}{4}\sum_{\mu,\nu=1}^{2n}K_{\mu\nu}\gamma_\mu\gamma_\nu\right),
\end{equation}
with $K$ a real antisymmetric matrix. Their adjoint action on Majorana operators is linear,
\begin{equation}
U_Q\gamma_\mu U_Q^\dagger=\sum_{\nu=1}^{2n}Q_{\nu\mu}\gamma_\nu,
\quad
Q=\exp(K)\in\mathrm{SO}(2n).
\label{eq:orthogonal}
\end{equation}
Upon adjoining a reflection, such as the Pauli operator $X_n$, this action extends from $\mathrm{SO}(2n)$ to the full orthogonal group $\mathrm{O}(2n)$.

As a secondary ensemble we also study the Clifford--matchgate group $\CM_n:=\MG_n\cap\mathcal{C}_n$, the intersection with the Clifford group $\mathcal{C}_n$ (unitaries mapping every Pauli string to another Pauli string up to a phase). 
Whereas matchgates realize a \textit{continuous} $\mathrm{O}(2n)$ symmetry, Clifford–matchgates act on Majorana operators through the discrete hyperoctahedral group of signed permutations $\mathbb{B}_n:=(\mathbb{Z}_2)^{2n}\rtimes\mathrm{S}_{2n}$~\footnote{Here, $\rtimes$ is the Wreath product, cf.~Ref.~\cite{graczyk2005hyperoctahedral}.}. 
Concretely, for $U\in \CM_n$,  $U_{\mathbf{s},\pi}\gamma_\mu U_{\mathbf{s},\pi}^\dagger=s_\mu\gamma_{\pi(\mu)}$, where $\mathbf{s}=(s_1,\dots,s_{2n})$ is a vector with $s_\mu\in\{\pm1\}$ and $\pi\in\mathrm{S}_{2n}$.
Our aim is to characterize, for both ensembles and for arbitrary $k$, the full space of invariant operators on $\mathcal{H}_n^{\otimes k}$.

The first step is to exploit the decomposition into Majorana weight sector $\Gamma_{r_1,\dots,r_k}$, cf. Sec.~\ref{sec:matchgate_commutant_pairing_tensors}. 
In the matchgate case, the orthogonal covariance of Eq.~\eqref{eq:orthogonal} implies that commutant elements can be built from inter-replica contractions of Majorana indices. 
This leads to a natural family of invariant operators, dubbed \textit{pairing operators} and denoted by $\tilde T^{(\{x_{ij}\})}$, where $x_{ij}$ records the number of pairings that connect replica $i$ and replica $j$, with $1\le i<j\le k$. 
For low replica number $k\leq3$, these pairing tensors reproduce the previous construction~\cite{Wan23}, and already provide a complete description of the matchgate commutant. 
At higher replica order, however, several independent commutant elements can belong to the same Majorana-weight sector, so a more refined organizing principle is needed.

This framework is provided by the \textit{bridge operators},
\begin{equation}
\Lambda_{ab}:=\sum_{\mu=1}^{2n}\gamma_\mu^{(a)}\gamma_\mu^{(b)},
\qquad a,b\in[k],
\end{equation}
namely the summed inter-replica Majorana bilinears. 
As we discuss in Sec.~\ref{sec:rep_theoretic_matchgate_commutant}, these generate a representation of the replica Lie algebra $\mathfrak{so}(k)$ on the commutant space. 

This is the key structural insight of the paper: \textit{the matchgate commutant decomposes into irreducible sectors of a replica $\mathfrak{so}(k)$ symmetry, and the multiplicities that arise at higher $k$ are resolved systematically through a Gelfand--Tsetlin construction}. 
This yields our main result, i.e., an explicit orthonormal basis of $\Com_k(\MG_n)$ adapted to the subgroup chain $\mathrm{SO}(k)\supset\cdots\supset\mathrm{SO}(2)$, valid for arbitrary replica number $k$ and mode number $n$ (Sec.~\ref{sec:rep_theoretic_matchgate_commutant}).
Counting the operators arising in this construction, we obtain the matchgate commutant dimension as
\begin{equation}
\dim\Com_k(\MG_n)
=
\prod_{1\le i\le j\le k-1}
\frac{2n+i+j-1}{i+j-1}.
\label{eq:dim_MG_overview}
\end{equation}
The structure is qualitatively different from the Clifford--matchgate case. For $\CM_n$, the invariant structures are labeled by even-cardinality subsets of replicas, and the commutant basis (derived in Sec.~\ref{sec:CM_commutant}) has dimension
\begin{equation}
\dim\Com_k(\CM_n)
=
\binom{2n+2^{k-1}-1}{2^{k-1}-1}.
\label{eq:dim_CM_overview}
\end{equation}
For fixed $k$, the matchgate commutant dimension~\eqref{eq:dim_MG_overview} grows polynomially in $n$ as $\dim\Com_k(\MG_n)=\mathcal{O}\left(n^{k(k-1)/2}\right)$, reflecting the continuous $\mathrm{O}(2n)$ symmetry. Meanwhile, the Clifford--matchgate dimension~\eqref{eq:dim_CM_overview} grows as $\dim\Com_k(\CM_n)= \mathcal{O}\left(n^{2^{k-1}-1}\right)$, a scaling governed by the combinatorics of even replica subsets. These two dimensions strictly coincide for $k\le 3$ but diverge at $k=4$, where discrete invariants absent from the continuous matchgate commutant emerge.

The explicit construction of the matchgate commutant basis has broad consequences across quantum information and many-body physics. In Sec.~\ref{sec:applications} we develop a suite of applications; the main results are the following.

\paragraph*{Twirling channels and Weingarten calculus.}
The commutant basis enables a complete evaluation of the matchgate twirling channel
\begin{equation}
\Twirl_{\MG_n}^{(k)}(W):=\int_{\mathrm{O}(2n)} dQ\,U_Q^{\otimes k}W(U_Q^\dagger)^{\otimes k}
\end{equation}
via the double-commutant theorem, and underpins the development of matchgate Weingarten calculus---a concrete computational toolbox for fermionic Gaussian integrals (Sec.~\ref{sec:twirlings}). The Clifford--matchgate analogue is discussed in Sec.~\ref{sec:cliff_twirlings}.

\paragraph*{Twirled vacuum projector and Gaussian state moments.}
A particularly useful special case is the $k$-copy twirl of the vacuum $|\mathbf{0}\rangle\langle\mathbf{0}|^{\otimes k}$, which yields a closed-form commutant element~(Sec.~\ref{subsec:MG_vacuum_twirl}):
\begin{equation}
\Twirl_{\MG_n}^{(k)}\!\left(\ketbra{ \mathbf{0} }{ \mathbf{0} }^{\otimes k}\right)=\frac{P_{\mathbf{0}}^{(k)}}{\Tr P_{\mathbf{0}}^{(k)}},\label{eq:definetti_overview}
\end{equation}
where $P_{\mathbf{0}}^{(k)}$ is the projector onto the trivial $\mathfrak{so}(k)$ sector.
This operator encodes all $k$-point moments of the fermionic Gaussian state orbit, and is directly relevant for fermionic shadow tomography~\cite{Wan23,Zhao21,Huang2022}: the measurement channel and variance bounds of matchgate shadow protocols are determined by the $k=2$ and $k=3$ commutant structure, while the twirled vacuum projector governs the shadow estimation of Gaussian-state fidelities (Sec.~\ref{subsec:shadow_tomography}).

\paragraph*{Frame potentials.}
The $k$-th unitary frame potential equals the commutant dimension~\eqref{eq:dim_MG_overview}. For the \emph{state} frame potential of the matchgate vacuum orbit, we obtain the closed-form expression (Sec.~\ref{sec:state_FP})
\begin{equation}
\mathcal{F}_{\MG_n}^{(k)}=\frac{1}{2}\prod_{1\le i<j\le n}\frac{2n-i-j}{k+2n-i-j}.
\label{eq:state_FP_overview}
\end{equation}

Furthermore, the state frame potential of the Clifford-matchgate vaccum orbit is given by (Sec.~\ref{sec:state_FP})
\begin{equation}
    \mathcal{F}^{(k)}_{\CM_n} = \frac{2^{n(2-k)}}{\binom{2n}{n}} \binom{n+2^{k-2}-1}{2^{k-2}-1}. \label{eq:CM_frame_potential_overview}
\end{equation}

\paragraph*{Matchgate designs.}
Comparing the frame potentials~\eqref{eq:state_FP_overview} and~\eqref{eq:CM_frame_potential_overview} shows that random Clifford--matchgate circuits do not form a matchgate design (Sec.~\ref{subsec:CliffMatchDesi}).

\paragraph*{Gaussian de Finetti theorem.}
We prove a fermionic Gaussian de Finetti theorem: any subsystem of a state of many replicas that is Gaussian-symmetric must be close to a mixture over Gaussian states (Sec.~\ref{sec:definetti}).

\paragraph*{Nonstabilizerness of Gaussian states.}
Via the twirled vacuum, Eq.~\eqref{eq:definetti_overview}, we compute in closed form the average stabilizer R\'enyi entropy of random fermionic Gaussian states---a quantity previously accessible only numerically (Sec.~\ref{sec:magicsre}).

\paragraph*{Measures of fermionic non-Gaussianity.}
Every commutant element $W\in\Com_k(\MG_n)$ defines a Gaussian-invariant functional
\begin{equation}
\varphi_W(\ket{\Psi})=\Tr\!\bigl[W\,(|\Psi\rangle\langle\Psi|)^{\otimes k}\bigr],
\end{equation}
yielding a systematic hierarchy of non-Gaussianity probes. These include fermionic antiflatness (expressible via the covariance matrix), projector-based invariants built from $P_{\mathbf{0}}^{(k)}$, and higher-order Casimir diagnostics (Sec.~\ref{subsec:nonGaussianity}).

Taken together, these results provide a complete algebraic foundation for symmetries of fermionic Gaussian unitaries at arbitrary replica order, with explicit formulas for twirling channels, frame potentials, non-Gaussianity measures, and design distances.

\section{Matchgate commutant: pairing operators}
\label{sec:matchgate_commutant_pairing_tensors}
In this section, we characterize the operators belonging to the $k$-th order matchgate commutant $\Com_k(\MG_n)$.
To that end, we introduce the suitable notation we will consider throughout the text.

For an ordered subset $S=\{\mu_1,\mu_2,\dots,\mu_r\}\subseteq[2n]$ with $\mu_1<\mu_2<\cdots<\mu_r$, we define the Majorana string
\begin{equation}
\gamma_S:= i^{|S|(|S|-1)/2} \gamma_{\mu_1}\cdots\gamma_{\mu_r},
\end{equation}
where $|S|$ denotes the cardinality of $S$, and  $\gamma_\varnothing:=\mathbb{1}$. We refer to $|S|$ as the weight of the Majorana operator $\gamma_S$. 

The operator space $\mathcal{B}(\mathcal{H}_n)$ is equipped with the Hilbert--Schmidt inner product $\langle A,B\rangle:=\Tr(A^\dagger B)$, with respect to which the family $\{\gamma_S\}_{S\subseteq[2n]}$ forms an orthogonal basis. 
Accordingly, every operator $A\in\mathcal{B}(\mathcal{H}_n)$ admits the unique expansion
\begin{equation}
A=\sum_{S\subseteq[2n]}A_S\,\gamma_S, \quad A_S=\frac{\Tr(A\gamma_S)}{2^n}.
\end{equation}

A useful grading is provided by the Majorana weight. We denote by $\Gamma_r:=\mathrm{span}\{\gamma_S:\ |S|=r\}$ the subspace of operators with fixed Majorana weight $r$. We also introduce the fermionic parity operator
\begin{equation}
\mathcal{P}=(-i)^n\prod_{\mu=1}^{2n}\gamma_\mu=\prod_{j=1}^n Z_j.
\end{equation}

Our central object is the replica theory. 
We consider $k$ copies of the system, $\mathcal{H}_n^{\otimes k}$, and refer to these copies as replicas. 
Replica labels are denoted by $a,b\in[k]:=\{1,\dots,k\}$. For a $k$-tuple of ordered subsets $\mathbf{S}:=(S^1,\dots,S^k)$, we write
\begin{equation}
\gamma_{\mathbf{S}}:=\bigotimes_{\ell=1}^k \gamma_{S^\ell}.
\end{equation}
Denoting $r_\ell=|S^\ell|$ for $\ell\in[k]$, then $\gamma_{\mathbf{S}}$ belongs to the Majorana weight sector $\Gamma_{r_1,\dots,r_k}:=\Gamma_{r_1}\otimes\cdots\otimes\Gamma_{r_k}$. 
Any operator $W\in\Gamma_{r_1,\dots,r_k}$ can therefore be expanded uniquely as
\begin{equation}
W=\sum_{\mathbf{S}} W_{\mathbf{S}}\,\gamma_{\mathbf{S}},
\qquad
W_{\mathbf{S}}=\frac{\Tr(W\,\gamma_{\mathbf{S}})}{2^{nk}},
\label{eq:O_sector_expansion}
\end{equation}
where the sum runs over all tuples $\mathbf{S}=(S^1,\dots,S^k)$ with $|S^\ell|=r_\ell$. 
Because of fermionic statistics, the coefficients $W_{\mathbf{S}}$ are antisymmetric within each replica block. 
The replicated Majorana basis $\{ \gamma_\mathbf{S}\}$ is the natural language for the commutant problem.

The first key observation is that the matchgate action preserves the Majorana weight in each replica separately. Therefore, the full replicated operator space decomposes into invariant sectors labeled by the Majorana weights $(r_1,\dots,r_k)$, and the commutant can be studied independently in each such sector:
\begin{equation}
\Com_k(\MG_n)
=
\bigoplus_{r_1,\dots,r_k=0}^{2n}
\Bigl(\Com_k(\MG_n)\cap \Gamma_{r_1,\dots,r_k}\Bigr).
\label{eq:comm_sector_decomp}
\end{equation}
This reduces the problem to a finite-dimensional question in invariant theory inside each fixed sector.

Let us therefore fix one sector $\Gamma_{r_1,\dots,r_k}$ and consider an arbitrary operator $W$ supported in it. Such an operator can be expanded in the basis of replicated Majorana strings, with coefficients $W_{\mathbf S}$ labeled by a $k$-tuple of ordered subsets $\mathbf S=(S^1,\dots,S^k)$, with $|S^\ell|=r_\ell$. 
Here, if $S^\ell=(S^\ell_1,\dots,S^\ell_{r_\ell})$, then the corresponding Majorana string in replica $\ell$ is totally antisymmetric in the entries of $S^\ell$. Accordingly, the coefficients $O_{\mathbf S}$ are naturally viewed as elements of
\begin{equation}
(\wedge^{r_1}\mathbb C^{2n})\otimes \cdots \otimes (\wedge^{r_k}\mathbb C^{2n}),
\end{equation}
where $\wedge^r\mathbb C^{2n}$ denotes the space of fully antisymmetric tensors spanned by Majorana strings of weight $r$. 
In other words, one may think of $W_{\mathbf S}$ as a tensor of total order $R:=\sum_{\ell=1}^k r_\ell$, whose indices are grouped into $k$ replica blocks of sizes $r_1,\dots,r_k$, and which is completely antisymmetric within each block.

The matchgate action induces an orthogonal transformation on Majorana operators [cf.~Eq.~\eqref{eq:orthogonal}]. For each $r\in\{0,\dots,2n\}$, let $\wedge^r Q$ denote the induced action of $Q\in \mathrm O(2n)$ on the antisymmetric space $\wedge^r\mathbb C^{2n}$. In the basis labeled by ordered subsets $S,\bar S\subset[2n]$ with $|S|=|\bar S|=r$, its matrix elements are
\begin{equation}
(\wedge^r Q)_{\bar S,S}
=
\det(Q_{\bar S,S}),
\end{equation}
where $Q_{\bar S,S}$ is the $r\times r$ submatrix of $Q$ with rows indexed by $\bar S$ and columns indexed by $S$.

Accordingly, in the sector $\Gamma_{r_1,\dots,r_k}$ the coefficients transform as
\begin{equation}
W_\mathbf{S}\mapsto\!\!\!\!\sum_{\bar S^1,\dots,\bar S^k}\!\!\!\!(\wedge^{r_1}Q^T)_{S^1,\bar S^1}\cdots(\wedge^{r_k}Q^T)_{S^k,\bar S^k}W_\mathbf{\bar{S}}\label{eq:coef_transform_blocks}
\end{equation}
Thus, $W$ belongs to the commutant if and only if its coefficients are invariant under this induced orthogonal action.

At this point, classical invariant theory gives a complete answer. Ignoring for a moment the replica-wise antisymmetry, one is led to the standard problem of invariant tensors for the orthogonal group acting on $R$ indices. 
By orthogonal Schur--Weyl duality, the commutant of $\mathrm O(2n)$ acting on $(\mathbb C^{2n})^{\otimes R}$ is generated by the image of the Brauer algebra $\mathcal B_R(2n)$~\cite{Brauer37,Goodman09symmetry}. 
Equivalently, by the first fundamental theorem of invariant theory for the orthogonal group~\cite{weyl1946classical}, every invariant tensor is obtained by contracting indices pairwise with the symmetric metric $\delta_{\mu\nu}$. 
Thus, the basic building blocks of the commutant are tensors associated with perfect matchings of the $R$ index positions.

This suggests a concrete basis construction. 
One starts from all possible pairings of the $R$ slots and associates to each pairing the corresponding tensor of Kronecker deltas. 
However, the replica-wise antisymmetry imposes a crucial restriction: if two indices belonging to the same replica are paired together, the resulting contraction vanishes identically, because a symmetric tensor $\delta_{\mu\nu}$ is contracted against an antisymmetric pair of indices. 
Therefore, only pairings that connect \emph{different} replicas can contribute.

This is the central simplification. 
The commutant is not generated by arbitrary Brauer pairings, but only by pairings that connect distinct replicas. 
In the following subsection, we make this construction explicit and show how these inter-replica pairing tensors provide a natural spanning set, and eventually a basis, for $\Com_k(\MG_n)$.

\subsection{Pairing operators as the matchgate commutant basis}
Here, we proceed to construct the basis of the matchgate commutant $\Com_k(\MG_n)$. 
Recall that $\mathbf{S}=(S^1,\dots,S^k)$ is a $k$-tuple of ordered subsets, and let $S^\ell_a$ denote the $a$-th element of $S^\ell$. Thus, the subset $S^\ell$ contributes $r_\ell$ labeled slots, for a total number of slots $R=\sum_{\ell=1}^k r_\ell$. To describe all possible contractions among these slots, it is convenient to introduce the set of index positions
$\mathcal{P}_{\mathbf r}:=\{(\ell,a):\ell\in [k],\ a\in [r_\ell]\}$. 
A pairing configuration $\pi$ is then a perfect matching of $\mathcal{P}_{\mathbf r}$, namely a partition of the $R$ slots into disjoint unordered pairs. 
Each pair $\{(\ell,a),(m,b)\} \in \pi$ specifies which two positions are identified with one another.  
We further restrict attention to \emph{inter-replica pairings}, where each pair in $\pi$ connects slots from different replicas:
$\ell \neq m$ for all $\{(\ell,a),(m,b)\} \in \pi$.
Given such a matching, the associated pairing tensor is
\begin{equation}
T_{\mathbf S}^{\pi}:=\prod_{\{(\ell,a),(m,b)\}\in\pi}\delta_{S^\ell_a,S^m_b}.
\end{equation}
This tensor simply enforces the identifications prescribed by $\pi$: each Kronecker delta requires that the two mode labels sitting in the paired positions coincide. 
Therefore, an operator $W$ belongs to the commutant if its coefficient tensor $W_{\mathbf S}$ can be expanded as a linear combination of these pairing tensors $W_{\mathbf S}=\sum_\pi c_\pi T_{\mathbf S}^{\pi}$, where the sum runs over all valid pairing configurations $\pi$. 

We now impose the fermionic antisymmetry within each replica. To this end, we define the replica-wise antisymmetrization projector
\begin{equation}
\mathcal{A}_{r_1,\dots,r_k}:=\prod_{a=1}^{k}\mathcal{A}^{(a)}_{r_a},
\end{equation}
where $\mathcal{A}^{(a)}_{r_a}$ antisymmetrizes the $r_a$ slots belonging to replica $a$. 
Explicitly, for a rank-$r$ tensor $v$,
\begin{equation}
(\mathcal{A}_r v)_{j_1\dots j_r}:=\frac{1}{r!}\sum_{\sigma\in\mathrm{S}_r}\mathrm{sgn}(\sigma)\,v_{j_{\sigma(1)}\dots j_{\sigma(r)}},
\label{eq:antisymm}
\end{equation}
with $\mathrm{S}_r$ the permutation group on $r$ elements and $\mathrm{sgn}(\sigma)\in \{\pm 1\}$ is the sign of the permutation $\sigma$. This is the natural projection onto the fully antisymmetric subspace, as required by the fermionic nature of the Majorana strings.

Applying this projector to a pairing tensor $T_{\mathbf S}^\pi$ gives its physical, fermionically antisymmetric version,
\begin{equation}
\tilde{T}_{\mathbf S}^{\pi}:=\mathcal{A}_{r_1,\dots,r_k}T_{\mathbf S}^{\pi}.
\end{equation}
At this stage, a major simplification occurs. Although different pairings $\pi$ may look combinatorially distinct, after antisymmetrization the resulting tensor depends only on how many pairings connect one replica to another, and not on the detailed choice of slots involved. 

More precisely, to each pairing configuration, we associate a symmetric matrix of nonnegative integers $x_{ij}=x_{ji}$, with $1\le i,j\le k$, where $x_{ij}$ counts the number of edges connecting replica $i$ to replica $j$. 
Since pairing two indices within the same replica is annihilated by antisymmetrization, only inter-replica pairings survive, and therefore $x_{ii}=0$.
Moreover, these numbers cannot be chosen arbitrarily: each replica $j$ contains exactly $r_j$ slots, and every slot must be paired exactly once. Therefore, their counts satisfy the constraints
\begin{equation}
r_j=\sum_{i=1}^{k}x_{ij},\quad \text{for all }j\in[k].
\label{eq:bridge_constraint}
\end{equation}
Thus, the matrix $(x_{ij})$ provides a complete bookkeeping for the surviving pairing data.

As shown in Appendix~\ref{app:pairingTensors}, all pairings $\pi$ with the same adjacency matrix $\{x_{ij}\}$ give the same antisymmetric tensor, up to an overall sign. We therefore denote by $\tilde{T}^{(\{x_{ij}\})}_{\mathbf S}$, the unique antisymmetric pairing tensor associated with the configuration $\{x_{ij}\}$.

These tensors can now be turned into operators by contracting them with the corresponding Majorana strings. For each admissible configuration $\{x_{ij}\}$, we define the corresponding pairing operator as
\begin{equation}
\tilde{T}^{(\{x_{ij}\})}:=\sum_{\mathbf S}\tilde{T}^{(\{x_{ij}\})}_{\mathbf S}\,\gamma_{\mathbf S},\label{eq:tttttt}
\end{equation}
where the sum runs over all $k$-tuples of ordered subsets $\mathbf{S}=(S^1,\dots,S^k)$ with $|S^\ell|=r_\ell$, and where $\gamma_{\mathbf S}$ denotes the corresponding tensor-product Majorana string. 
By construction, this operator is $\mathrm{O}(2n)$-invariant.

The collection of all such operators~\eqref{eq:tttttt} spans the subspace of $\Com_k(\MG_n)$ inside the fixed replica-weight sector $\Gamma_{r_1,\dots,r_k}$. Equivalently, the problem of constructing a basis of the commutant in this sector is reduced to the much simpler combinatorial problem of classifying all symmetric nonnegative integer matrices $(x_{ij})$ with vanishing diagonal and row sums fixed by Eq.~\eqref{eq:bridge_constraint}. 
We now turn to the explicit construction of this basis.

\subsubsection{$k=1$} 
For $k=1$, there is only a single replica, with weight sectors $\Gamma_{r_1}$ labeled by $r_1=0,\dots,2n$. In this case, any pairing would necessarily connect two indices within the same replica. However, such intra-replica pairings vanish after antisymmetrization, since the symmetric contraction $\delta_{\mu\nu}$ is incompatible with the antisymmetry of the Majorana indices. Therefore, no nontrivial invariant tensor exists in any sector with $r_1>0$. 
The only surviving sector is $\Gamma_0$, which is one-dimensional and corresponds to the identity operator; hence $\Com_1(\MG_n)=\mathrm{span}\{\mathbb{1}\}$. 

\subsubsection{$k=2$}
For $k=2$, there are two replicas, with sectors labeled by the pair of Majorana weights $(r_1,r_2)$. 
Since only one inter-replica pairings survive, every index in replica $1$ must be paired with an index in replica $2$. This is possible if and only if $r_1=r_2=:r$ and $x_{12}=r$. 
Thus, in each sector $\Gamma_{r,r}$ there is a unique admissible configuration, and hence a unique invariant tensor up to normalization. 
We denote the corresponding antisymmetric pairing tensor by $\tilde{T}^{(r)}_{\mathbf S}
:=
\tilde{T}^{(x_{12}=r)}_{\mathbf S}$ with $r=0,1,\dots, 2n$. 
Its explicit construction is given in Appendix~\ref{app:pairingTensorsBas2}. 
Contracting this tensor with the replicated Majorana strings yields the associated commutant operator
\begin{equation}
\tilde{T}^{(r)} =
\sum_{\mathbf S}\tilde{T}^{(r)}_{\mathbf S}\,\gamma_{\mathbf S} = \frac{1}{N_r}\sum_{|S|=r}\gamma_S\otimes \gamma_S,
\label{eq:antiK2}
\end{equation}
where $N_r=\binom{2n}{r}^{1/2}$ is a convenient normalization factor. 
Since each operator $\tilde{T}^{(r)}$ belongs to a different sector $\Gamma_{r,r}$, the family $\{\tilde{T}^{(r)}\}_{r=0}^{2n}$ is mutually orthogonal. 
Therefore, these operators form an orthogonal basis of the two-replica matchgate commutant; in particular, $\dim \Com_2(\MG_n)=2n+1$.

\subsubsection{$k=3$}
\label{subsec:k3anti}
For $k=3$, the inter-replica pairing numbers $x_{ij}$ are uniquely determined  via a system of three linear equations. by the integers $r_1, r_2, r_3$ specifying the Majorana weight sector $\Gamma_{r_1, r_2, r_3}$.
The corresponding operator in the commutant  $\Com_3(\MG_n)$ can be written as 
\begin{equation}
\tilde{T}^{(r_1, r_2, r_3)} = \sum_{|S^{1}|=r_1} \sum_{|S^{2}|=r_2} \sum_{|S^{3}|=r_3}  \tilde{T}_{\mathbf{S}}^{\pi} \,\gamma_{S^{1}} \! \otimes \gamma_{S^{2}} \! \otimes \gamma_{S^{3}}.
\end{equation}
Constructing the antisymmetric pairing tensor $\tilde{T}_{\mathbf{S}}^{\pi}$, see Appendix~\ref{app:pairingTensorsBas3}, we obtain that
\begin{equation}
\begin{split}
\tilde{T}^{(r_1, r_2, r_3)}& =\frac{1}{N_{r_1,r_2,r_3}}  
\sum_{\substack{
A_{12},A_{13},A_{23}
}} \\
&\times \qquad 
\gamma_{A_{12}\cup A_{13}}
\otimes
\gamma_{A_{12}\cup A_{23}}
\otimes
\gamma_{A_{13}\cup A_{23}},
\end{split}
\label{eq:antiK3}
\end{equation}
where the subsets $ A_{12},A_{13},A_{23}\subseteq[2n]$ are disjoint, have fixed sizes $|A_{12}|=x_{12}$, $|A_{13}|=x_{13}$, $|A_{23}| = x_{23}$, and $N_{r_1,r_2,r_3}$ is a normalization constant. Since each Majorana weight subspace $\Gamma_{r_1,r_2,r_3}$ corresponds to a unique commutant element $\tilde{T}^{(r_1, r_2, r_3)}$, these operators provide an orthonormal basis of  $\Com_3(\MG_n)$, with the orthonormality relation $\Tr[(\tilde{T}^{(r_1, r_2, r_3))})^\dag (\tilde{T}^{(r'_1, r'_2, r'_3))}) ] =\delta_{r_1,r'_1} \delta_{r_2,r'_2} \delta_{r_3,r'_3}$. Since the basis operators are indexed by three non-negative integers such that $r_1+r_2+r_3 \leq 2n$, the total number of independent operators, i.e. the dimension of the $k=3$ matchgate commutant is $\dim\bigl(\Com_3(\MG_n)\bigr) = \binom{2n+3}{3}$.

\subsubsection{Beyond $k=3$}
The operator families in Eqs.~\eqref{eq:antiK2} and \eqref{eq:antiK3} reproduce the explicit bases of $\Com_2(\MG_n)$ and $\Com_3(\MG_n)$ introduced in Ref.~\cite{Wan23}. For $k=2$ and $k=3$, the construction based on antisymmetric pairing tensors $\tilde{T}^{(\{x_{ij}\})}_{\mathbf S}$ is especially effective: in each admissible Majorana weight sector there is a unique pairing pattern, so the resulting operators form a natural orthonormal basis of the commutant.

This simple picture breaks down for $k\ge 4$. In that case, a given Majorana weight sector $\Gamma_{r_1,\dots,r_k}$ may admit several distinct pairing patterns $\{x_{ij}\}$. As a result, the corresponding operators $\tilde{T}^{(\{x_{ij}\})}$ belonging to the same sector are, in general, not mutually orthogonal. Moreover, the full family of pairing operators becomes overcomplete, so it should be regarded as a spanning set rather than as a basis of $\Com_k(\MG_n)$. An explicit example already appears for $k=4$, see Appendix~\ref{app:pairingTensorsBas4}.

\section{Algebraic construction of the matchgate commutant basis}
\label{sec:rep_theoretic_matchgate_commutant}
In principle, the family of operators $\tilde{T}^{(\{x_{ij}\})}$ obtained from antisymmetric pairing tensors can be orthogonalized, for instance by a Gram-Schmidt procedure~\cite{Cheney09linear}, at a fixed replica number $k\ge 4$ and any system size $n$. 
In practice, however, this approach quickly becomes unmanageable: within a given Majorana weight sector $\Gamma_{r_1,\dots,r_k}$, the number of linearly dependent and generally non-orthogonal operators grows rapidly with $n$. 
In other words, a direct orthogonalization is not a useful route to a general construction of an orthogonal basis of $\Com_k(\MG_n)$.
In this section, we instead exploit the representation theory of the orthogonal group and its commutant algebra to construct an orthonormal basis of the matchgate commutant $\Com_k(\MG_n)$ for arbitrary replica number $k$ and system size $n$.

\subsection{Bridge operators and the replica Lie algebra}

Throughout this section, $\gamma^{(a)}_\mu$ denotes the Majorana fermion operator acting on the
$a$th replica of the $k\ge 2$ copy Hilbert space $\mathcal H_n^{\otimes k}$. We define the
\emph{bridge operators}~\cite{Bravyi05flo}
\begin{equation}
\Lambda_{ab}:=\sum_{\mu=1}^{2n}\gamma^{(a)}_\mu\gamma^{(b)}_\mu,
\qquad
1\le a<b\le k.
\label{eq:Sab_def}
\end{equation}
These operators couple replicas $a$ and $b$ by contracting the Majorana index $\mu$ across the two copies. The operator $\Lambda$ was previously introduced to characterize Gaussian states: a state $\rho$ is a fermionic Gaussian state if and only if it satisfies~\cite{Bravyi05flo}
\begin{equation} \label{eq:Lambda_condition_mixed}
    [\Lambda, \rho \otimes \rho] = 0.
\end{equation}
As we show here, however, they play a
more fundamental role: \textit{they generate the symmetry algebra underlying the replica structure of matchgate
circuits.}

Using the canonical anticommutation relations $\{\gamma^{(a)}_\mu,\gamma^{(b)}_\nu\}=2\delta_{ab}\delta_{\mu\nu}\mathbb{1}$, one readily finds that the bridge operators close under commutation
\begin{equation}
[\Lambda_{ab},\Lambda_{cd}]
=
2\bigl(
\delta_{bc}\Lambda_{ad}
-\delta_{ac}\Lambda_{bd}
-\delta_{bd}\Lambda_{ac}
+\delta_{ad}\Lambda_{bc}
\bigr).
\label{eq:so_k_comm}
\end{equation}
Thus, the operators $\Lambda_{ab}$ furnish a representation of the orthogonal Lie algebra
$\mathfrak{so}(k)$, namely the Lie algebra of antisymmetric $k\times k$ matrices~\cite{georgi2000lie}.

We will be interested in the associative algebra generated by these operators, namely the image of the universal enveloping algebra under this representation:
\begin{equation}
\mathrm{Im}\  \mathfrak{U}(\mathfrak{so}(k))
=
\Span\!\left\{
\mathbb{1},\;
\Lambda_{a_1b_1},\;
\Lambda_{a_1b_1}\Lambda_{a_2b_2},\;
\dots
\right\}.
\label{eq:imageU}
\end{equation}
Equivalently, this is the full operator algebra on $\mathcal H_n^{\otimes k}$ generated by the bridge operators.

Importantly, as shown in Appendix~\ref{app:indu}, the algebra $\mathrm{Im}\  \mathfrak{U}(\mathfrak{so}(k))$ coincides with the span of the pairing operators
$\tilde{T}^{(\{x_{ij}\})}$. Such an $\mathfrak{so}(k)$ Lie algebra has been previously identified in the context of certain matchgate ensembles~\cite{Bao21enriched, Swann23spacetime}, but one of our main results is that it fully exhausts the commutant for arbitrary replica number:
\begin{equation}
\Com_k(\MG_n)=\mathrm{Im}\  \mathfrak{U}(\mathfrak{so}(k)).
\end{equation}
The operators $\tilde{T}^{(\{x_{ij}\})}$ provide a natural spanning set of $\Com_k(\MG_n)$, but for $k\ge 4$ this set is generally overcomplete and non-orthogonal. The Lie-algebraic formulation, in contrast, gives a more structured description of the commutant and will allow us to construct an explicit orthonormal basis for arbitrary replica number $k\geq2$, and thus to obtain the complete characterization of the matchgate commutant.

\subsection{Block structure of the commutant}

The bridge operators generate a representation of the Lie algebra $\mathfrak{so}(k)$ on the
$k$-replica Hilbert space. A basic principle of representation theory is that this action decomposes into irreducible sectors, namely invariant subspaces that cannot be decomposed further. We denote these irreducible $\mathfrak{so}(k)$ representations by $V_\nu$, where $\nu$ is the corresponding highest-weight label.

At finite $n$, only finitely many irreducible sectors appear; we denote their set by
$\mathcal I_{n,k}$. Since the matchgate commutant coincides with the image of the enveloping algebra of this $\mathfrak{so}(k)$ action, it decomposes accordingly into independent blocks
\begin{equation}
\Com_k(\MG_n)
\cong
\bigoplus_{\nu\in\mathcal I_{n,k}} \End(V_\nu).
\label{eq:block_decomp}
\end{equation}
Thus, each irreducible sector $V_\nu$ contributes one full matrix block $\End(V_\nu)$ to the commutant.

This decomposition enables us to immediately calculate the dimension of the matchgate commutant. Since, $\End(V_\nu)$ is the space of all linear operators acting on $V_\nu$, we have $\dim \End(V_\nu)=(\dim V_\nu)^2$ because $\End(V_\nu)$ is isomorphic to $V_\nu\otimes V_\nu^\star$ and the dual space $V_\nu^\star$ is isomorphic to $V_\nu$. 
Summing over all irreducible sectors, we obtain
\begin{equation}
\dim \Com_k(\MG_n)
=
\sum_{\nu\in\mathcal I_{n,k}} (\dim V_\nu)^2.
\label{eq:dim_sum_squares}
\end{equation}
Therefore, counting operators in the matchgate commutant reduces to two steps: (i) identifying which irreducible $\mathfrak{so}(k)$ representations appear in the bridge-operator representation, and (ii) computing their dimensions. 
The remaining task is constructive. For each sector $V_\nu$, we seek an explicit basis of the corresponding block $\End(V_\nu)$ written in terms of polynomials in the bridge operators $\Lambda_{ab}$.

A natural approach is to mimic the standard construction familiar from angular-momentum theory: choose a Cartan subalgebra, identify raising and lowering operators, construct a highest-weight vector in each irreducible representation, and generate the remaining states by repeated lowering. For $k\leq 4$, this strategy is sufficient, since the relevant weight spaces are multiplicity-free: each admissible set of Cartan eigenvalues identifies at most one state.

For $k\geq 5$, however, the situation changes qualitatively. The rank of $\mathfrak{so}(k)$ is at least two, and irreducible representations generically contain weight spaces of dimension larger than one. Several linearly independent states may then share the same Cartan eigenvalues, a situation known as the \textit{missing-label problem}~\cite{Cramp__2023}. 
Repeated application of lowering operators still generates the full irreducible representation, but the resulting vectors are neither uniquely labeled nor naturally orthogonal. 
Orthogonalizing them a posteriori is always possible, but it obscures the symmetry structure. 
A more systematic resolution is provided by the Gelfand--Tsetlin construction, which we develop in the following.

\subsection{Gelfand--Tsetlin basis from the subgroup chain}
\label{subsec:gt_chain}

The central idea of the Gelfand--Tsetlin (GT) construction~\cite{Gelfand50finite,zhelobenko1973compact,molev06gelfand} is very simple: instead of analyzing an irreducible sector $V_\nu$ only with respect to the full symmetry group $\mathrm{SO}(k)$, one studies how it decomposes step by step along a nested subgroup chain. 
Concretely, the subgroup chain relevant to our problem is
\begin{equation}
\mathrm{SO}(k)\supset \mathrm{SO}(k-1)\supset \mathrm{SO}(k-2)\supset \cdots \supset \mathrm{SO}(3)\supset \mathrm{SO}(2).
\label{eq:gt_chain}
\end{equation}
In our setting, the subgroup $\mathrm{SO}(m)$ is generated by the bridge operators $\Lambda_{ab}$ with $1\le a<b\le m$. Thus, for each $m$, we have a natural action of $\mathrm{SO}(m)$ on the same replica space, obtained simply by restricting attention to the first $m$ replicas.
The advantage of the chain \eqref{eq:gt_chain} is that it provides additional commuting observables. 
These extra observables refine the highest-weight labeling and resolve the degeneracies that appear for $k\ge 5$. Concretely, for each subgroup $\mathrm{SO}(m)$ we consider its Casimir operators. These are the group-theoretic analogs of the familiar operator $\mathbf J^2$ in angular-momentum theory: they commute with the whole $\mathrm{SO}(m)$ action and therefore take a fixed value on each irreducible $\mathrm{SO}(m)$ representation.
The reader may find it helpful to consult Appendix~\ref{app:GT_angular_momentum}, where the entire GT machinery is illustrated in the familiar setting of angular-momentum coupling for quantum-mechanical spins, using the chain $\mathrm{SO}(4)\supset\mathrm{SO}(3)\supset\mathrm{SO}(2)$.

For each $\mathrm{SO}(m)$, let $\mathcal C^{(m)}_j$, with $j=1,\dots,\lfloor m/2\rfloor$, denote a complete set of algebraically independent Casimir operators. The simplest example is the quadratic Casimir,
\begin{equation}
\mathcal C^{(m)}_1
=
\frac14\sum_{1\le a<b\le m}\Lambda_{ab}^2,
\qquad m=2,\dots,k,
\label{eq:quadratic_casimir}
\end{equation}
where the factor of $1/4$ compensates for the normalization chosen for the bridge operators. This operator plays the same role as total angular momentum squared ($\mathbf J^2$): it distinguishes different irreducible $\mathrm{SO}(m)$ sectors, but in general it is not enough by itself to separate all states.

More generally, the Casimirs can be constructed as invariant polynomials of the antisymmetric generator matrix~\footnote{Each generator $\Lambda_{ab}$ is Hermitian, but the matrix of operators $\mathbf\Lambda^{(m)}$ is antisymmetric in the indices, $\Lambda_{ab}=-\Lambda_{ba}$. In this sense $\mathbf\Lambda^{(m)}$ is an antisymmetric matrix whose entries are operators.} 
\begin{equation}
\mathbf\Lambda^{(m)}=(\Lambda_{ab})_{1\le a,b\le m}.
\label{eq:lambdaMAT}
\end{equation}
Because $\mathbf\Lambda^{(m)}$ is antisymmetric, all odd trace invariants vanish identically,
$\Tr\!\left[(\mathbf\Lambda^{(m)})^{2\ell+1}\right]=0,$
so only even powers can contribute. 
For odd $m=2r+1$, all primitive Casimirs may be chosen as even trace invariants
\begin{equation}
\mathcal C^{(2r+1)}_j \propto
\Tr\!\left[\left(\frac{\mathbf\Lambda^{(2r+1)}}{2}\right)^{2j}\right],
\qquad j=1,\dots,r.
\label{eq:casi1}
\end{equation}
For even $m=2r$, the first $r-1$ primitive Casimirs can again be chosen in this way,
\begin{equation}
\mathcal C^{(2r)}_j \propto
\Tr\!\left[\left(\frac{\mathbf\Lambda^{(2r)}}{2}\right)^{2j}\right],
\qquad j=1,\dots,r-1,
\label{eq:casi2}
\end{equation}
but the last one is different: it is not another trace invariant, but rather the Pfaffian of the antisymmetric generator matrix,
\begin{equation}
\mathcal C^{(2r)}_r \propto
\Pf\!\left(\frac{\mathbf\Lambda^{(2r)}}{2}\right).
\label{eq:pfafianCasimir}
\end{equation}
Thus, for $\mathfrak{so}(2r)$ the highest-degree primitive Casimir has a qualitatively different form from the lower ones.

The GT basis is obtained by diagonalizing these Casimirs simultaneously along the whole subgroup chain. More precisely, inside a fixed irreducible sector $V_\nu$, one considers the commuting family
\begin{equation}
\left\{
\mathcal C^{(k)}_j,\,
\mathcal C^{(k-1)}_j,\,
\dots,\,
\mathcal C^{(3)}_j,\,
\mathcal C^{(2)}
\right\}.
\label{eq:gt_commuting_family}
\end{equation}
A GT basis vector is then defined as a joint eigenvector of all these operators. In this way, one does not label a state only by the highest weight $\nu$ of the full $\mathrm{SO}(k)$ representation, but also by the sequence of irreducible representations that appears when restricting step by step along the chain
\begin{equation}
\mathrm{SO}(k)\downarrow \mathrm{SO}(k-1)\downarrow \cdots \downarrow \mathrm{SO}(2).
\label{eq:SOchain}
\end{equation}
Here, the symbol $\downarrow$ denotes restriction of representations along the subgroup chain. 
That is, an irreducible representation of $\mathrm{SO}(k)$, when viewed as a representation of the subgroup $\mathrm{SO}(k-1)\subset \mathrm{SO}(k)$, generally decomposes into a direct sum of irreducible $\mathrm{SO}(k-1)$ representations, and the same procedure is then iterated down to $\mathrm{SO}(2)$.

This sequence is encoded by an orthogonal GT pattern. Physically, the GT pattern supplies the additional quantum numbers needed to distinguish states that would otherwise share the same Cartan eigenvalues. 
For the purpose of the matchgate commutant basis determination, this construction has three important advantages. First, it provides a canonical symmetry-adapted basis in each irreducible sector $V_\nu$, and therefore in each block $\End(V_\nu)$ of the commutant. Second, it resolves the missing-label problem by supplementing the highest-weight label with the intermediate subgroup labels. Third, because GT basis vectors are joint eigenvectors of a commuting family of Hermitian operators, the resulting basis operators in the matchgate commutant are automatically mutually orthogonal.

\subsection{Constructing explicit operators in the commutant}
\label{subsec:gt_basis_elems}

We now turn the preceding representation-theoretic discussion into an explicit construction of operators in the commutant $\Com_k(\MG_n)$. The construction proceeds in two stages: first, we build \emph{diagonal} operators (projectors) that select individual GT sectors within each irreducible block, and then we build \emph{off-diagonal} operators that connect different GT sectors and complete the basis.

\paragraph{Step 1: Diagonal projectors.}
For each irreducible sector $V_\nu$, the GT construction provides a family of projectors
\begin{equation}
P_{\nu,\mathfrak m},
\qquad
\mathfrak m\in \GT(\nu),
\label{eq:GT_projectors}
\end{equation}
where $\mathfrak m$ labels a GT pattern---equivalently, a complete set of intermediate subgroup labels in the chain~\eqref{eq:SOchain}---and therefore specifies a one-dimensional GT sector inside $\End(V_\nu)$. Because these projectors correspond to distinct joint eigenspaces of the commuting Casimir family~\eqref{eq:gt_commuting_family}, they are mutually orthogonal:
\begin{equation}
P_{\nu,\mathfrak m}P_{\nu',\mathfrak n}
=
\delta_{\nu\nu'}\delta_{\mathfrak m\mathfrak n}\,P_{\nu,\mathfrak m}.
\label{eq:GT_projector_relations}
\end{equation}
Concretely, each projector can be realized as a polynomial in the commuting Casimir operators along the subgroup chain, so they are explicitly constructible once the Casimirs are known. Together, they resolve each block $\End(V_\nu)$ into one-dimensional sectors adapted to the GT decomposition.

\paragraph{Step 2: Off-diagonal operators.}
The projectors alone span only the diagonal part of each block $\End(V_\nu)$. To obtain the full block, one must also construct off-diagonal operators connecting different GT labels $\mathfrak m \neq \mathfrak n$ within the same irreducible representation (\textit{irrep} for short).
The key idea is to use the \emph{ladder operators} of the nested algebras $\mathfrak{so}(m)$ along the GT chain~\eqref{eq:gt_chain} (for a self-contained illustration of this procedure in the familiar setting of angular momentum, see Appendix~\ref{app:GT_angular_momentum}). For each $\mathfrak{so}(m)$, choosing a Cartan subalgebra splits the remaining generators into raising and lowering operators that shift weight labels by elementary steps. Since a GT pattern records the full sequence of intermediate labels, any two patterns $\mathfrak m$ and $\mathfrak n$ inside the same irrep $V_\nu$ can be connected by a suitable ordered product $Y^{(\nu)}_{\mathfrak m,\mathfrak n}$ of such ladder operators. Sandwiching this product between the corresponding projectors yields
\begin{equation}
X^{(\nu)}_{\mathfrak m,\mathfrak n}
\;\propto\;
P_{\nu,\mathfrak m}\,
Y^{(\nu)}_{\mathfrak m,\mathfrak n}\,
P_{\nu,\mathfrak n}.
\label{eq:Xmn_sandwich}
\end{equation}
Whenever the right-hand side is nonzero, $X^{(\nu)}_{\mathfrak m,\mathfrak n}$ maps the GT sector $\mathfrak n$ to the sector $\mathfrak m$ and annihilates all others. After fixing the overall normalization, these operators furnish the matrix units spanning the full block $\End(V_\nu)$. Their mutual Hilbert--Schmidt orthogonality is guaranteed because they connect orthogonal GT sectors: if either the source or the target labels differ, the inner product vanishes.

\paragraph{The complete commutant basis.}
Combining both diagonal and off-diagonal operators across all sectors, we arrive at
\begin{equation}
\Com_k(\MG_n)
=
\Span\!\left\{
X^{(\nu)}_{\mathfrak m,\mathfrak n}
\;:\;
\nu\in\mathcal I_{n,k},\;
\mathfrak m,\mathfrak n\in\GT(\nu)
\right\}.
\label{eq:commutant_transition_basis}
\end{equation}
This converts the abstract block decomposition~\eqref{eq:block_decomp} into an explicit operator basis adapted to the GT decomposition. Constructing such a basis therefore reduces to two tasks: identifying the set $\mathcal I_{n,k}$ of irreducible $\mathfrak{so}(k)$ sectors present at finite $n$, and constructing the associated GT basis within each sector.

In practice, the procedure consists of four steps: (i) organize the bridge operators $\Lambda_{ab}$ into a convenient realization of $\mathfrak{so}(k)$; (ii) isolate each irreducible sector $V_\nu$ using the Casimirs of the full $\mathrm{SO}(k)$ action; (iii) resolve the GT sectors inside $V_\nu$ by simultaneously diagonalizing the commuting family associated with the subgroup chain~\eqref{eq:gt_chain}; and (iv) construct the off-diagonal operators $X^{(\nu)}_{\mathfrak m,\mathfrak n}$ that complete the block $\End(V_\nu)$. 
Before carrying out this program explicitly for a few representative replica orders $k$, we first show in the next subsection how the representation-theoretic data identified above, the set of admissible highest weights $\mathcal I_{n,k}$ and the dimensions of the corresponding irreducible sectors, already suffice to determine the full dimension of the matchgate commutant.

\subsection{Dimension of the matchgate commutant from irreducible sectors}

The block decomposition~\eqref{eq:block_decomp}
reduces the problem of finding $\dim \Com_k(\MG_n)$ to a purely representation-theoretic counting: for each irreducible
$\mathfrak{so}(k)$ sector $V_\nu$ that appears in the bridge representation, we need its dimension $\dim V_\nu$. This is provided by the Weyl dimension formula~\cite{fulton1991representation}.
Moreover, we need to enumerate all admissible highest weight sectors $\mathcal I_{n,k}$.

For $\mathfrak{so}(k)$, the set $\mathcal I_{n,k}$ is fixed by the parity of $k$. Writing $k=2r$ ($k=2r+1$) for even (odd) $k$, the rank of the algebra is $r$, and an irreducible representation is labeled by a dominant
highest weight
\begin{equation}
\nu=(\nu_1,\dots,\nu_r).
\end{equation}
In our finite-$n$ setting, the allowed highest weights $\nu_j$ are integers that satisfy
\begin{equation}
n\ge \nu_1\ge \nu_2\ge \cdots \ge \nu_r\ge 0,
\qquad k=2r+1,
\label{eq:highest_weights_Br}
\end{equation}
for the odd orthogonal algebra, and
\begin{equation}
n\ge \nu_1\ge \nu_2\ge \cdots \ge |\nu_r|,
\qquad k=2r,
\label{eq:highest_weights_Dr}
\end{equation}
for the even orthogonal algebra.

Let $\mathfrak g$ be a simple Lie algebra of rank $r$, and let $\Phi^+$ denote a choice of
positive roots. In the standard realization of the classical Lie algebras, the roots are
identified with vectors in $\mathbb R^r$ equipped with the Euclidean inner product. For a
dominant highest weight $\nu=(\nu_1,\dots,\nu_r)$, the Weyl dimension formula reads
\begin{equation}
\dim V_\nu
=
\prod_{\alpha\in\Phi^+}
\frac{\langle \nu+\rho,\alpha\rangle}{\langle \rho,\alpha\rangle},
\qquad
\rho:=\frac12\sum_{\alpha\in\Phi^+}\alpha,
\label{eq:weyl_dim_general}
\end{equation}
where $\rho$ is the Weyl vector.
For the orthogonal algebras relevant here, Eq.~\eqref{eq:weyl_dim_general} reduces to compact product formulas, fixed by the Weyl vector and sets of the roots $\Phi^+$, provided explicitly in Appendix~\ref{app:Weylfromulas}.

Finally, the problem of counting operators in the matchgate commutant reduces to two pieces of data: the set $\mathcal I_{n,k}$ of allowed highest weights [cf. Eq.~\eqref{eq:highest_weights_Br} and \eqref{eq:highest_weights_Dr}], and the dimensions of the corresponding irreducible sectors $\dim V_\nu$ [cf. Eq.~\eqref{eq:weyl_Dr_final} and \eqref{eq:weyl_Br}]. These ingredients, through Eq.~\eqref{eq:dim_sum_squares}, specify the dimension of the matchgate commutant $\dim( \Com_k(\MG_n) )$ at any system size $n$ and replica number $k\geq 2$. 

To evaluate the sum in Eq.~\eqref{eq:dim_sum_squares} explicitly, we reduce the Weyl dimension formulas to Vandermonde-type determinants, and note that the sum over dominant weights reduces to a discrete Hankel determinant.
This is a standard mechanism in determinant calculus and orthogonal polynomial theory, see, \textit{e.g.},~Ref.~\cite{Krattenthaler1999, deBruijn1955}.
In Appendix~\ref{app:matchgateDIM}, we detail this calculation, which yields the following compact formula for the matchgate commutant dimension at any system size $n \geq 1$ and replica number $k \geq2$
\begin{equation}
\dim \Com_k(\MG_n) = \prod_{1 \le i \le j \le k-1} \frac{2n+i+j-1}{i+j-1}.
\label{eq:rmt_tableau_prod2}
\end{equation}
An alternative viewpoint on this result is provided in Appendix~\ref{app:RMT}, where we provide a random matrix theory derivation of the matchgate commutant dimension.
In the following explicit constructions, we obtain variants of this formula at fixed $2\leq k \leq 5$.

\subsection{Explicit construction of the commutant basis at fixed $k$}
\label{sec:fixk}

In the following, we apply the general framework developed above to construct the matchgate
commutant basis explicitly for the first few nontrivial cases, $k=2,3,4,5$. These examples
illustrate how the bridge-algebra and GT constructions become concrete at fixed
replica number $k$.

\subsubsection{$k=2$}

We start with the simplest nontrivial case of two replicas and make the GT structure fully explicit. For
$k=2$, the subgroup chain \eqref{eq:gt_chain} terminates immediately at $\mathrm{SO}(2),$
so the GT construction involves only a single step. Correspondingly, there is only one bridge
operator,
\begin{equation}
\Lambda_{12}=\sum_{\mu=1}^{2n}\gamma^{(1)}_\mu\gamma^{(2)}_\mu,
\label{eq:Lambda12_def}
\end{equation}
which generates the Abelian Lie algebra $\mathfrak{so}(2)$. In this case, the commuting GT family
contains only the operator associated with this single $\mathrm{SO}(2)$ step, and the GT basis is
therefore just the eigenbasis of the Cartan generator
\begin{equation}
H=\frac{i}{2}\Lambda_{12}.
\end{equation}

The key simplification enabling finding the eigenbasis of $H$ is that the individual terms
$i\,\gamma^{(1)}_\mu\gamma^{(2)}_\mu$ entering $i\Lambda_{12}$ are Hermitian, mutually
commuting, and square to the identity. Hence they may be simultaneously diagonalized, and each
contributes an eigenvalue $\pm1$. Since $i\Lambda_{12}$ is their sum, its spectrum is
\begin{equation}
\Spec(i\Lambda_{12})=\{\,2r-2n:\ r=0,\dots,2n\,\}.
\label{eq:k2_spectrum}
\end{equation}
Equivalently, the Cartan generator $H$ has eigenvalues
\begin{equation}
\Spec(H)=\{-n,-n+1,\dots,n\}.
\end{equation}

This immediately identifies the irreducible sectors. Since $\mathfrak{so}(2)$ is Abelian, all
its irreducible representations are one-dimensional and are labeled by the eigenvalue $\nu$ of
$H$. Thus the admissible highest weights are
\begin{equation}
\mathcal I_{n,2}=\{-n,-n+1,\dots,n\},
\label{eq:k2_allowed_weights}
\end{equation}
in agreement with the general finite-$n$ truncation discussed above, cf.
Eq.~\eqref{eq:highest_weights_Dr}.
Because the subgroup chain has only one step, there are no nontrivial intermediate subgroup labels. Accordingly, for each $\nu\in\mathcal I_{n,2}$, the corresponding GT set contains a single element,
\begin{equation}
\GT(\nu)=\{\varnothing\},
\end{equation}
so the GT label $\mathfrak m$ is trivial and may be omitted. Consequently, for $k=2$ there are no off-diagonal operators $X^{(\nu)}_{\mathfrak m,\mathfrak n}$, cf.~\eqref{eq:Xmn_sandwich}, and the matchgate commutant $\Com_2(\MG_n)$ is spanned solely by the orthogonal projectors $P_{\nu}$~\eqref{eq:GT_projectors}.

Since the eigenvalues in Eq.~\eqref{eq:k2_spectrum} are all distinct, the projector onto the sector
with highest weight $\nu=r-n$ is simply the corresponding spectral projector of $i\Lambda_{12}$,
\begin{equation}
P_{\nu}=
\prod_{\substack{s=0\\ s\neq r}}^{2n}
\frac{i\Lambda_{12}-(2s-2n)}{2(r-s)},
\qquad
r=0,\dots,2n.
\label{eq:k2_projectors}
\end{equation}
These projectors satisfy
\begin{equation}
P_\nu P_{\nu'}=\delta_{\nu\nu'}P_\nu,
\end{equation}
and therefore form the GT basis of the $k=2$ commutant. Equivalently, they are the minimal
idempotents in the polynomial algebra generated by $\Lambda_{12}$. This basis of $\Com_2(\MG_n)$ is distinct from the basis generated by the pairing
operators, cf. Eq.~\eqref{eq:antiK2}.

From the representation-theoretic viewpoint, each projector $P_\nu$ selects one
one-dimensional irreducible $\mathfrak{so}(2)$ sector. Since all such sectors have dimension one, the general counting formula reduces to
\begin{equation}
\begin{split}
    \dim \Com_2(\MG_n)
&=\sum_{\nu\in\mathcal I_{n,2}}\! (\dim V_\nu)^2=\sum_{\nu=-n}^{n}1=2n+1.
\end{split}
\end{equation}
Summarizing, for $k=2$, the  GT decomposition becomes completely explicit: the subgroup chain has only one step, the irreducible sectors are labeled by the $\mathfrak{so}(2)$ weights $\nu=-n,\dots,n$, each sector carries a trivial GT pattern, and the commutant is spanned by the spectral projectors of the single bridge operator $\Lambda_{12}$.

\subsubsection{$k=3$}
The case $k=3$ is the first one in which the bridge algebra becomes non-Abelian. The three
bridge operators $\Lambda_{ab}$, with $1\le a<b\le 3$, furnish a representation of
$\mathfrak{so}(3)$. A convenient basis is
\begin{equation}
M_1:=\tfrac12\Lambda_{23},\qquad
M_2:=-\tfrac12\Lambda_{13},\qquad
M_3:=\tfrac12\Lambda_{12},
\label{eq:k3_Mi_def}
\end{equation}
for which the standard $\mathfrak{so}(3)$ commutation relations hold.

In the notation of the previous sections, $k=3$ corresponds to the rank-one algebra
$B_1\simeq \mathfrak{so}(3)$. The admissible highest weights are therefore
\begin{equation}
\mathcal I_{n,3}=\{0,1,\dots,n\},
\label{eq:k3_allowed_weights}
\end{equation}
in agreement with Eq.~\eqref{eq:highest_weights_Br}. The Weyl dimension formula, see Appendix~\ref{app:Weylfromulas}, reduces to
\begin{equation}
\dim V_\nu=2\nu+1.
\end{equation}
Consequently, the counting formula~\eqref{eq:dim_sum_squares} gives
\begin{equation}
\dim \Com_3(\MG_n)
=
\sum_{\nu\in\mathcal I_{n,3}}(\dim V_\nu)^2
=
\binom{2n+3}{3},
\label{eq:k3_dim_comm}
\end{equation}
consistently with the commutant dimension obtained in
Sec.~\ref{subsec:k3anti} by counting antisymmetric pairing tensors.

For $k=3$, the GT chain~\eqref{eq:gt_chain} reduces to the single nontrivial restriction
\begin{equation}
\mathrm{SO}(3)\downarrow \mathrm{SO}(2).
\label{eq:k3_GT_chain}
\end{equation}
Thus, for each fixed irrep $V_\nu$, a GT pattern consists only of the final
$\mathrm{SO}(2)$ weight. In our notation, the highest-weight label is still a single integer $\nu$, while the GT label is
\begin{equation}
\begin{split}
   \mathfrak m=m,\quad
m=-\nu,-\nu+1,\dots,\nu. 
\end{split}
\end{equation}
Hence
\begin{equation}
\begin{split}
    \GT(\nu)&=\{-\nu,-\nu+1,\dots,\nu\},
\\
\#\GT(\nu)&=2\nu+1=\dim V_\nu.
\end{split}
\label{eq:k3_GT_set}
\end{equation}
Since $k=3$ is rank one, the pair $(\nu,\mathfrak m):=(\nu,m)$ already provides a complete
set of labels.

To isolate the irreducible sectors explicitly, we use the quadratic Casimir of the
$\mathrm{SO}(3)$ action. This is the $m=3$, $j=1$ instance of the general family
$\mathcal C^{(m)}_j$ introduced in Eq.~\eqref{eq:casi1}, 
\begin{equation}
\mathcal C^{(3)}_1
=
M_1^2+M_2^2+M_3^2
=
\tfrac14\bigl(\Lambda_{12}^2+\Lambda_{13}^2+\Lambda_{23}^2\bigr) := K,
\label{eq:k3_Casimir}
\end{equation}
which commutes with the full $\mathrm{SO}(3)$ action. In our anti-Hermitian convention, its
eigenvalues are $-\nu(\nu+1)$, with $\nu=0,1,\dots,n$. The corresponding projectors onto the
irreducible sectors $V_\nu$ are
\begin{equation}
P_\nu
=
\prod_{\substack{\nu'=0\\ \nu'\neq \nu}}^{n}
\frac{K+\nu'(\nu'+1)}{-\nu(\nu+1)+\nu'(\nu'+1)}.
\label{eq:k3_projectors}
\end{equation}
Inside the image of $P_\nu$, the Cartan generator $M_3$ resolves the basis further according to
the $\mathrm{SO}(2)$ weights $m\in\GT(\nu)$. Since $M_3$ has eigenvalues $i\,m$, the
corresponding one-dimensional GT projectors, cf. Eq.~\eqref{eq:GT_projectors}, are
\begin{equation}
P_{\nu,\mathfrak m}
:=
P_{\nu,m}
=
P_\nu
\prod_{\substack{m'=-\nu\\ m'\neq m}}^{\nu}
\frac{M_3-i\,m'}{i(m-m')}.
\label{eq:k3_GT_projectors}
\end{equation}
These projectors resolve the block $\End(V_\nu)$ into its one-dimensional GT sectors.

The full block $\End(V_\nu)$ is then spanned by th GT transition operators, cf. Eq.~\eqref{eq:Xmn_sandwich}, 
\begin{equation}
\begin{split}
    X^{(\nu)}_{\mathfrak m,\mathfrak n}
&:=
X^{(\nu)}_{m,m'},
\\
\mathfrak m=m,\ \mathfrak n=m',
&\quad 
m,m'\in\GT(\nu),
\end{split}
\label{eq:k3_transition_ops}
\end{equation}
which connect the GT sector labeled by $m'$ to the one labeled by $m$. The
diagonal operators $X^{(\nu)}_{\mathfrak m,\mathfrak m}$ coincide with the one-dimensional GT projectors $P_{\nu,\mathfrak m}$, while the off-diagonal operators $X^{(\nu)}_{\mathfrak m,\mathfrak n}$, with
$\mathfrak m\neq \mathfrak n$, act as transition operators between distinct $\mathrm{SO}(2)$
weight sectors inside the same $\mathrm{SO}(3)$ irrep.

A convenient explicit choice for the off-diagonal operators is obtained by sandwiching suitable powers of the ladder operators
between the one-dimensional GT projectors:
\begin{equation}
X^{(\nu)}_{m,m'}
=
\mathcal N_{\nu;m,m'}\,
P_{\nu,m}\,
 M_{m,m'}\,
P_{\nu,m'},
\label{eq:k3_transition_explicit}
\end{equation}
where
\begin{equation}
 M_{m,m'}
:=
\begin{cases}
M_+^{\,m-m'}, & m\ge m',\\[1mm]
M_-^{\,m'-m}, & m<m',
\end{cases}
\qquad
M_\pm:=M_1\pm iM_2.
\label{eq:k3_Mmmprime}
\end{equation}
The normalization factor may be chosen as
\begin{equation}
\mathcal N_{\nu;m,m'}
=
\sqrt{
\frac{(\nu-m_>)!\,(\nu+m_<)!}
     {(\nu-m_<)!\,(\nu+m_>)!}
},
\label{eq:k3_transition_norm_explicit}
\end{equation}
where
\begin{equation}
m_>:=\max(m,m'),
\qquad
m_<:=\min(m,m').
\end{equation}
With this choice, the operators are orthonormal with respect to the Hilbert--Schmidt inner
product,
\begin{equation}
\Tr\!\left[
\bigl(X^{(\nu)}_{\mathfrak m,\mathfrak n}\bigr)^\dagger
X^{(\nu')}_{\mathfrak m',\mathfrak n'}
\right]
=
\delta_{\nu,\nu'}\,
\delta_{\mathfrak m,\mathfrak m'}\,
\delta_{\mathfrak n,\mathfrak n'}.
\label{eq:k3_orthonormality}
\end{equation}

Collecting all irreducible sectors, we obtain the explicit GT-adapted basis
\begin{equation}
\!\Com_3(\MG_n)
=
\Span\!\left\{
X^{(\nu)}_{\mathfrak m,\mathfrak n}
\,:\,
\nu\in\mathcal I_{n,3},\;
\mathfrak m,\mathfrak n\in\GT(\nu)
\right\}.
\label{eq:k3_basis}
\end{equation}
The total number of operators $X^{(\nu)}_{\mathfrak m,\mathfrak n}$ is given by
Eq.~\eqref{eq:k3_dim_comm}.

Thus, for $k=3$, the GT construction takes a particularly simple form. The highest-weight label
$\nu$ selects the $\mathrm{SO}(3)$ irrep, while the GT labels
$\mathfrak m,\mathfrak n\in\GT(\nu)$ reduce to the final $\mathrm{SO}(2)$ weights
$m,m'=-\nu,\dots,\nu$. Via the isomorphism
$\mathfrak{so}(3)\simeq \mathfrak{su}(2)$, the same construction can be rephrased in the
language of standard angular-momentum theory; we keep the discussion in the $\mathrm{SO}$
framework for uniformity with the higher-$k$ cases.

\subsubsection{$k=4$}

The case $k=4$ is the first one in which the symmetry algebra has rank two. The six bridge
operators $\Lambda_{ab}$, with $1\le a<b\le 4$, furnish a representation of
$\mathfrak{so}(4)$. In the notation of the previous sections, the admissible highest weights are
\begin{equation}
\mathcal I_{n,4}
=
\left\{
\nu=(\nu_1,\nu_2):
\ n\ge \nu_1\ge |\nu_2|,
\quad
\nu_1,\nu_2\in\mathbb Z
\right\},
\label{eq:k4_allowed_weights}
\end{equation}
in agreement with Eq.~\eqref{eq:highest_weights_Dr}. For each $\nu\in\mathcal I_{n,4}$, the
Weyl dimension formula for $D_2\simeq \mathfrak{so}(4)$, see Appendix~\ref{app:Weylfromulas},
reduces to
\begin{equation}
\dim V_\nu
=
(\nu_1-\nu_2+1)(\nu_1+\nu_2+1).
\label{eq:k4_irrep_dim}
\end{equation}
Consequently, the general counting formula gives
\begin{equation}
\begin{split}
    \dim \Com_4(\MG_n)
&=
\sum_{\nu\in\mathcal I_{n,4}}(\dim V_\nu)^2\\
&=
\binom{2n+6}{6}+\binom{2n+5}{6}.
\end{split}
\label{eq:k4_dim_closed}
\end{equation}

For $k=4$, the GT chain takes the form
\begin{equation}
\mathrm{SO}(4)\downarrow \mathrm{SO}(3)\downarrow \mathrm{SO}(2).
\label{eq:k4_GT_chain}
\end{equation}
Accordingly, the construction proceeds in three steps, corresponding to the highest-weight label
$\nu$, the intermediate $\mathrm{SO}(3)$ label $s$, and the final $\mathrm{SO}(2)$ weight $m$.

The first step is to isolate the irreducible $\mathrm{SO}(4)$ sector $V_\nu$. For this, we use
the two primitive Casimirs of $\mathfrak{so}(4)$, i.e. the specialization of
Eqs.~\eqref{eq:quadratic_casimir} and \eqref{eq:pfafianCasimir} to $m=4$. The first is the
quadratic Casimir
\begin{equation}
\mathcal C^{(4)}_1
:=
\frac14\sum_{1\le a<b\le 4}\Lambda_{ab}^2,
\label{eq:k4_C2}
\end{equation}
and the second is the Pfaffian-type Casimir
\begin{equation}
\mathcal C^{(4)}_{2}
:=
\Pf\!\left(\frac{\mathbf\Lambda^{(4)}}{2}\right)
=
\frac18\Bigl(
\Lambda_{12}\Lambda_{34}
-\Lambda_{13}\Lambda_{24}
+\Lambda_{14}\Lambda_{23}
\Bigr),
\label{eq:k4_CPf}
\end{equation}
where $\mathbf\Lambda^{(4)}$ is the $4\times4$ antisymmetric generator matrix introduced above.
With our anti-Hermitian convention, their eigenvalues on $V_\nu$ are
\begin{equation}
c^{(4)}_1(\nu)
=
-\bigl[\nu_1(\nu_1+2)+\nu_2^2\bigr],
\quad
c^{(4)}_2(\nu)
=
\nu_2(\nu_1+1).
\label{eq:k4_Casimir_eigenvalues}
\end{equation}
Since the quadratic Casimir alone does not always separate all irreducible sectors, we first
project onto the eigenspace of $\mathcal C^{(4)}_1$,
\begin{equation}
P^{(1)}_\nu
=
\prod_{\substack{c \in \{c^{(4)}_1(\nu') \,:\, \nu' \in \mathcal I_{n,4}\}\\ c \neq c^{(4)}_1(\nu)}}
\frac{\mathcal C^{(4)}_1-c}{c^{(4)}_1(\nu)-c},
\label{eq:k4_projector_C2}
\end{equation}
where the product runs over all distinct eigenvalues of $\mathcal C^{(4)}_1$ present in the
allowed set. Any remaining accidental degeneracies are then resolved by $\mathcal C^{(4)}_2$,
which yields the full projector onto the $\nu$-block,
\begin{equation}
P_\nu
=
P^{(1)}_\nu
\prod_{\substack{\nu'\in\mathcal I_{n,4}\\ c^{(4)}_1(\nu') = c^{(4)}_1(\nu)\\ \nu'\neq \nu}}
\frac{\mathcal C^{(4)}_{2}-c^{(4)}_{2}(\nu')}{c^{(4)}_{2}(\nu)-c^{(4)}_{2}(\nu')}.
\label{eq:k4_projector_nu}
\end{equation}
This completes the first GT step: the highest-weight label $\nu$ is fixed.

The second GT step is the restriction
$\mathrm{SO}(4)\downarrow \mathrm{SO}(3)$. The branching is multiplicity-free and takes the form
\begin{equation}
V_\nu\downarrow \mathrm{SO}(3)
\simeq
\bigoplus_{s=|\nu_2|}^{\nu_1} V_s,
\label{eq:k4_branching_SO4_SO3}
\end{equation}
where $s$ increases in unit steps. To isolate the corresponding $\mathrm{SO}(3)$ sectors, we use
the quadratic Casimir of the $\mathrm{SO}(3)$ subgroup acting on the first three replicas.
Introducing
\begin{equation}
M_1:=\tfrac12\Lambda_{23},\qquad
M_2:=-\tfrac12\Lambda_{13},\qquad
M_3:=\tfrac12\Lambda_{12},
\label{eq:k4_Mi}
\end{equation}
we obtain the standard $\mathfrak{so}(3)$ commutation relations. The associated quadratic Casimir
is the specialization of \eqref{eq:quadratic_casimir} to $m=3$,
\begin{equation}
K^{(3)}
:=
M_1^2+M_2^2+M_3^2
=
\tfrac14\bigl(\Lambda_{12}^2+\Lambda_{13}^2+\Lambda_{23}^2\bigr),
\label{eq:k4_K3}
\end{equation}
with eigenvalues $-s(s+1)$. Accordingly, inside the image of $P_\nu$, the projector onto the
$\mathrm{SO}(3)$ irrep labeled by $s$ is
\begin{equation}
P_{\nu,s}
=
P_\nu
\prod_{\substack{s'=|\nu_2|\\ s'\neq s}}^{\nu_1}
\frac{K^{(3)}+s'(s'+1)}{-s(s+1)+s'(s'+1)}.
\label{eq:k4_projector_s}
\end{equation}

The third and final GT step is the restriction
$\mathrm{SO}(3)\downarrow \mathrm{SO}(2)$. Each $\mathrm{SO}(3)$ irrep $V_s$ is resolved by the label
\begin{equation}
m=-s,-s+1,\dots,s,
\end{equation}
determining the eigenvalue of the Cartan generator $M_3$.
Thus, for fixed $\nu$, the GT set is
\begin{equation}
\!\!\GT(\nu)
=
\left\{
\mathfrak m=(s,m):
\!s=|\nu_2|,\dots,\nu_1;
\ 
m=-s,\dots,s
\right\}.
\label{eq:k4_GT_set}
\end{equation}
The corresponding count is
\begin{equation}
\begin{split}
    \#\GT(\nu)
&=
\sum_{s=|\nu_2|}^{\nu_1}(2s+1)\\&
=
(\nu_1-|\nu_2|+1)(\nu_1+|\nu_2|+1)
=
\dim V_\nu,
\end{split}
\label{eq:k4_GT_count}
\end{equation}
in agreement with Eq.~\eqref{eq:k4_irrep_dim}. 
Therefore the one-dimensional GT projectors, cf. Eq.~\eqref{eq:GT_projectors}, are
\begin{equation}
P_{\nu,\mathfrak m}
:=
P_{\nu,s,m}
=
P_{\nu,s}
\prod_{\substack{m'=-s\\ m'\neq m}}^{s}
\frac{M_3-i\,m'}{i(m-m')},
\qquad
\mathfrak m=(s,m).
\label{eq:k4_GT_projectors}
\end{equation}
This completes the GT construction of the projectors on the $\End(V_\nu)$ block.

The block $\End(V_\nu)$ is then spanned by the GT transition operators
\begin{equation}
X^{(\nu)}_{\mathfrak m,\mathfrak n},
\qquad
\mathfrak m,\mathfrak n\in\GT(\nu),
\label{eq:k4_transition_ops}
\end{equation}
defined as in Eq.~\eqref{eq:Xmn_sandwich}. The diagonal operators
$X^{(\nu)}_{\mathfrak m,\mathfrak m}$ are precisely the one-dimensional GT projectors $P_{\nu,\mathfrak m}$, while the
off-diagonal operators connect distinct GT sectors inside the same $\nu$-block.

For $k=4$, these off-diagonal GT operators may be constructed explicitly from the one-dimensional
GT projectors. At fixed $s$, transitions between different $\mathrm{SO}(2)$ weights are obtained by
sandwiching powers of the ladder operators $M_\pm=M_1\pm iM_2$ between the projectors
$P_{\nu,s,m}$. Moreover, to connect different values of $s$, one uses the rank-one $\mathrm{SO}(3)$ tensor
built from the bridges involving the fourth replica,
\begin{equation}
T_{\pm 1}=\tfrac12(\pm\Lambda_{14}+i\Lambda_{24}),\,\,\,\,
T_{0}=\tfrac12\Lambda_{34},\,\,\,\,
\label{eq:Tops}
\end{equation}
and sandwiches these operators between the corresponding GT projectors $P_{\nu,s,m}$. In this
way, all
$X^{(\nu)}_{\mathfrak m,\mathfrak n}$ may be generated explicitly from the bridge algebra; see
Appendix~\ref{app:k4_detailsT} for details.

Collecting all irreducible sectors, we obtain the explicit GT-adapted basis
\begin{equation}
\Com_4(\MG_n)
=
\Span\!\left\{
X^{(\nu)}_{\mathfrak m,\mathfrak n}
\;:\;
\nu\in\mathcal I_{n,4},\;
\mathfrak m,\mathfrak n\in \GT(\nu)
\right\},
\label{eq:k4_basis}
\end{equation}
consisting of $\dim \Com_4(\MG_n)$, cf. Eq.~\eqref{eq:k4_dim_closed}, orthogonal operators
$X^{(\nu)}_{\mathfrak m,\mathfrak n}$.

Summarizing, for $k=4$, the GT construction resolves the commutant in three stages. The highest-weight
label $\nu=(\nu_1,\nu_2)$ selects the $\mathrm{SO}(4)$ irrep, the intermediate label $s$ selects
the $\mathrm{SO}(3)$ sector in the multiplicity-free branching
$\mathrm{SO}(4)\downarrow \mathrm{SO}(3)$, and the final label $m$ resolves the
$\mathrm{SO}(2)$ weight. Equivalently, the GT labels are $\mathfrak m=(s,m)$. 
We conclude remarking the existence of a special $\mathfrak{so}(4)\simeq \mathfrak{su}(2)_L\oplus \mathfrak{su}(2)_R$ which provide an alternative description. For uniformity of the presentation, we decided to keep the discussion entirely within the GT framework.

\subsubsection{$k=5$}

We finally consider the first genuinely rank-two odd orthogonal case. The ten bridge operators
$\Lambda_{ab}$, with $1\le a<b\le 5$, furnish a representation of
$\mathfrak{so}(5)\simeq B_2$. In the notation of the previous sections, the admissible highest
weights are
\begin{equation}
\mathcal I_{n,5}
=
\left\{
\nu=(\nu_1,\nu_2):
\ n\ge \nu_1\ge \nu_2\ge 0,\ 
\nu_1,\nu_2\in\mathbb Z
\right\},
\label{eq:k5_allowed_weights}
\end{equation}
in agreement with Eq.~\eqref{eq:highest_weights_Br}. For each such $\nu$, the Weyl dimension
formula for $B_2$, see Appendix~\ref{app:Weylfromulas}, reduces to
\begin{equation}
\dim V_\nu
=
\frac{1}{6}
(2\nu_1+3)(2\nu_2+1)(\nu_1-\nu_2+1)(\nu_1+\nu_2+2).
\label{eq:k5_irrep_dim}
\end{equation}
The general counting formula \eqref{eq:dim_sum_squares} then yields
\begin{equation}
\dim \Com_5(\MG_n)
=
\frac{(n+2)(2n+3)(2n+5)}{30}\binom{2n+7}{7},
\label{eq:k5_dim_closed}
\end{equation}
consistently with Eq.~\eqref{eq:rmt_tableau_prod2} for $k=5$.

For $k=5$, the GT chain is
\begin{equation}
\mathrm{SO}(5)\downarrow \mathrm{SO}(4)\downarrow \mathrm{SO}(3)\downarrow \mathrm{SO}(2).
\label{eq:k5_GT_chain}
\end{equation}
Accordingly, the GT construction proceeds in four steps, corresponding to the highest-weight label
$\nu$, the intermediate $\mathrm{SO}(4)$ highest weight $\mu$, the $\mathrm{SO}(3)$ spin $s$, and
the final $\mathrm{SO}(2)$ weight $m$.

The first step is to isolate the irreducible $\mathrm{SO}(5)$ sector $V_\nu$. Since
$\mathfrak{so}(5)\simeq B_2$ has rank two, two primitive Casimirs are needed. A convenient
choice is the quadratic Casimir, i.e. the specialization of \eqref{eq:quadratic_casimir} to
$m=5$,
\begin{equation}
\mathcal C^{(5)}_1
:=
\frac14\sum_{1\le a<b\le 5}\Lambda_{ab}^2,
\label{eq:k5_C2}
\end{equation}
together with a quartic Casimir $\mathcal C^{(5)}_2$, cf. Eq.~\eqref{eq:casi1}, normalized so that its eigenvalue on
$V_\nu$ is
\begin{equation}
c^{(5)}_2(\nu)
=
\left(\nu_1+\frac32\right)^2
\left(\nu_2+\frac12\right)^2
-\frac{9}{16},
\label{eq:k5_c4_eigenvalue}
\end{equation}
while the quadratic Casimir has eigenvalue
\begin{equation}
c^{(5)}_1(\nu)
=
-\bigl[\nu_1(\nu_1+3)+\nu_2(\nu_2+1)\bigr].
\label{eq:k5_c2_eigenvalue}
\end{equation}
The ordered pair $\bigl(c^{(5)}_1(\nu),c^{(5)}_2(\nu)\bigr)$ uniquely determines
$\nu\in\mathcal I_{n,5}$. To avoid possible degeneracies of the quadratic Casimir, we first
project onto its eigenspace,
\begin{equation}
P^{(1)}_\nu
=
\prod_{\substack{c \in \{c^{(5)}_1(\nu') \,:\, \nu' \in \mathcal I_{n,5}\}\\ c \neq c^{(5)}_1(\nu)}}
\frac{\mathcal C^{(5)}_1-c}{c^{(5)}_1(\nu)-c},
\label{eq:k5_projector_C2}
\end{equation}
where the product runs over the distinct quadratic-Casimir eigenvalues present in
$\mathcal I_{n,5}$. Any remaining degeneracies are then resolved by $\mathcal C^{(5)}_2$,
yielding the full projector onto the $\nu$-block,
\begin{equation}
P_\nu
=
P^{(1)}_\nu
\prod_{\substack{\nu'\in\mathcal I_{n,5}\\ c^{(5)}_1(\nu') = c^{(5)}_1(\nu)\\ \nu'\neq \nu}}
\frac{\mathcal C^{(5)}_2-c^{(5)}_2(\nu')}{c^{(5)}_2(\nu)-c^{(5)}_2(\nu')}.
\label{eq:k5_projector_nu}
\end{equation}
This completes the first GT step: the highest-weight label $\nu$ is fixed.

The second GT step is the multiplicity-free restriction
$\mathrm{SO}(5)\downarrow \mathrm{SO}(4)$. The admissible intermediate weights
$\mu=(\mu_1,\mu_2)$ satisfy
\begin{equation}
\nu_1\ge \mu_1\ge \nu_2\ge |\mu_2|,
\qquad
\mu_1,\mu_2\in\mathbb Z.
\label{eq:k5_SO5_SO4_branching}
\end{equation}
To isolate the corresponding $\mathrm{SO}(4)$ sector, we use the two primitive Casimirs of
$\mathfrak{so}(4)$, i.e. the specialization of Eqs.~\eqref{eq:quadratic_casimir} and
\eqref{eq:pfafianCasimir} to $m=4$,
\begin{equation}
\mathcal C^{(4)}_1
:=
\frac14\sum_{1\le a<b\le 4}\Lambda_{ab}^2,
\label{eq:k5_C2_4}
\end{equation}
and
\begin{equation}
\mathcal C^{(4)}_2
:=
\Pf\!\left(\frac{\mathbf\Lambda^{(4)}}{2}\right)
=
\frac18\Bigl(
\Lambda_{12}\Lambda_{34}
-\Lambda_{13}\Lambda_{24}
+\Lambda_{14}\Lambda_{23}
\Bigr),
\label{eq:k5_CPf_4}
\end{equation}
where $\mathbf\Lambda^{(4)}$ is the $4\times 4$ antisymmetric generator matrix. Their
eigenvalues on the $\mathrm{SO}(4)$ irrep labeled by $\mu$ are
\begin{equation}
c^{(4)}_1(\mu)
=
-\bigl[\mu_1(\mu_1+2)+\mu_2^2\bigr],
\qquad
c^{(4)}_2(\mu)
=
\mu_2(\mu_1+1).
\label{eq:k5_Casimir4_eigenvalues}
\end{equation}
Inside the image of $P_\nu$, we first isolate the eigenspace of $\mathcal C^{(4)}_1$,
\begin{equation}
P^{(1)}_{\nu,\mu}
=
P_\nu
\prod_{\substack{c \in \{c^{(4)}_1(\mu')\}\\ c \neq c^{(4)}_1(\mu)}}
\frac{\mathcal C^{(4)}_1-c}{c^{(4)}_1(\mu)-c},
\label{eq:k5_projector_mu_C2}
\end{equation}
where $\mu'$ runs over the allowed intermediate weights obeying
\eqref{eq:k5_SO5_SO4_branching} for fixed $\nu$, and the product runs over the distinct
eigenvalues of $\mathcal C^{(4)}_1$ present in that set. Any remaining accidental degeneracies
are then resolved by $\mathcal C^{(4)}_2$, yielding
\begin{equation}
P_{\nu,\mu}
=
P^{(1)}_{\nu,\mu}
\prod_{\substack{\mu'\\ c^{(4)}_1(\mu') = c^{(4)}_1(\mu)\\ \mu'\neq \mu}}
\frac{\mathcal C^{(4)}_2-c^{(4)}_2(\mu')}{c^{(4)}_2(\mu)-c^{(4)}_2(\mu')}.
\label{eq:k5_projector_mu}
\end{equation}
This completes the second GT step: the intermediate $\mathrm{SO}(4)$ label $\mu$ is fixed.

The third GT step is the multiplicity-free restriction
$\mathrm{SO}(4)\downarrow \mathrm{SO}(3)$. For fixed $\mu$, the admissible
$\mathrm{SO}(3)$ spins are
\begin{equation}
s=|\mu_2|,\ |\mu_2|+1,\dots,\mu_1.
\label{eq:k5_SO4_SO3_branching}
\end{equation}
To isolate the corresponding $\mathrm{SO}(3)$ sector, we use the quadratic Casimir of the
subgroup generated by $\Lambda_{12},\Lambda_{13},\Lambda_{23}$, namely the specialization of
\eqref{eq:quadratic_casimir} to $m=3$,
\begin{equation}
K^{(3)}
=
\frac14\bigl(\Lambda_{12}^2+\Lambda_{13}^2+\Lambda_{23}^2\bigr),
\label{eq:k5_SO3_Casimir}
\end{equation}
whose eigenvalues are $-s(s+1)$. Hence, inside the image of $P_{\nu,\mu}$, the projector onto
the $\mathrm{SO}(3)$ irrep labeled by $s$ is
\begin{equation}
P_{\nu,\mu,s}
=
P_{\nu,\mu}
\prod_{\substack{s'=|\mu_2|\\ s'\neq s}}^{\mu_1}
\frac{K^{(3)}+s'(s'+1)}{-s(s+1)+s'(s'+1)}.
\label{eq:k5_projector_s}
\end{equation}

The fourth and final GT step is the restriction
$\mathrm{SO}(3)\downarrow \mathrm{SO}(2)$. The last GT label is the $\mathrm{SO}(2)$ weight
\begin{equation}
m=-s,-s+1,\dots,s,
\label{eq:k5_SO3_SO2_branching}
\end{equation}
resolved by the Cartan generator
\begin{equation}
M_3:=\tfrac12\Lambda_{12},
\label{eq:k5_M3}
\end{equation}
whose eigenvalues are $i\,m$. The full GT set is therefore
\begin{equation}
\GT(\nu)
=
\left\{
\mathfrak m=(\mu;s,m):
\begin{array}{l}
\nu_1\ge \mu_1\ge \nu_2\ge |\mu_2|,\\
s=|\mu_2|,\dots,\mu_1,\\
m=-s,\dots,s
\end{array}
\right\},
\label{eq:k5_GT_set}
\end{equation}
where $\mu=(\mu_1,\mu_2)$. The corresponding one-dimensional GT projectors are
\begin{equation}
P_{\nu,\mathfrak m}
:=
P_{\nu,\mu,s,m}
=
P_{\nu,\mu,s}
\prod_{\substack{m'=-s\\ m'\neq m}}^{s}
\frac{M_3-i\,m'}{i(m-m')}.
\label{eq:k5_GT_projectors}
\end{equation}
Counting these GT labels reproduces the irrep dimension,
\begin{equation}
\#\GT(\nu)
=
\sum_{\mu_1=\nu_2}^{\nu_1}
\sum_{\mu_2=-\nu_2}^{\nu_2}
\sum_{s=|\mu_2|}^{\mu_1}(2s+1)
=
\dim V_\nu,
\label{eq:k5_GT_count}
\end{equation}
in agreement with Eq.~\eqref{eq:k5_irrep_dim}. This completes the GT resolution of the
$\nu$-block.

The block $\End(V_\nu)$ is then spanned by the GT transition operators
\begin{equation}
X^{(\nu)}_{\mathfrak m,\mathfrak n},
\qquad
\mathfrak m,\mathfrak n\in\GT(\nu),
\label{eq:k5_transition_ops}
\end{equation}
defined as in Eq.~\eqref{eq:commutant_transition_basis}. The diagonal operators
$X^{(\nu)}_{\mathfrak m,\mathfrak m}$ are precisely the one-dimensional GT projectors, while the
off-diagonal operators connect distinct GT sectors inside the same $\nu$-block.

For $k=5$, the off-diagonal GT operators may be constructed explicitly by sandwiching suitable
products of bridge operators between the one-dimensional GT projectors,
\begin{equation}
X^{(\nu)}_{\mathfrak m,\mathfrak n}
\;\propto\;
P_{\nu,\mathfrak m}\,\mathcal O_{\mathfrak m,\mathfrak n}\,P_{\nu,\mathfrak n},
\end{equation}
where $\mathcal O_{\mathfrak m,\mathfrak n}$ is chosen to connect the source GT sector
$\mathfrak n$ to the target sector $\mathfrak m$. Concretely, changes in the final
$\mathrm{SO}(2)$ weight $m$ are generated within the $\mathrm{SO}(3)$ subgroup by the
corresponding ladder operators, changes in the intermediate spin $s$ are generated by operators
transforming as vectors under that $\mathrm{SO}(3)$, and changes in the $\mathrm{SO}(4)$ label
$\mu$ are generated by bridge operators outside the $\mathrm{SO}(4)$ subalgebra, namely those
involving the fifth replica. Appendix~\ref{app:offk5} gives the detailed construction.

Collecting all irreducible sectors, we obtain the explicit GT-adapted basis
\begin{equation}
\Com_5(\MG_n)
=
\Span\!\left\{
X^{(\nu)}_{\mathfrak m,\mathfrak n}
\;:\;
\nu\in\mathcal I_{n,5},\;
\mathfrak m,\mathfrak n\in \GT(\nu)
\right\},
\label{eq:k5_basis}
\end{equation}
consisting of $\dim \Com_5(\MG_n)$, cf. Eq.~\eqref{eq:k5_dim_closed}, orthogonal operators
$X^{(\nu)}_{\mathfrak m,\mathfrak n}$.

Summarizing, for $k=5$, the GT construction resolves the commutant in four stages: the
highest-weight label $\nu=(\nu_1,\nu_2)$ selects the $\mathrm{SO}(5)$ irrep, the intermediate
weight $\mu=(\mu_1,\mu_2)$ resolves the restriction to $\mathrm{SO}(4)$, the spin $s$
resolves the further restriction to $\mathrm{SO}(3)$, and the weight $m$ resolves the final
$\mathrm{SO}(2)$ sector.

\subsection{Generalization to arbitrary $k$}
\label{subsec:general_k}

The explicit constructions for $k\leq 5$ detailed in the preceding subsections reveal a systematic algebraic framework that readily generalizes to arbitrary replica number $k$. For any $k$, the commutant $\Com_k(\MG_n)$ is spanned by the GT transition operators $X^{(\nu)}_{\mathfrak m,\mathfrak n}$, which are determined by the canonical subgroup chain~\eqref{eq:gt_chain}.

At each level $m$ of this chain ($2\leq m\leq k$), the algebra $\mathfrak{so}(m)$ has rank
$r_m=\lfloor m/2\rfloor$. Consequently, an irreducible representation of $\mathrm{SO}(m)$ is specified by
$r_m$ highest-weight labels. In our notation, the top-level $\mathrm{SO}(k)$ irrep is labeled by
$\nu$, while the GT pattern $\mathfrak m$ records the intermediate highest weights appearing in the successive restrictions
\[
\mathrm{SO}(k)\downarrow \mathrm{SO}(k-1)\downarrow\cdots\downarrow \mathrm{SO}(2).
\]
More explicitly, for fixed $\nu$, the GT label $\mathfrak m$ consists of the collection of intermediate weights
$\mu^{(m)}$ for $m=2,\dots,k-1$, subject to the standard multiplicity-free interlacing conditions for orthogonal groups. Thus the full label of a one-dimensional GT sector is the pair $(\nu,\mathfrak m)$, where $\nu$ specifies the top-level irrep and $\mathfrak m$ resolves the intermediate subgroup data.

To isolate these one-dimensional sectors explicitly from the bridge algebra, one constructs a sequence of nested projectors along the subgroup chain. At each level $m$, isolating the chosen $\mathrm{SO}(m)$ sector requires exactly $r_m$ algebraically independent Casimir operators. As discussed in Sec.~\ref{subsec:gt_chain}, for odd $m=2r+1$, these may be chosen as the even-degree trace invariants $\mathcal C^{(m)}_j$, $j=1,\dots,r$. For even $m=2r$, the first $r-1$ primitive Casimirs are again trace invariants, while the highest-degree one is replaced by the Pfaffian invariant $\mathcal C^{(2r)}_r$.

Crucially, because the eigenvalues of these Casimirs are coupled polynomial functions of the highest-weight labels, one cannot in general project the labels independently. Instead, the projection must proceed sequentially by isolating joint eigenspaces. As demonstrated explicitly for $k=4$ and $k=5$, one first uses the lowest-degree Casimir, typically the quadratic Casimir $\mathcal C^{(m)}_1$, to isolate a given eigenspace. If distinct allowed weights share the same quadratic eigenvalue, the higher-degree Casimirs---or, in the even case, the Pfaffian, which carries the additional chiral sign information---are then applied successively to resolve the accidental degeneracies. In this way, one constructs the nested projectors that ultimately lead to the one-dimensional GT projectors $P_{\nu,\mathfrak m}$. This strategy relies only on one-dimensional Lagrange interpolation polynomials and avoids the need to solve high-degree characteristic equations or introduce nonlinear operator combinations.

Once the one-dimensional GT projectors $P_{\nu,\mathfrak m}$ have been constructed down to the $\mathrm{SO}(2)$ level, the full set of off-diagonal transition operators $X^{(\nu)}_{\mathfrak m,\mathfrak n}$ is obtained by sandwiching suitable tensor operators---built from the bridge algebra at the appropriate level of the subgroup chain---between the corresponding source and target projectors. Collecting these operators over all admissible highest weights $\nu\in\mathcal I_{n,k}$ yields the complete orthogonal GT-adapted basis~\eqref{eq:commutant_transition_basis} of $\Com_k(\MG_n)$ for arbitrary replica number $k$ and system size $n$.

\section{Clifford--matchgate commutant}
\label{sec:CM_commutant}

We now turn to the Clifford--matchgate group $\CM_n\subset\MG_n$, introduced in Sec.~\ref{sec:notation}.
Recall that $U_{\mathbf{s},\pi}\in \CM_n$ acts on the Majorana operators as
\begin{equation}
U_{\mathbf{s},\pi}\gamma_\mu U_{\mathbf{s},\pi}^\dagger=s_\mu\,\gamma_{\pi(\mu)},
\qquad
s_\mu\in\{\pm1\},
\quad
\pi\in \mathrm S_{2n}.
\label{eq:CM_action_majorana}
\end{equation}
As aforementioned, the action on Majorana operators is that of the group of signed permutations $\mathbb{B}_n\equiv (\mathbb{Z}_2)^{2n}\rtimes\mathrm{S}_{2n}$, i.e., is generated by just two ingredients: arbitrary permutations and independent sign flips of the $2n$ modes. This simplicity allows the commutant $\Com_k(\CM_n)$ to be characterized in a particularly transparent way using \emph{replica patterns}.

\subsection{Construction of the basis}

Consider an operator on $k$ replicas expanded in the Majorana basis,
\begin{equation}
W=\sum_{S_1,\dots,S_k\subseteq[2n]}
W_{S_1,\dots,S_k}\,
\gamma_{S_1}\otimes\cdots\otimes\gamma_{S_k}.
\label{eq:CM_majorana_expansion}
\end{equation}
For each Majorana label $\mu\in[2n]$ and each term in the expansion, we define the \textit{replica pattern}
\begin{equation}
    I(\mu):=\{\,j\in[k]\;| \; \mu\in S_j\,\}\subseteq[k],
\end{equation}
\textit{i.e.}, the set of replicas in which $\gamma_\mu$ appears.
Let also $x_I = |\{\mu\in [2n]\,|\, I(\mu) = I\}|$ be the \emph{occupation number}, counting how many Majorana labels realize a given pattern $I$.
The two generators of $\CM_n$, sign flips and permutations, impose complementary constraints on any operator in $\Com_k(\CM_n)$.

\paragraph{Sign-flip constraint.}
Invariance under the individual reflection $\gamma_\mu\mapsto -\gamma_\mu$ requires that each $\gamma_\mu$ appear an even number of times across all replicas, i.e., $|I(\mu)|:= 0\;\mathrm{mod}\; 2$.
Consequently, $x_I$ may be nonzero only for subsets $I\subseteq [k]$ of even cardinality.

\paragraph{Permutation constraint.}
A permutation $\pi\in \mathrm{S}_{2n}$ acts by relabeling Majorana operators,
$\gamma_\mu \;\mapsto\; \gamma_{\pi(\mu)}$, which reshuffles Majorana labels within each monomial but preserves the replica structure.
Two monomials that realize the same collection of replica patterns $\{I(\mu)\}_{\mu=1}^{2n}$ are therefore related by such a relabeling. An invariant operator must assign equal coefficients to all monomials in the same permutation orbit, and hence cannot distinguish \emph{which} specific Majorana realizes a given pattern, only \emph{how many} do.

Combining both constraints, any operator in $\Com_k(\CM_n)$ is completely specified by the occupation numbers $\{x_I\}_{I\in\mathcal{I}_k}$, where all microscopic label information has been erased.

Let $\mathcal{I}_k=\{I\subseteq[k]\;:\;|I|=0\;\mathrm{mod}\; 2\}$. The total number of Majoranas gives the single constraint
\begin{equation}
    \sum_{I\in \mathcal{I}_k} x_I=2n\;,
\label{eq:CM_sum_constraint}
\end{equation}
Here $x_{\varnothing}$ accounts for Majorana labels that do not appear in any replica (physically, the spectator modes) and is determined by
\begin{equation}
    x_\varnothing=2n-\sum_{\substack{I\in \mathcal{I}_k \\ I\neq \varnothing}} x_I\;.
\end{equation}
The number of free (nonempty) occupation numbers is $2^{k-1}-1$, since $|\mathcal{I}_k\setminus\{\varnothing\}| = 2^{k-1}-1$.
For each choice of nonnegative integers $\{x_I\}$ satisfying the constraint~\eqref{eq:CM_sum_constraint}, there exists a unique operator in $\Com_k(\CM_n)$, obtained by uniformly summing over all monomials whose Majorana labels realize the prescribed pattern counts.
Since monomials with distinct occupation data are orthogonal under the Hilbert--Schmidt inner product, the resulting operators are automatically linearly independent.

We now construct the basis operators explicitly. For a subset $I\subseteq[k]$,
we define the \emph{pattern operator}
\begin{equation}
\gamma_\mu^{(I)}
:=
\bigotimes_{j=1}^k A_j,
\qquad
A_j=
\begin{cases}
\gamma_\mu, & j\in I,\\
\mathbb 1, & j\notin I.
\end{cases}
\label{eq:CM_gammaI_def}
\end{equation}
Equivalently, $\gamma_\mu^{(I)}$ places $\gamma_\mu$ on the replicas belonging to $I$ and
the identity on the remaining replicas. For any choice of nonnegative integers
$\{x_I\}_{I\in\mathcal I_k}$ satisfying Eq.~\eqref{eq:CM_sum_constraint}, choose once and for all
an assignment $\mu\mapsto I(\mu)$ realizing these occupation numbers, and define
\begin{equation}
\Omega(\{x_I\})
:= \frac{1}{\prod_{I\in\mathcal{I}_k} x_I!}
\sum_{\pi\in \mathrm S_{2n}}
\prod_{\mu=1}^{2n}
\gamma_{\pi(\mu)}^{(I(\mu))},
\label{eq:CM_basis_operator}
\end{equation}
where the product is taken in increasing order.
By construction, $\Omega(\{x_I\})$ is invariant under both permutations and sign flips, and hence
belongs to $\Com_k(\CM_n)$. Moreover, different occupation data $\{x_I\}$ correspond to
disjoint Majorana sectors, so the operators $\Omega(\{x_I\})$ are linearly independent.

Thus the family
\begin{equation}
\bigl\{\Omega(\{x_I\})\;:\;x_I\ge 0,\ \sum_{I\in\mathcal I_k}x_I=2n\bigr\}
\label{eq:CM_basis_family}
\end{equation}
forms a basis of $\Com_k(\CM_n)$. Its orthogonality follows directly from the orthogonality of
distinct Majorana monomials in the Hilbert--Schmidt inner product. More precisely, one finds
\begin{equation}
\Tr\!\left[\Omega(\{x_I\})^\dagger \Omega(\{y_I\})\right]
=
\delta_{\{x_I\},\{y_I\}}\,
2^{kn}\,
\binom{2n}{\{x_I\}},
\label{eq:CM_basis_orthogonality}
\end{equation}
where
\begin{equation}
\binom{2n}{\{x_I\}}
:=
\frac{(2n)!}{\prod_{I\in\mathcal I_k}x_I!}
\label{eq:CM_multinomial}
\end{equation}
is the multinomial coefficient associated with the occupation data. 
In particular,
\begin{equation}
\|\Omega(\{x_I\})\|_2^2
=
2^{kn}\binom{2n}{\{x_I\}}.
\label{eq:CM_basis_norm}
\end{equation}

The dimension of the commutant now follows immediately. The number of allowed pattern classes is
\begin{equation}
|\mathcal I_k|
=
\sum_{\ell=0}^{\lfloor k/2\rfloor}\binom{k}{2\ell}
=
2^{k-1},
\label{eq:CM_num_patterns}
\end{equation}
since half of the $2^k$ subsets of $[k]$ have even cardinality. Therefore,
$\dim\Com_k(\CM_n)$ is simply the number of weak compositions of $2n$ into $2^{k-1}$ parts:
\begin{equation}
\dim\Com_k(\CM_n)
=
\binom{2n+2^{k-1}-1}{2^{k-1}-1}.
\label{eq:CM_dim_general}
\end{equation}

This completes the characterization of the Clifford--matchgate commutant. In summary, an invariant
operator is obtained by specifying how many of the $2n$ Majorana labels realize each even replica
pattern, and then summing uniformly over all microscopic realizations of that pattern data.

For $k\le 3$, every even subset $I\subseteq[k]$ is necessarily a pair, so the Clifford--matchgate and full matchgate commutants coincide:
\begin{equation}
\Com_k(\CM_n)=\Com_k(\MG_n),
\qquad k\le 3,
\label{eq:CM_equals_MG_smallk}
\end{equation}
consistently with the commutant construction at $k=2,3$ in Ref.~\cite{Wan23}. Our construction applies to arbitrary  $k\geq 1$.
The two commutants first diverge at $k=4$, where the four-element pattern $I=\{1,2,3,4\}$ becomes available in addition to the six pair patterns. Consequently, the Clifford--matchgate commutant at $k=4$ can be generated by a single additional generator added to the matchgate commutant:
 \begin{equation}
\Omega{\{(x_{\{1234\}}=1,x_\varnothing=2n-1)\}} = \sum_{j} (\gamma_j)^{\otimes 4}.
\end{equation}
This is reminiscent of the case of the Clifford group, which only requires one additional operator to generate the Clifford commutant from the Haar commutant for $k=4$~\cite{zhu2016clifford,gross2021schurweylduality}.

This extra generator enlarges the Clifford--matchgate commutant: Eq.~\eqref{eq:CM_dim_general} gives
\begin{equation}
\dim\Com_4(\CM_n)=\binom{2n+7}{7},
\label{eq:CM_dim_k4}
\end{equation}
which is strictly larger than the matchgate value $\dim\Com_4(\MG_n)=\binom{2n+6}{6}+\binom{2n+5}{6}$ at the same replica order, cf.\ Eq.~\eqref{eq:k4_dim_closed}. Physically, the additional invariants reflect the coarser symmetry of Clifford--matchgates: signed permutations form a discrete subgroup of the continuous orthogonal rotations generated by full matchgates, leaving more operators invariant. In Sec.~\ref{subsec:CliffMatchDesi} we further analyze the relation between the state ensembles generated by $\CM_n$ and $\MG_n$.

\section{Applications}
\label{sec:applications}
In this section, we showcase some applications of our matchgate commutant constructions in a wide range of subfields from quantum information to many-body physics. 

\subsection{The matchgate twirl in GT basis}
\label{sec:twirlings}

The twirl over a subgroup $\mathrm{G}$ of the unitary group is defined by
\begin{equation}
\Twirl_\mathrm{G}^{(k)}(W)
:=
\int_\mathrm{G} dU\,
U^{\otimes k} W (U^\dagger)^{\otimes k},
\label{eq:twirl_def_append}
\end{equation}
where $dU$ is the Haar measure for continuous groups and the uniform counting measure for discrete groups. 
By construction, $\Twirl_\mathrm{G}^{(k)}$ is the Hilbert--Schmidt projector onto the
commutant $\Com_k(\mathrm{G})$.

Here, we consider the matchgate group $\mathrm{G} = \MG_n$.
Once an orthonormal GT-adapted basis of the matchgate commutant $\Com_k(\MG_n)$ is constructed, see Sec.~\ref{sec:rep_theoretic_matchgate_commutant}, the twirl becomes completely explicit. 
The twirl of an arbitrary operator $W \in \mathcal{B}(\mathcal{H}_n^{\otimes k}) $ acting on $k$ replicas of the system, is simply the orthogonal projection of $W$ onto the commutant basis:
\begin{equation}
\Twirl_{\MG_n}^{(k)}(W)
=
\sum_{\nu\in\mathcal I_{N,k}}
\sum_{\mathfrak m,\mathfrak n\in\GT(\nu)}
\Tr\!\left[
\bigl(X^{(\nu)}_{\mathfrak m,\mathfrak n}\bigr)^\dagger W
\right]
X^{(\nu)}_{\mathfrak m,\mathfrak n}.
\label{eq:twirl_GT_trace}
\end{equation}

This formula makes the block structure of the twirl transparent. Since the operators
$X^{(\nu)}_{\mathfrak m,\mathfrak n}$ span the block $\End(V_\nu)$, the twirl acts by first
discarding all components of $W$ orthogonal to the commutant, and then resolving the remaining
part into the irreducible blocks labeled by $\nu$. In particular, one may write
\begin{equation}
\Twirl_{\MG_n}^{(k)}(W)
=
\sum_{\nu\in\mathcal I_{N,k}}
\Twirl_{\MG_{n,\nu}}^{(k)}(W),
\end{equation}
where
\begin{equation}
\Twirl_{\MG_{n,\nu}}^{(k)}(W)
:=
\sum_{\mathfrak m,\mathfrak n\in\GT(\nu)}
\Tr\!\left[
\bigl(X^{(\nu)}_{\mathfrak m,\mathfrak n}\bigr)^\dagger W
\right]
X^{(\nu)}_{\mathfrak m,\mathfrak n}
\label{eq:twirl_block_nu}
\end{equation}
is the contribution of the $\nu$-block.

As an example, consider the case $k=4$. The relevant irreducible sectors are labeled
by
\begin{equation}
\nu=(\nu_1,\nu_2)\in\mathcal I_{n,4},
\qquad
n\ge \nu_1\ge |\nu_2|,
\end{equation}
and the GT labels are $\mathfrak m=(s,m)$, given by Eq.~\eqref{eq:k4_GT_set}. Hence the fourth-order matchgate twirl decomposes as
\begin{equation}
\Twirl_{\MG_n}^{(4)}(W)
=
\sum_{\nu\in\mathcal I_{n,4}}
\Twirl_{\MG_{n,\nu}}^{(4)}(W),
\label{eq:k4_twirl_decomp}
\end{equation}
with
\begin{equation}
\Twirl_{\MG_{n,\nu}}^{(4)}(W)
= \!\!\!
\sum_{s,s'=|\nu_2|}^{\nu_1}
\sum_{m=-s}^{s}
\sum_{m'=-s'}^{s'}
c^{\nu}_{s,m,s',m'}(W) \,
X^{(\nu_1,\nu_2)}_{(s,m),(s',m')},
\label{eq:k4_twirl_block}
\end{equation}
where
\begin{equation}
c^{\nu}_{s,m,s',m'}(W)
:=
\Tr\!\left[
\bigl(X^{(\nu_1,\nu_2)}_{(s,m),(s',m')}\bigr)^\dagger W
\right].
\label{eq:k4_twirl_coeff}
\end{equation}
Thus, at $k=4$, the matchgate twirl is reduced to a finite sum over the GT labels
$(\nu_1,\nu_2;s,m;s',m')$. Once the GT-adapted basis operators
$X^{(\nu)}_{\mathfrak m,\mathfrak n}$ have been constructed, the computation of
$\Twirl_{\MG_n}^{(4)}(W)$ amounts simply to evaluating their Hilbert--Schmidt overlaps with the input operator $W$ and summing the corresponding basis elements.

\subsection{Clifford--matchgate twirl}
\label{sec:cliff_twirlings}
The Clifford--matchgate twirl is defined in complete analogy with the matchgate twirl,
\begin{equation}
\Twirl_{\CM_n}^{(k)}(W)
:=
\int_{\CM_n} dU\,
U^{\otimes k} W (U^\dagger)^{\otimes k},
\label{eq:CM_twirl_def}
\end{equation}
where $dU$ is the uniform counting measure on the finite group $\CM_n$. By construction,
$\Twirl_{\CM_n}^{(k)}$ is the Hilbert--Schmidt projector onto the commutant
$\Com_k(\CM_n)$.

Since the operators $\Omega(\{x_I\})$ introduced above form an orthogonal basis of
$\Com_k(\CM_n)$, the twirl admits an explicit expansion in that basis.
Using Eq.~\eqref{eq:CM_basis_norm}, we define normalized operators $ \bar{\Omega}(\{x_I\}) = \Omega(\{x_I\})/\|\Omega(\{x_I\})\|_2$. 
For an arbitrary operator
$W\in\mathcal B(\mathcal H_n^{\otimes k})$, one has
\begin{equation}
\Twirl_{\CM_n}^{(k)}(W)
=
\sum_{\{x_I\}}
\Tr\!\left[W\, \bar{\Omega}(\{x_I\})\right]
\,
\bar \Omega(\{x_I\}),
\label{eq:CM_twirl_basis}
\end{equation}
where the sum runs over all valid replica patterns $\{x_I\}_{I\in\mathcal I_k}$.

This formula is the Clifford--matchgate analogue of the GT expansion of the matchgate twirl. The
difference is that the role of the GT basis is now played by the pattern operators
$ \bar \Omega(\{x_I\})$, which are labeled directly by the replica-pattern occupancies $\{x_I\}$.
Accordingly, the computation of the Clifford--matchgate twirl reduces to evaluating the overlaps
$\Tr[W\, \bar \Omega(\{x_I\})]$ and summing the corresponding basis elements.

\subsection{Twirl of the vacuum projector and moments of fermionic Gaussian states}
\label{subsec:MG_vacuum_twirl}

A useful application of the matchgate twirl is the computation of moments of
states obtained by acting with matchgates on the vacuum, i.e. of pure fermionic Gaussian states. Indeed, every such state is of the form
\begin{equation}
\ket{\psi_U}=U\ket{\mathbf 0},
\qquad
U\in \mathrm{G},
\end{equation}
where $\ket{\mathbf 0}=\ket{0}^{\otimes n}$ denotes the fermionic vacuum state and we consider $U$ to be unitary taken with uniform probability from the group $\mathrm{G}= \MG_n$ or $\mathrm{G} = \CM_n$. In this section we compute the average $k$-copy state $\ket{\psi_U} \bra{\psi_U}^{\otimes k}$ using the matchgate twirl found in Sec.~\ref{sec:twirlings} and Clifford matchgate twirl described in Sec.~\ref{sec:cliff_twirlings}.

\subsubsection{Matchgate group}

We start by considering fermionic Gaussian states
\begin{equation}
\ket{\psi_U}=U\ket{\mathbf 0},
\end{equation}
where $U$ is taken with Haar measure~\cite{Braccia2025flo} over the matchgate group $\MG_n$.
We aim to compute the expectation value of the $k$-copy density matrix of the uniformly sampled fermionic Gaussian state
\begin{equation}
\mathbb E_{U\in \MG_n}[  \ket{\psi_U} \bra{\psi_U}^{\otimes k} ] := \int_{\MG_n} dU\,
\bigl(U\ketbra{\mathbf 0}{\mathbf 0}U^\dagger\bigr)^{\otimes k}.
\label{eq:MG_vacuum_moment_def}
\end{equation}
Expanding the tensor product, we immediately verify that 
\begin{equation}
\mathbb E_{U\in \MG_n}[  \ket{\psi_U} \bra{\psi_U}^{\otimes k} ] = \Twirl_{\MG_n}^{(k)}( \ket{\mathbf{0}}\bra{\mathbf{0}}^{\otimes k}),
\label{eq:MG_vacuum_moment_tw}
\end{equation}
or, in other words, that the $k$-th moment of fermionic Gaussian state corresponds to the matchgate twirl~\eqref{eq:twirl_GT_trace} of the replicated vacuum $\ket{\mathbf{0}}\bra{ \mathbf{0}}^{\otimes k}$.
The GT expansion of the matchgate twirl simplifies considerably for the replicated vacuum. Using
\begin{equation}
\gamma_{2j-1}=c_j+c_j^\dagger,
\qquad
\gamma_{2j}=i(c_j-c_j^\dagger),
\end{equation}
one finds, for every $1\le a<b\le k$,
\begin{equation}
\Lambda_{ab}
=
\sum_{\mu=1}^{2n}\gamma_\mu^{(a)}\gamma_\mu^{(b)}
=
2\sum_{j=1}^n
\Bigl(
c_j^{(a)\dagger}c_j^{(b)}
+
c_j^{(a)}c_j^{(b)\dagger}
\Bigr).
\label{eq:bridge_c_operators_vac}
\end{equation}
Since $c_j\ket{\mathbf 0}=0$, this implies that each bridge operator annihilates the replicated vaccum state
\begin{equation}
\Lambda_{ab}\ket{\mathbf 0}^{\otimes k}=0,
\qquad
1\le a<b\le k.
\label{eq:vacuum_bridge_annihilation}
\end{equation}
Hence, $\ket{\mathbf 0}^{\otimes k}$ belongs to the trivial $\mathrm{SO}(k)$ sector, corresponding to the
highest weight $\nu=\mathbf 0$. The GT sum in \eqref{eq:twirl_GT_trace} therefore reduces to the
$\nu=\mathbf 0$ contribution; since the corresponding irrep is one-dimensional, the associated GT
label set contains a single element, and the sums over $\mathfrak m$ and $\mathfrak n$ collapse
as well. The GT sum in \eqref{eq:twirl_GT_trace} therefore reduces
to
\begin{equation}
\Twirl_{\MG_n}^{(k)}\!\left(\ketbra{\mathbf 0}{\mathbf 0}^{\otimes k}\right)
=
\frac{P_{\mathbf 0}^{(k)}}{\Tr P_{\mathbf 0}^{(k)}},
\label{eq:MG_vacuum_twirl_general}
\end{equation}
where $P_{\mathbf 0}^{(k)}$ is the projector onto the trivial $\mathrm{SO}(k)$ sector selected by the GT
construction. The normalization follows from trace preservation of the twirl.

The condition in Eq.~\eqref{eq:vacuum_bridge_annihilation} is closely related to a characterization
of fermionic Gaussian states: a pure state $\ket{\psi}^{\otimes 2}$ is annihilated by
$\Lambda$ if and only if $\ket{\psi}$ is a fermionic Gaussian state~\cite{Bravyi05flo,Dias24classical}. Thus the trivial $\mathrm{SO}(k)$ sector naturally captures the Gaussian
structure of the matchgate ensemble.

For $k=2$, it was already known that the matchgate twirl of the vacuum yields a projector onto the
Gaussian sector~\cite{melo2013}. Equation~\eqref{eq:MG_vacuum_twirl_general} generalizes this observation to
arbitrary replica number $k$, providing a unified representation-theoretic explanation.

More concretely, the projector $P_{\mathbf 0}^{(k)}$ projects onto the linear span of states of
the form $\ket{\psi}^{\otimes k}$, where $\ket{\psi}$ is a fermionic Gaussian state. This
subspace lies inside the symmetric subspace of the $k$-replica Hilbert space, and we refer
to it as the \emph{Gaussian-symmetric subspace}.

The projector $P_{\mathbf 0}^{(k)}$ is already determined by the quadratic Casimir
\begin{equation}
\mathcal C_2^{(k)}
:=
\frac14\sum_{1\le a<b\le k}\Lambda_{ab}^2.
\label{eq:MG_quad_casimir}
\end{equation}
For $k=2r+1$, the quadratic Casimir eigenvalue on the $SO(k)$ irrep of highest weight
$\nu=(\nu_1,\dots,\nu_r)$ is \cite{Goodman09symmetry, fulton1991representation}
\begin{equation}
c_2^{(2r+1)}(\nu)
=
-\sum_{j=1}^{r}\nu_j\bigl(\nu_j+2r-2j+1\bigr),
\label{eq:B_r_casimir_general}
\end{equation}
while for $k=2r$ it is
\begin{equation}
c_2^{(2r)}(\nu)
=
-\sum_{j=1}^{r}\nu_j\bigl(\nu_j+2r-2j\bigr).
\label{eq:D_r_casimir_general}
\end{equation}
Hence the trivial sector is uniquely isolated by Lagrange interpolation in
$\mathcal C_2^{(k)}$, which gives
\begin{equation}
P_{\mathbf 0}^{(k)}
=
\prod_{\substack{\nu\in\mathcal I_{n,k}\\ \nu\neq \mathbf 0}}
\frac{\mathcal C_2^{(k)}-c_2^{(k)}(\nu)}
{-\,c_2^{(k)}(\nu)}.
\label{eq:vacuum_projector_general}
\end{equation}
Combining \eqref{eq:MG_vacuum_twirl_general} and \eqref{eq:vacuum_projector_general}, we obtain
\begin{equation}
\Twirl_{\MG_n}^{(k)}\!\left(\ketbra{\mathbf 0}{\mathbf 0}^{\otimes k}\right)
=
\frac{1}{\Tr P_{\mathbf 0}^{(k)}}
\prod_{\substack{\nu\in\mathcal I_{n,k}\\ \nu\neq \mathbf 0}}
\frac{\mathcal C_2^{(k)}-c_2^{(k)}(\nu)}
{-\,c_2^{(k)}(\nu)},
\label{eq:vacuum_twirl_general_compact}
\end{equation}
where the allowed highest weights are those in Eqs.~\eqref{eq:highest_weights_Dr} and
\eqref{eq:highest_weights_Br}, and the Casimir eigenvalues are given by
Eqs.~\eqref{eq:B_r_casimir_general} and \eqref{eq:D_r_casimir_general}.
Finally, the normalization constant reads
\begin{equation}
\Tr P_{\mathbf 0}^{(k)}
=
2\prod_{1\le i<j\le n}
\frac{k+2n-i-j}{2n-i-j}.
\label{eq:ZnK_product}
\end{equation}
as shown in Appendix~\ref{app:P0norm}.

\subsubsection{Clifford matchgate group}
When the state of interest is a stabilizer free-fermion state
\begin{equation}
\ket{\psi_U}=U\ket{\mathbf 0},
\end{equation}
with $U$ taken with the uniform counting measure~\cite{Braccia2025flo} over the Clifford matchgate group $\CM_n$, the $k$-th moment of the state can be analogously calculated as the Clifford matchgate twirl~\eqref{eq:CM_twirl_basis} of the replicated vacuum state:
\begin{equation}
\mathbb E_{U\in \CM_n}[  \ket{\psi_U} \bra{\psi_U}^{\otimes k} ] = \Twirl_{\CM_n}^{(k)}( \ket{\mathbf{0}}\bra{\mathbf{0}}^{\otimes k}).
\label{eq:CM_vacuum_moment_tw}
\end{equation}
Explicitly
\begin{equation}
\Twirl_{\CM_n}^{(k)}\!\left(\ketbra{ \mathbf{0} }{\mathbf{ 0}}^{\otimes k}\right)
=
\sum_{\{x_I\}}
\Tr\!\left[\ketbra{\mathbf{0}}{\mathbf{0}}^{\otimes k} \bar\Omega(\{x_I\})\right]
\,
\bar \Omega(\{x_I\}).
\label{eq:CM_moment_general}
\end{equation}
Evaluating the overlap with the vacuum projector yields
\begin{equation}
\Twirl_{\CM_n}^{(k)}\!\left(\ketbra{ \mathbf{0} }{\mathbf{0}}^{\otimes k}\right)
=
2^{-kn}
\sum_{\text{even }\{x_I\}}
\frac{
\binom{n}{\{x_I\}/2}
}{
\binom{2n}{\{x_I\}}
}
\,
\bar \Omega(\{x_I\}),
\label{eq:stab_ff_moments}
\end{equation}
where the sum runs over occupation data for which every $x_I$ is even, and $\binom{2n}{\{x_I\}}$ is defined in Eq.~\eqref{eq:CM_multinomial}. 
Thus, only even occupation sectors contribute to the moment operator of the stabilizer
free-fermion ensemble.

\subsection{Connection to fermionic shadow tomography}
\label{subsec:shadow_tomography}

The commutant structure developed in this work provides the algebraic foundation for classical shadow protocols based on matchgate circuits~\cite{Wan23,Zhao21}, which we now make explicit.

In a matchgate shadow protocol, one samples a random Gaussian unitary $U_Q\in\MG_n$~\cite{Helsen2022}, applies it to an unknown state $\rho$, and measures in the computational basis to obtain an outcome $\ket{b}$. The resulting classical shadow is $\hat\rho = \mathcal{M}^{-1}(U_Q^\dagger\ketbra{b}{b}U_Q)$, where the measurement channel $\mathcal{M} = \mathbb{E}_{U\in\MG_n}[U^\dagger(\cdot)U]$ is precisely the $k=1$ single-copy twirl. From the commutant at $k=2$, Ref.~\cite{Wan23} showed that the inverse measurement channel takes the explicit form
\begin{equation}
\mathcal{M}^{-1}
=
\sum_{\ell=0}^{n}
\binom{2n}{2\ell}\binom{n}{\ell}^{-1}\,\mathcal P_{2\ell},
\label{eq:shadow_inverse_channel}
\end{equation}
where $\mathcal P_{2\ell}$ is the projector onto the subspace $\Gamma_{2\ell}$ of products of $2\ell$ Majorana operators. This formula follows directly from the $k=2$ commutant basis in Eq.~\eqref{eq:antiK2}: each basis element $\Upsilon^{(2)}_r$ selects a Majorana-weight sector, and the eigenvalues of the twirl channel in that sector are determined by the pairing-tensor overlaps with the identity.

The variance of any shadow estimator for the expectation value $\Tr(O\rho)$ is controlled by the third moment of the ensemble~\cite{Huang2022}. Specifically, the single-sample variance satisfies
\begin{equation}
\mathrm{Var}[\hat o]
\le
\mathbb{E}_{U,b}\!\left[\bigl(\Tr\bigl[O\,\mathcal{M}^{-1}(U^\dagger\ketbra{b}{b}U)\bigr]\bigr)^2\right],
\end{equation}
which is expressible as a trace over three replicas involving the third-fold twirl channel $\mathcal{E}^{(3)}_{\MG_n}$. The explicit $k=3$ commutant basis of Eq.~\eqref{eq:antiK3} therefore determines the variance structure of matchgate shadows. Combined with the matchgate 3-design property (Corollary~1 in Ref.~\cite{Wan23})---which is encoded in our framework by the equality $\Com_3(\MG_n)=\Com_3(\CM_n)$---these results show that the Clifford--matchgate subgroup produces identical measurement channels and variance bounds as the full matchgate group.

In particular, the twirled vacuum projector~\eqref{eq:MG_vacuum_twirl_general} plays a direct role in shadow estimation of Gaussian-state properties. Estimating the fidelity $\Tr(\varrho\rho)$ between an unknown state $\rho$ and a fermionic Gaussian state $\varrho$ reduces to evaluating the expectation value of the Gaussian density operator with respect to the shadow samples. The variance of this fidelity estimator is bounded by the state frame potential (Sec.~\ref{sec:state_FP}), which we compute in closed form below. More broadly, the full commutant toolkit---the GT basis, the twirl formulas, and the design characterization---provides the ingredients needed to extend the matchgate shadow formalism to higher-order estimation tasks, such as nonlinear functionals of $\rho$ that require $k\ge 4$ replicas, where the commutant structure genuinely goes beyond the results available at low replica order.

\subsection{Frame potentials}
\label{sec:state_FP}

Frame potentials are scalar diagnostics that probe how uniformly a group of unitaries explores the space of quantum operations or the space of quantum states~\cite{mele2024introductiontohaar}. Given an ensemble $\mathrm{G}$ of unitaries acting on an $n$-qubit Hilbert space, one can define two complementary quantities. The \emph{unitary frame potential}~\cite{Gross07} measures the spread of $\mathrm{G}$ in unitary space by averaging the $2k$th power of the overlap between two independent random unitaries drawn from $\mathrm{G}$,
\begin{equation}
\mathfrak{F}_{\mathrm{G}}^{(k)}
=
\mathbb E_{U,V\in \mathrm{G}}
\!\left[
\bigl|\Tr(U^\dagger V)\bigr|^{2k}
\right].
\label{eq:unitary_fp_def}
\end{equation}
The \emph{state frame potential} instead probes how uniformly the orbit of a fixed reference state $\ket{\phi}$ under $\mathrm{G}$ covers state space:
\begin{equation}
\mathcal{F}_{\mathrm{G}}^{(k)}
=
\mathbb E_{U,V\in \mathrm{G}}
\!\left[
\bigl|\!\bra{\phi}U^\dagger V\ket{\phi}\bigr|^{2k}
\right].
\label{eq:state_fp_def}
\end{equation}
A small frame potential signals a highly uniform distribution, whereas a large one indicates clustering. The minimum of the unitary frame potential over all ensembles is achieved by the Haar measure over the full unitary group, for which $\mathfrak{F}_{\mathrm{Haar}}^{(k)}=k!$ (coinciding with the number of permutations on $k$ elements). When a finite group $\mathrm{G}$ saturates this bound, i.e., $\mathfrak{F}_{\mathrm{G}}^{(k)}=k!$, the ensemble is called a unitary $k$-design~\cite{Dankert09, Brandao2016}. Analogous design conditions can be formulated for state ensembles by comparing $\mathcal{F}_{\mathrm{G}}^{(k)}$ to the state frame potential of the Haar orbit.

In the remainder of this section, we evaluate both types of frame potential for the matchgate and Clifford--matchgate ensembles, using the commutant machinery developed above.

\subsubsection{Unitary frame potential}
\label{sec:unitary_FP}

By Haar invariance, Eq.~\eqref{eq:unitary_fp_def} reduces to the single-unitary average
\begin{equation}
\mathfrak{F}_{\mathrm{G}}^{(k)}
=
\mathbb E_{U\in \mathrm{G}}
\!\left[
|\Tr(U)|^{2k}
\right]
=
\Tr\!\left[
\Twirl_{\mathrm{G}}^{(k)}\!\left(\mathbf{1}^{\otimes k}\right)
\right].
\label{eq:unitary_fp_twirl}
\end{equation}
For the matchgate ensemble, the twirl projects onto the commutant, and the identity is decomposed over the commutant basis. In other words, the unitary frame potential simply equals the dimension of the $k$th-order commutant. For the matchgate group, we thus obtain
\begin{equation}
\mathfrak{F}_{\MG_n}^{(k)} = \dim \Com_k(\MG_n) = \prod_{1\le i\le j\le k-1}\frac{2n+i+j-1}{i+j-1}.
\label{eq:unitary_fp_result}
\end{equation}
In Appendix~\ref{app:RMT}, we provide a complementary random matrix theory derivation of this result.
Analogously for the Clifford matchgate group $G = \CM_n$, using the twirl~\eqref{eq:CM_twirl_basis} we find
\begin{equation}
\mathfrak{F}_{\CM_n}^{(k)} = \dim\Com_k(\CM_n)
=
\binom{2n+2^{k-1}-1}{2^{k-1}-1},
\label{eq:cmg_fp_result}
\end{equation}
with $\mathfrak{F}_{\CM_n}^{(k)} > \mathfrak{F}_{\MG_n}^{(k)} $ for $k\geq4$, and equality at $k\leq3$. Consequently, $\CM_n$ forms an exact 3-design for the continuous $\MG_n$ group. For $k \geq 4$, approaching the statistical properties of $\MG_n$—and moving closer to the unitary limit—requires supplementing the discrete $\CM_n$ operations with non-Clifford matchgate resources, such as $T$-gates~\cite{Casas2026matchgates}, see also Sec.~\ref{subsec:CliffMatchDesi}.

\subsubsection{State frame potential}
\label{sec:state_FP_sub}

We now turn to the state frame potential. For the fermionic Gaussian ensembles, the reference state is the vacuum $\ket{\mathbf{0}}=\ket{0}^{\otimes n}$, and the ensemble consists of pure states
\begin{equation}
\ket{\psi_U}=U\ket{\mathbf 0},
\qquad U\in \mathrm{G}.
\end{equation}
The $k$th state frame potential is
\begin{equation}
\mathcal F_{\mathrm{G}}^{(k)}
=
\mathbb E_{U,V\in \mathrm{G}}
\left[
\left|
\braket{\psi_U}{\psi_V}
\right|^{2k}
\right].
\label{eq:frame_potential_overlap_general}
\end{equation}
The key step is to rewrite the overlap in replica form. Introducing the density matrix $\rho_U:=\ketbra{\psi_U}{\psi_U}= U\ketbra{\mathbf 0}{\mathbf 0}U^\dagger$, we have
\begin{equation}
\left|
\braket{\psi_U}{\psi_V}
\right|^{2k}
=
\bigl(\Tr[\rho_U\rho_V]\bigr)^k
=
\Tr\!\left[\rho_U^{\otimes k}\rho_V^{\otimes k}\right].
\label{eq:frame_potential_replica_identity}
\end{equation}
The state frame potential therefore admits the replica representation
\begin{equation}
\mathcal F_{\mathrm{G}}^{(k)}
=
\mathbb E_{U,V\in \mathrm{G}}
\Tr\!\left[\rho_U^{\otimes k}\rho_V^{\otimes k}\right] = \Tr\!\left[
\left(
\mathbb E_{U\in \mathrm{G}}\rho_U^{\otimes k}
\right)^{\!2}
\right],
\label{eq:frame_potential_replica_general}
\end{equation}
where the latter equality holds because $U$ and $V$ are drawn independently.

For the matchgate group $\mathrm{G}=\MG_n$, the average in Eq.~\eqref{eq:frame_potential_replica_general} is the matchgate twirl of the vacuum projector,
\begin{equation}
\mathbb E_{U\in\MG_n}\rho_U^{\otimes k}
=
\Twirl_{\MG_n}^{(k)}\!\left(\ketbra{\mathbf 0}{\mathbf 0}^{\otimes k}\right),
\end{equation}
which, using the vacuum twirl formula~\eqref{eq:vacuum_twirl_general_compact} and the projector property $(P_{\mathbf 0}^{(k)})^2 = P_{\mathbf 0}^{(k)}$, yields
\begin{equation}
\mathcal F_{\MG_n}^{(k)}
=
\frac{\Tr P_{\mathbf 0}^{(k)}}{\bigl(\Tr P_{\mathbf 0}^{(k)}\bigr)^2}
= \frac{1}{\Tr P_{\mathbf 0}^{(k)}}
= \frac{1}{2}\prod_{1\le i<j\le n}
\frac{2n-i-j}{k+2n-i-j}.
\label{eq:FG_frame_potential_projector_eval}
\end{equation}
In Appendix~\ref{app:RMTfp}, we provide a complementary random matrix theory derivation of this result.

For the Clifford--matchgate ensemble $\mathrm{G}= \CM_n$, we use the moment operator~\eqref{eq:stab_ff_moments} to obtain
\begin{equation}
    \mathcal{F}^{(k)}_{\CM_n} = \Tr\!\left[\Twirl_{\CM_n}^{(k)}\!\left(\ketbra{\mathbf 0}{\mathbf 0}^{\otimes k}\right)^{\!2} \right] = 2^{-kn} \sum_{\{x_I\}} \frac{\binom{n}{\{x_I\}}^2}{\binom{2n}{2\{x_I\}}},
\end{equation}
where the sum runs over nonnegative integers $\{x_I\}_{I\in\mathcal I_k}$ satisfying $\sum_{I}x_I=n$. Using the identity $\sum_{\{x_I\}}\prod_{I}(2x_I)!/x_I!^2 = 4^n\binom{n+2^{k-2}-1}{2^{k-2}-1}$, which follows from the generating function $(1-4x)^{-1/2}=\sum_{j\geq 0}\binom{2j}{j}x^j$, we obtain
\begin{equation}
    \mathcal{F}^{(k)}_{\CM_n} = \frac{2^{n(2-k)}}{\binom{2n}{n}} \binom{n+2^{k-2}-1}{2^{k-2}-1}.
    \label{eq:CM_frame_potential_final}
\end{equation}

\subsection{Matchgate state designs}
\label{subsec:CliffMatchDesi}

A central question in the theory of randomness in quantum systems is whether a given ensemble forms an approximate unitary or state $k$-design~\cite{mele2024introductiontohaar}. An ensemble $\mathcal{E}$ is a state $k$-design if its $k$th moment operator matches that of the full symmetry group---in our case, the matchgate group $\MG_n$. Since the Clifford--matchgate group $\CM_n$ is a finite subgroup of $\MG_n$, a natural question is whether $\CM_n$ can serve as an efficient $k$-design for the matchgate ensemble.

For $k\le 3$, the commutants of the two groups coincide~\cite{Wan23}, $\Com_k(\CM_n)=\Com_k(\MG_n)$, which implies that the Clifford--matchgate ensemble is an \emph{exact} matchgate $k$-design at these orders. This mirrors the situation for the usual Clifford group, whose moments up to $k=3$ already reproduce those of the Haar ensemble. Given this strong agreement at the second and third moments, it is natural to ask whether the set of stab
ilizer free fermions might also approximately reproduce the fourth moment of free fermions, as happens in the non-free-fermionic setting. In particular, although the Clifford group has a large relative error as a unitary $4$-design~\cite{Helsen23}, the set of stabilizer states nevertheless forms a state $4$-design with only constant relative error~\cite{Bittel25commutant}.

Clearly, at $k=4$, the commutants diverge: $\dim\Com_4(\CM_n)=\binom{2n+7}{7}$ exceeds $\dim\Com_4(\MG_n)=\binom{2n+6}{6}+\binom{2n+5}{6}$, so the Clifford--matchgate ensemble ceases to be an exact design. The key question is therefore whether $\CM_n$ nevertheless provides a good \emph{approximate} matchgate state $4$-design. To quantify this, we consider the notion of relative error~\cite{mele2024introductiontohaar}.

An ensemble $\mathcal{E}$ is a relative-error $\epsilon$-approximate matchgate state $k$-design if
\begin{equation} \label{eq:rel_error_state_design}
    (1\!-\epsilon)\Twirl_{\MG_n}^{(k)}\!\bigl(\ketbra{\mathbf 0}{\mathbf 0}^{\otimes k}\bigr) \!\leq \! \Twirl_{\mathcal{E}}^{(k)}\!\bigl(\ketbra{\mathbf 0}{\mathbf 0}^{\otimes k}\bigr) \leq(1\!+\epsilon)\Twirl_{\MG_n}^{(k)} \!\bigl(\ketbra{\mathbf 0}{\mathbf 0}^{\otimes k}\bigr),
\end{equation}
in the operator ordering. A convenient diagnostic is the anticoncentration parameter $I_\mathcal{E}^{(k)} = \Tr\bigl(\ketbra{\mathbf 0}{\mathbf 0}^{\otimes k} \Twirl_\mathcal{E}^{(k)}(\ketbra{\mathbf 0}{\mathbf 0}^{\otimes k})\bigr)$, which for both $\MG_n$ and $\CM_n$ equals the state frame potential. Using the asymptotic expansion of the Clifford--matchgate state frame potential~\eqref{eq:CM_frame_potential_final} and the asymptotics of the matchgate state frame potential, we obtain $I_{\CM_n}^{(4)} \simeq n^{7/2}\, 2^{-4n}$ and $I_{\MG_n}^{(4)} \simeq n^{3}\, 2^{-4n}$. Combining these yields a relative deviation 
\begin{equation}
    \left\vert\frac{I_{\CM_n}^{(4)}-I_{\MG_n}^{(4)}}{I_{\MG_n}^{(4)}} \right\vert \sim \sqrt{n}.
    \label{eq:rel_error_sqrt_N}
\end{equation}
Thus the relative error \emph{grows} with system size (at least as $\sqrt{n}$)---in stark contrast to the analogous non-matchgate setting, where the relative error for state design is constant~\cite{Bittel25commutant}. This divergence reflects the growing structural mismatch between the discrete signed-permutation symmetry of $\CM_n$ and the continuous $\mathrm{O}(2n)$ symmetry of $\MG_n$ at $k=4$.

\subsection{Gaussian de Finetti theorem}
\label{sec:definetti}

The quantum de Finetti theorem~\cite{Christandl2007} asserts that a quantum state that is symmetric under permutation of many copies must be well approximated, on a small number of copies, by convex combinations of independent and identically distributed (i.i.d.) product states. It has been adapted to the stabilizer~\cite{gross2021schurweylduality} and bosonic Gaussian setting~\cite{Leverrier_2009,Leverrier_2018}. Using the formalism developed in this work, we prove a new version of de Finetti theorems adapted to natural notion of Gaussian symmetry, encoded by the operators $\Lambda_{ij}$. Our de Finetti theorem shows that states which are “Gaussian-symmetric” across many copies are well approximated by convex combinations of tensor powers of fermionic Gaussian states. 

Concretely, a state $\rho$ on $\mathcal{H}_n^{\otimes k}$ is \emph{Gaussian-symmetric} if $\Lambda_{ab}\rho =0$ for all $1\leq a<b\leq k$, meaning it lies in the trivial sector of the replica $\mathrm{SO}(k)$ action. Such a state is supported on the Gaussian-symmetric subspace, i.e.\ the image of the projector $P_{\mathbf 0}^{(k)}$ onto the trivial GT sector $\nu=(0,\ldots,0)$.

\paragraph{Pure-state version.}
Let $\ket{\psi}\in\mathcal{H}_n^{\otimes k}$ satisfy $\Lambda_{ab}\ket{\psi} =0$ for all $a<b$. Then there exists a probability measure $\mu$ on pure Gaussian states $\ket{\phi}$ such that the reduced state on any $\ell$ copies satisfies
\begin{equation} \label{eq:gaussian_de_finetti_pure_main}
    \frac{1}{2}\bigl\|\psi_{1\dots \ell} - \int d\mu(\phi)\, \phi^{\otimes \ell} \bigr\|_1 \leq \frac{\ell\, n(n-1)}{k+1}.
\end{equation}
In particular, the probability measure is given by $d\mu(\phi) = \Tr(\phi^{\otimes k}\psi) d\phi$.

The proof follows directly from the standard de Finetti argument~\cite{Christandl2007}, where the error bound is governed by the ratio $R_G = \Tr P_{\mathbf 0}^{(k-\ell)}/\Tr P_{\mathbf 0}^{(k)}$. Using the explicit formula~\eqref{eq:dim_proj_gsym}, this ratio takes the product form
\begin{equation} \label{eq:ratio_main}
R_G = \prod_{j=0}^{n-1} \frac{k-\ell+2n-i-j}{k+2n-i-j} = \prod_{j=0}^{n-1} \left(1 - \frac{\ell}{k+2n-i-j} \right).
\end{equation}
Applying the Weierstrass product inequality $\prod_i(1 - s_i) \ge 1-\sum_i s_i$ and bounding each denominator from below by $k+1$ (since $i+j \le 2n-1$ for $1\le i<j\le n$), we arrive at
\begin{equation}
    R_G \ge 1- \frac{\ell\, n(n-1)}{2(k+1)},
\end{equation}
from which Eq.~\eqref{eq:gaussian_de_finetti_pure_main} follows via $\epsilon = 2(1-R_G)$.

\paragraph{Mixed-state extension.}
The result extends to mixed states by spectral decomposition. If $\rho = \sum_i a_i \ketbra{\psi_i}{\psi_i}$ satisfies $\Lambda_{ab}\rho=0$, then each eigenvector satisfies $\Lambda_{ab}\ket{\psi_i}=0$ individually (since $\Lambda_{ab}\rho\ket{\psi_i}=a_i\Lambda_{ab}\ket{\psi_i}=0$). Applying the pure-state bound to each $\ket{\psi_i}$ and invoking convexity of the trace norm yields
\begin{equation}
\frac{1}{2}\bigl\|
\rho_{1\dots \ell}
-
\sigma^{(\ell)}
\bigr\|_1
\le
\frac{\ell\, n(n-1)}{k+1},
\label{eq:gaussian_de_finetti_mixed_main}
\end{equation}
where $\sigma^{(\ell)}=\sum_i a_i \int d\mu_i(\phi)\,\phi^{\otimes \ell}$ is a convex combination of Gaussian product states.

\paragraph{Comparison with prior work.}
A version of this bound was previously established in Ref.~\cite{Vershynina2014} for $\ell=1$, with an error scaling as $\mathcal{O}(n\, 2^{4n}\, k^{-1/3})$. Our result provides an exponential improvement in the dependence on system size, reducing it to the polynomial bound $\mathcal{O}(n^2/k)$. Moreover, Ref.~\cite{melo2013} introduced a criterion for identifying convex-Gaussian states based on the existence of a Gaussian-symmetric extension. Our de Finetti theorem further implies that if a state is $\epsilon$-far from the set of convex-Gaussian states, then the criterion needs to be verified only up to $\mathcal{O}(n^2 \epsilon^{-1})$ replicas.

\subsection{Nonstabilizerness of fermionic Gaussian states}
\label{sec:magicsre}

A natural application of the matchgate commutant at $k=4$ is the evaluation of the stabilizer R\'enyi entropy (SRE)~\cite{leone2022stabilizerrenyientropy,leone2024stabilizer} for random fermionic Gaussian states. The SRE quantifies how far a quantum state is from the set of stabilizer states---a property known as nonstabilizerness, or magic---and plays a central role in quantum resource theory. It is nonnegative, vanishes precisely on stabilizer states, is invariant under Clifford unitaries, and is additive under tensor products.

For a pure $n$-qubit state $\ket{\psi}$, the second SRE ($q=2$) is defined as
\begin{equation}
\mathcal{M}(\ket{\psi})
:=
-\log_2\!\left(
\frac{1}{2^n}
\sum_{P\in\mathcal P_n}
\langle \psi|P|\psi\rangle^4
\right),
\label{eq:M2_def_note}
\end{equation}
where $\mathcal P_n$ denotes the $4^n$ Pauli strings on $n$ qubits (products of $\mathbb{1},X,Y,Z$, modulo phases).

We consider random fermionic Gaussian states $\ket{\psi_O}=U_O\ket{\mathbf 0}$ generated by Haar-random orthogonal matrices $O\in \mathrm{O}(2n)$, and define the annealed SRE as
\begin{equation}
\mathcal{M}_{\MG_n}^{\mathrm{ann}}
:=
-\log_2  \left[ \mathbb E_{O\sim \mathrm O(2n)}
\!\left(
\frac{1}{2^n}
\sum_{P\in\mathcal P_n}
\langle \psi_O|P|\psi_O\rangle^4
\right) \right].
\label{eq:annealed_M2_def_note}
\end{equation}

The key step is to introduce the four-replica operator
\begin{equation}
Q_4 = \prod_{j=1}^n \left( \sum_{P \in \{I, X, Y, Z\}} P_j^{\otimes 4} \right),
\end{equation}
introduced in the context of the Clifford group~\cite{zhu2016clifford, roth2018recoveringquantumgates, leone2021quantum}, Eq.~\eqref{eq:annealed_M2_def_note} may be rewritten as
\begin{equation}
M_{\MG_n}^{\mathrm{ann}}
=
-\log_2 \left[
\frac{1}{2^n}
\Tr\!\left(
\Twirl_{\MG_n}^{(4)}
\!\left(\ketbra{\mathbf 0}{\mathbf 0}^{\otimes 4}\right)
Q_4
\right)
\right].
\label{eq:M2_twirl_formula_note}
\end{equation}
Thus, for $k=4$, the problem reduces to evaluating the overlap of $Q_4$ with the projector $P_{\mathbf 0}^{(4)}/\Tr P_{\mathbf 0}^{(k)}$~\eqref{eq:vacuum_twirl_general_compact} onto the trivial GT sector of the matchgate-twirled vacuum.

In Appendix~\ref{app:sre} we show that $Q_4$ can be written in Majorana form as
\begin{equation}
Q_4 = \prod_{\mu=1}^{2n} \bigl(I^{\otimes 4} + \gamma_\mu^{\otimes 4}\bigr),
\label{eq:Q4_product_gamma_final}
\end{equation}
see also Ref.~\cite{Bera2025SYK}. Equivalently,
\begin{equation}
Q_4 = 2^{2n} P_+,
\qquad
P_+ := \prod_{\mu=1}^{2n}\frac{1+\chi_\mu}{2},
\label{eq:Q4_Pplus_note}
\end{equation}
where $P_+$ projects onto the simultaneous $+1$ eigenspace of the commuting local chirality operators $\chi_\mu := \gamma_\mu^{\otimes 4}$.

To evaluate $\Tr(P_{\mathbf 0}^{(4)}P_+)$, we decompose the $SO(3)$ generators $M_i$ of Eq.~\eqref{eq:k4_Mi} into contributions from individual Majorana labels,
$
M_i = \sum_{\mu=1}^{2n} M_i^{(\mu)},
\label{eq:Mi_local_sum}
$
with
\begin{equation}
M_1^{(\mu)}\!\!=\tfrac12\gamma_\mu^{(2)}\gamma_\mu^{(3)}, \!\!
\quad
M_2^{(\mu)}\!\!=-\tfrac12\gamma_\mu^{(1)}\gamma_\mu^{(3)},\!\!
\quad
M_3^{(\mu)}\!\!=\tfrac12\gamma_\mu^{(1)}\gamma_\mu^{(2)}.
\label{eq:Mi_local}
\end{equation}
For each fixed $\mu$, the operators $M_i^{(\mu)}$ satisfy the $\mathfrak{so}(3)$ commutation relations and commute with the local chirality operator $\chi_\mu$.
Hence the subspace selected by
\[
P_\mu^+:=\frac{1+\chi_\mu}{2}
\]
is invariant under the local $SO(3)$ action. The corresponding local Casimir is
\begin{equation}
K_\mu^{(3)}:=\bigl(M_1^{(\mu)}\bigr)^2+\bigl(M_2^{(\mu)}\bigr)^2+\bigl(M_3^{(\mu)}\bigr)^2.
\label{eq:K3_local}
\end{equation}
Using $(\gamma_\mu^{(a)}\gamma_\mu^{(b)})^2=-I$, we obtain
\begin{equation}
K_\mu^{(3)}=-\frac34\,I.
\label{eq:K3_local_value}
\end{equation}
Since the $SO(3)$ Casimir eigenvalues are $-s(s+1)$, this identifies the image of $P_\mu^+$ as a spin-$\tfrac12$ representation of the local $SO(3)$ algebra.

Because $P_+ = \prod_\mu P_\mu^+$, the full image of $P_+$ is therefore equivalent, with respect to the generators $M_i$, to a tensor product of $2n$ spin-$\tfrac12$ representations. On the other hand, $P_{\mathbf 0}^{(4)}$ projects onto the trivial GT sector $\nu=(0,0)$, which along the chain
\[
SO(4)\downarrow SO(3)\downarrow SO(2)
\]
corresponds uniquely to $s=0$ and $m=0$. Therefore $\Tr(P_{\mathbf 0}^{(4)}P_+)$ is exactly the multiplicity of the total-spin singlet in $(\mathbb C^2)^{\otimes 2n}$, namely
\begin{equation}
\Tr\!\left(P_{\mathbf 0}^{(4)}P_+\right)
=
\binom{2n}{n}-\binom{2n}{n+1}
=
C_n,
\qquad
C_n=\frac{1}{n+1}\binom{2n}{n}.
\label{eq:P0Pplus_Catalan}
\end{equation}
Using Eq.~\eqref{eq:Q4_Pplus_note}, this gives
\begin{equation}
\Tr\!\left(P_{\mathbf 0}^{(4)}Q_4\right)=2^{2n}C_n.
\label{eq:TrP0Q4_Catalan}
\end{equation}
Finally, dividing by the normalization of the vacuum sector,
we arrive at
\begin{equation}
\Tr\!\left(
\Twirl_{\MG_n}^{(4)}
\!\left(\ketbra{\mathbf 0}{\mathbf 0}^{\otimes 4}\right)
Q_4
\right)
=
\frac{\Tr(P_{\mathbf 0}^{(4)}Q_4)}{\Tr P_{\mathbf 0}^{(4)}}
=
\frac{2^{2n}}{C_{n+1}}.
\label{eq:twirled_trace_final_note}
\end{equation}
Substituting this into Eq.~\eqref{eq:M2_twirl_formula_note}, we obtain
\begin{equation}
M_{\MG_n}^{\mathrm{ann}}
=
-\log_2\!\left(\frac{2^n}{C_{n+1}}\right).
\label{eq:M2ann_final_note}
\end{equation}
Using the Stirling expansion of the Catalan numbers, we obtain
\begin{equation}
\mathcal{M}_{\MG_n}^{\mathrm{ann}} 
=
n+2-\frac{3}{2}\log_2 n-\frac{1}{2}\log_2\pi
-\frac{21}{8\ln 2}\frac{1}{n}
+\mathcal O(n^{-2}).
\label{eq:M2_asymp_note}
\end{equation}
This completes the $k=4$ evaluation of the annealed SRE. The leading behavior is linear in system size $n$, with subleading logarithmic corrections. This scaling is consistent with numerically obtained averages for fermionic Gaussian states in the fixed particle-number sector~\cite{Collura24fermionicgaussian}.

\subsection{Measures of fermionic non-Gaussianity}
\label{subsec:nonGaussianity}

The matchgate commutant provides a systematic framework for constructing measures of fermionic non-Gaussianity, in direct analogy with how the Clifford commutant underlies the stabilizer R\'enyi entropy (SRE)~\cite{leone2022stabilizerrenyientropy}. The guiding principle is that any operator $W\in\Com_k(\MG_n)$ defines a Gaussian-invariant quantity
\begin{equation}
    \varphi_W(\ket{\Psi}) := \Tr\!\bigl[W\,(|\Psi\rangle\langle\Psi|)^{\otimes k}\bigr],
    \label{eq:nongauss_general}
\end{equation}
since $[W,U_G^{\otimes k}]=0$ for all $U_G\in\MG_n$ by definition. The explicit commutant bases constructed in this work therefore furnish a complete catalog of candidate non-Gaussianity measures at each replica number $k$.

A first natural class of such measures is obtained from the fermionic antiflatness (FAF), introduced in Ref.~\cite{Sierant25faf}.
The FAF is defined by $\mathcal{F}_k(\ket{\Psi}) = n - \frac{1}{2}\Tr[(\CorrMat^T \CorrMat)^k]$, where $\CorrMat$ is the $2n\times 2n$ antisymmetric covariance matrix with entries $\CorrMat_{mn}=-\frac{i}{2}\langle[\gamma_m,\gamma_n]\rangle$.
In the present language, $\mathcal{F}_k$ is built from a specific element of $\Com_{2k}(\MG_n)$ corresponding to the maximally symmetric pairing sector $r_1=\cdots=r_{2k}=1$; it involves only the two-point Majorana correlators and can be efficiently computed as a function of $\CorrMat$ alone~\cite{Sierant25faf}.

The commutant framework, however, reveals measures that go beyond the FAF construction. Two further classes are particularly natural.
First, the projector $P_{\mathbf{0}}^{(k)}$ onto the trivial $\mathrm{SO}(k)$ sector of the replicated Hilbert space belongs to $\Com_k(\MG_n)$ (cf.\ Sec.~\ref{subsec:MG_vacuum_twirl}). Its expectation value on the replicated state,
\begin{equation}
    \varphi_0^{(k)}(\ket{\Psi}) := \Tr\!\bigl[P_{\mathbf{0}}^{(k)}\,(|\Psi\rangle\langle\Psi|)^{\otimes k}\bigr],
    \label{eq:nongauss_projector}
\end{equation}
measures the overlap of $\ket{\Psi}^{\otimes k}$ with the Gaussian-symmetric subspace, and equals unity if and only if $\ket{\Psi}$ is a fermionic Gaussian state.
Second, the Casimir operators $\mathcal{C}^{(k)}_j$ of the replica $\mathrm{SO}(k)$ action commute with all matchgate unitaries and hence lie in $\Com_k(\MG_n)$. For generic non-Gaussian states, these Casimirs take values distinct from those on the Gaussian manifold, providing independent probes of non-Gaussianity that are sensitive to different aspects of the state's higher-order correlations.

From a practical standpoint, all measures of the form~\eqref{eq:nongauss_general} share the key advantage of requiring no optimization or minimization---in contrast to the fermionic rank or the Gaussian extent~\cite{Hebenstreit19, Lyu24NGE}. When $\ket{\Psi}$ admits an efficient tensor-network representation, the replica trace can be evaluated by contracting the tensor network on $k$ copies, at a cost polynomial in system size for fixed $k$. Since the commutant elements are polynomial in the bridge operators $\Lambda_{ab}=\sum_\mu\gamma_\mu^{(a)}\gamma_\mu^{(b)}$, the unitaries $e^{\frac{\theta}{2}\Lambda_{ab}}$ provide a natural gate set for implementing these measures experimentally, see Sec.~\ref{sec:rep_theoretic_matchgate_commutant}. 
For example, $e^{i\frac{\theta}{2}\Lambda_{12}}$ is precisely the \emph{fermionic beam splitter} acting on two copies of the system, generating the fermionic convolution introduced in Ref.~\cite{Lyu24NGE}. The Gaussianity test of Ref.~\cite{Lyu24NGE} uses three copies of $\ket{\Psi}$, applies $e^{i\frac{\pi}{8}\Lambda_{12}}$ to mix two of them, and then measures the overlap with the third---a protocol that directly probes the relevant commutant element and is achievable with standard fermionic linear-optics operations.

\section{Conclusion}
\subsection{Summary}

We have presented a complete algebraic theory of the commutants generated by matchgate and Clifford--matchgate ensembles acting on $k$ replicas of a system of $n$ fermionic modes. For the continuous matchgate group, the bridge operators $\Lambda_{ab}$ generate a replica $\mathfrak{so}(k)$ Lie algebra that organizes the commutant into irreducible sectors. Using a Gelfand--Tsetlin construction adapted to this $\mathrm{SO}(k)$ symmetry, we obtained an orthonormal basis of the commutant that is explicitly constructive at every $k$ and $n$, together with the closed-form dimension formula $\dim\Com_k(\MG_n)=\prod_{1\le i\le j\le k-1}\frac{2n+i+j-1}{i+j-1}$.

A distinctive feature of the matchgate commutant, compared with its Clifford counterpart, is the absence of a tensor-product decomposition over physical qubits. In the Clifford case, the $n$-qubit structure induces a natural factorization of the commutant~\cite{Bittel25commutant,gross2021schurweylduality}, keeping the Gram matrix of basis operators manageable even for non-orthogonal bases. 
For matchgates, the $\mathrm{SO}(2n)$ action couples all $2n$ Majorana modes globally, and the pairing-operator basis would naturally develop nontrivial overlaps starting at $k=4$. 
The orthonormality of the GT basis is therefore
essential: it is the way to make exact twirl calculations tractable at arbitrary replica number $k$ and system size $n$.

Analyzing the discrete Clifford--matchgate group, we provided a full characterization of the Clifford matchgate commutant basis, with operators indexed by replica occupation patterns. This enabled finding the exact dimension of Clifford--matchgate commutant, $\dim\Com_k(\CM_n)=\binom{2n+2^{k-1}-1}{2^{k-1}-1}$, and derivation of the closed-form frame potentials at all $k$. Comparing the two commutants reveals the precise ``design gap'' between the continuous matchgate and discrete Clifford matchgate groups, that manifests itself in commutants which diverge for $k\geq 4$ replicas. 

As applications of our matchgate commutant construction, we derived explicit matchgate and Clifford matchgate twirl formulas, and used them to compute the $k$-copy twirled vacuum state. The latter, enabled us to derive closed form formulas for the matchgate and Clifford matchgate state frame potentials and to characterize the nonstabilizerness of random fermionic Gaussian states.
To further showcase the power of the commutant framework, we demonstrated a fermionic de~Finetti theorem, and discussed a systematic catalog of Gaussian-invariant quantities encompassing both the fermionic antiflatness and other probes of fermionic non-Gaussianity.

\subsection{Outlook}

\textbf{\textit{Fermionic Weingarten calculus and design depth.}}---The commutant bases developed here provide the essential ingredients for a fermionic Weingarten calculus---the analogue of the integration formulas that have proved transformative for the Haar~\cite{Weingarten78,Collins2006} and Clifford~\cite{gross2021schurweylduality,Bittel25commutant} groups. A natural next step is to develop computationally efficient implementations of these formulas at higher $k$, where the GT basis structure can be exploited to avoid costly Gram-matrix inversions. Combined with the exact frame potentials derived here, such a calculus would sharpen finite-size estimates for the circuit depth required to form approximate matchgate $k$-designs~\cite{Braccia2025flo,haferkamp2022linear}, with implications for the typicality and decoupling properties of parity-preserving fermionic dynamics.

\textbf{\textit{Non-Gaussianity measures in many-body physics.}}---While we have established a complete catalog of Gaussian-invariant measures via the commutant framework, a key open question is how these measures behave in physically relevant many-body scenarios. It remains to be understood how the projector-based measure $\varphi_0^{(k)}$, the Casimir-based probes $\mathcal{C}^{(k)}_j$, and higher-order commutant elements evolve under quench dynamics~\cite{calabrese2006timedependenceof,polkovnikov2011colloquium}, across quantum phase transitions~\cite{Sachdev_1993}, or in the presence of dissipation~\cite{passarelli2024chaosmagicdissipativequantum}. Of particular interest is whether these measures can diagnose non-Gaussian features in interacting fermionic systems more sensitively than the fermionic antiflatness~\cite{Sierant25faf}, which is based only on two-point Majorana correlation functions. Efficient evaluation strategies---for instance, exploiting the polynomial structure of commutant elements in the bridge operators combined with tensor-network techniques~\cite{Schollwock2011,Banuls23,ran2020tensor}---would make such investigations computationally feasible.

\textbf{\textit{Non-stabilizerness and fermionic non-Gaussianity.}}---The commutant framework naturally connects to the resource-theoretic study of magic and non-stabilizerness in fermionic systems~\cite{Sierant25faf,Coffman25magic,Collura24fermionicgaussian}. While fermionic Gaussian states are ``free'' from the perspective of classical simulation, they generically possess high stabilizer R\'enyi entropy~\cite{leone2022stabilizerrenyientropy} that quantifies their complexity with respect to the Clifford group~\cite{turkeshi2024magic,leone2021quantum}, as shown in Sec.~\ref{sec:magicsre}.  
Clifford matchgate commutant provides a natural bridge between the resource theories of non-Gaussianity and non-stabilizerness, which have so far been developed largely in parallel.

\textbf{\textit{Doped matchgate circuits.}}---Realistic fermionic circuits often deviate from the free-fermion limit through the inclusion of non-Gaussian ``doping'' gates, such as number-conserving interactions or non-matchgate entangling operations~\cite{Bravyi16,Oszmaniec22,ReardonSmith24improved}. Similarly to the Clifford case~\cite{leone2021quantum, Haferkamp2023,magni2025anticoncentrationcliffordcircuitsbeyond}, the matchgate commutant theory provides a natural starting point for analyzing such doped circuits: the matchgate commutant describes the invariant structure of the free-fermion part, and non-Gaussian gates can be treated perturbatively as symmetry-breaking deformations. Understanding how the commutant structure evolves under controlled doping could lead to sharper classical simulation bounds for near-free-fermion circuits~\cite{Dias24classical,ReardonSmith24improved,Hebenstreit19,Reardon24extent}, and to efficient certification protocols for the onset of genuine quantum advantage in fermionic systems~\cite{Oszmaniec22,Bittel24fermionic}.

\textbf{\textit{Testing fermionic Gaussian states.}}---
A natural direction for future work concerns applications to testing fermionic Gaussian states. In particular, the commutant perspective offers a natural setting for constructing observables tailored to matchgate symmetries, which could potentially be used to test whether a state is fermionic Gaussian or far from any fermionic Gaussian state~\cite{Lyu24NGE,Bittel24fermionic}. Such tests could be useful for verifying the preparation of fermionic Gaussian states in quantum computer, as well as for detecting non-Gaussian resources that enable quantum advantage beyond matchgate circuits.

\textbf{\textit{Commutant framework for generalized symmetries.}}---The replica symmetry approach developed here is not specific to matchgates and can in principle be extended to other physically motivated unitary ensembles with non-standard symmetry structures. The commutant formalism has recently been applied to study the interplay between symmetries and quantum information in systems with generalized or unconventional symmetries, such as dipole-conserving circuits or systems with subsystem symmetries~\cite{Moudgalya2022,Moudgalya23symm,Lastres2026nonuniversality}. It would be interesting to explore whether the bridge-algebra and GT-basis techniques introduced here---which exploit the Lie-algebraic structure of the replica space---can be adapted to characterize the commutants of other sub-universal circuit families, including number-conserving unitaries~\cite{Lyu2024gdo}, symplectic circuits, or circuits respecting non-Abelian symmetries. More broadly, the structural parallels between the matchgate commutant (governed by $\mathrm{SO}(k)$ representation theory) and the Clifford commutant (governed by permutation groups) suggest a deeper classification program for commutant algebras of restricted quantum circuits, with potential connections to invariant theory and algebraic combinatorics.
We leave these open
questions for future investigation.

\begin{acknowledgments}
\textbf{Acknowledgements.---}
We acknowledge discussions with B. Casas, N. Dowling, M. Heinrich, and S. Moudgalya. P.S.T thanks B. Kraus, F. Pollmann, S.H. Lin, B. Jobst, R. Morral-Yepes, and M. Langer for discussions and collaboration on related topics. 
X.T. acknowledges support from DFG under Germany's Excellence Strategy – Cluster of Excellence Matter and Light for Quantum Computing (ML4Q) EXC 2004/2 – 390534769, and DFG Collaborative Research Center (CRC) 183 Project No. 277101999 - project B01, and DFG Emmy Noether Programme proposal ``Digital Quantum Matter Ouf-of-Equilibrium'' No. 560726973. 
P.S. acknowledges fellowship within the “Generación D” initiative, Red.es, Ministerio para la Transformación Digital y de la Función Pública, for talent attraction (C005/24-ED CV1), funded by the European Union NextGenerationEU funds, through PRTR. P.S.T. acknowledges funding from the European Research
Council (ERC) under the European Union (ERC, DynaQuant, No. 101169765).

\textbf{Authors Contribution.---}
XT proposed the family of operators in $\Com_k(\MG_n)$ introduced in \cite{Sierant25faf}. Building on this, PS conducted numerical investigations and conjectured the dimensions of both the matchgate and Clifford-matchgate commutants. PST developed the antisymmetric pairing tensor approach for the matchgate commutant, demonstrated that these operators coincide with the algebra of bridge operators, and together with XT formalized the Clifford-matchgate commutant. PS formalized the bridge operators as the underlying principle of the $\mathfrak{so}(k)$ approach and developed the Gelfand-Tsetlin construction of the matchgate commutant. PS and XT contributed equally to drafting the manuscript, while all authors contributed to the analysis, physical interpretation of the applications, and final manuscript revisions.

\end{acknowledgments}

\appendix

\section{Details of the pairing tensors construction}
\label{app:pairingTensors_0}
\subsection{Antisymmetric pairing tensors}
\label{app:pairingTensors}
To rigorously justify that the antisymmetrized pairing tensors $T_{\mathbf{S}}^{\pi}$ are completely determined by the pairing counts $\{x_{ij}\}$, we analyze the action of the local symmetric groups. 

Let $\mathrm{S}_{r_i}$ denote the symmetric group acting on the $r_i$ indices of replica $i$. The antisymmetrization over each replica is executed by the projector $A = \bigotimes_{i=1}^k A_i$, which possesses the property $\mathcal{A}_i \sigma_i = \mathrm{sgn}(\sigma_i) \mathcal{A}_i$ for any local permutation $\sigma_i \in \mathrm{S}_{r_i}$. 

Two pairings $\pi$ and $\pi'$ with the same pairing data $\{x_{ij}\}$ are related by permuting endpoints within each replica, i.e., they lie in the same orbit under the local group $G_{\mathrm{loc}} = \mathrm{S}_{r_1} \times \dots \times \mathrm{S}_{r_k}$.
Thus, there exists $\sigma = (\sigma_1, \dots, \sigma_k) \in G_{\mathrm{loc}}$ such that
\begin{equation}
\tilde{T}^{\pi'} = \mathcal{A}(T^{\pi'}) = \mathcal{A}(\sigma \cdot T^\pi) = \left(\prod_{i=1}^k \mathrm{sgn}(\sigma_i)\right) \mathcal{A}(T^\pi) = \pm \tilde{T}^\pi.
\end{equation}
Hence, all antisymmetric tensors corresponding to the same $\{x_{ij}\}$ are equivalent up to a sign, and we may denote this equivalence class by
$\tilde{T}^{(\{x_{ij}\})}_{\mathbf{S}}$.

\subsection{Basis from pairing tensors}
\label{app:pairingTensorsBas}

\subsubsection{$k=2$}
\label{app:pairingTensorsBas2}

Here, we derive the explicit form of the commutant basis~\eqref{eq:antiK2} at $k=2$ replicas.
The matrix $x_{ij}$ specifies how many indices in replica $i$ are
paired to indices in replica $j$. For matchgate invariants, the intra-replica pairings
vanish, so $x_{11}=x_{22}=0$ and the only nontrivial entry is $x_{12}=x_{21}$, which specifies the Majorana weight sector $\Gamma_{r_1}$ as $r_1=x_{12}$.

The canonical perfect matching $\pi$ pairs the $a$-th indices of both replicas, defining the
pairing tensor
\begin{equation}
T^\pi_{S^{1},S^{2}}
=
\prod_{a=1}^r \delta_{S^{1}_a,\, S^{2}_a}.
\end{equation}
Applying the antisymmetrizer $\mathcal{A}_{r,r}$~\eqref{eq:antisymm} projects this into the
physical fermionic subspace:
\begin{equation}
\begin{split}
\tilde T^{(r)}_{S^{1},S^{2}} &
= 
(\mathcal{A}_{r,r} T^\pi)_{S^{1},S^{2}} \\
=&
\frac{1}{(r!)^2}
\sum_{\sigma_1,\sigma_2\in \mathrm{S}_r}
\sgn(\sigma_1)\sgn(\sigma_2)
\prod_{a=1}^r
\delta_{S^{1}_{\sigma_1(a)},\,S^{2}_{\sigma_2(a)}}.
\end{split}
\end{equation}
This antisymmetric tensor is nonvanishing if and only if the ordered sets $S^{1}$ and $S^{2}$
consist of the exact same subset of indices. Let $S = \{\mu_1 < \dots < \mu_r\}$ and
$S' = \{\nu_1 < \dots < \nu_r\}$ denote the sorted, underlying sets of $S^{1}$ and $S^{2}$,
respectively. Parametrizing the orderings as $S^{1} = \sigma_1(S)$ and $S^{2} = \sigma_2(S')$
for $\sigma_1, \sigma_2 \in \mathrm{S}_r$, the Kronecker delta in the product enforce
$S=S'$ (i.e., $\mu_i=\nu_i$ for all $i$), while the antisymmetrizer contributes precisely the relative sign $\sgn(\sigma_1)\sgn(\sigma_2)$ associated with the two orderings.

The corresponding commutant operator is formed by contracting $\tilde T^{(r)}_{S^{1},S^{2}}$ with Majorana
strings
\begin{equation}
\tilde T^{(r)}
=
\sum_{|S^{1}|=r}\sum_{|S^{2}|=r}
\tilde T^{(r)}_{S^{1},S^{2}}\;
\gamma_{S^{1}}\otimes \gamma_{S^{2}}.
\label{eq:taa1}
\end{equation}
Due to Majorana anticommutation, reordering a string into the canonical sorted order
$S$ generates the permutation parity, $\gamma_{\sigma(S)}=\sgn(\sigma)\gamma_{S}$. When
substituting this into the contraction, the antisymmetrizer signs cancel the reordering signs, while the constraint $S=S'$ reduces the double sum in~\eqref{eq:taa1} to a single sum, reproducing Eq.~\eqref{eq:antiK2}.

\subsubsection{$k=3$}
\label{app:pairingTensorsBas3}

For $k=3$, the replica weights $r_1,r_2,r_3$ uniquely determine the number of bridge operators between
each pair of replicas. Solving the degree constraints
\begin{equation}
r_i=\sum_{j\neq i}x_{ij},
\qquad i\in\{1,2,3\},
\end{equation}
gives
\begin{equation}\begin{split}
x_{12} = (r_1 + r_2 - r_3)/2, \\
x_{13} = (r_1 - r_2 + r_3)/2, \\
x_{23} = (-r_1 + r_2 + r_3)/2.
\label{eq:k3_bridge operators}
\end{split}
\end{equation}
Hence each replica splits into two groups of indices, according to which of the other two
replicas they are paired with: replica~1 into $x_{12}$ and $x_{13}$ indices, replica~2 into
$x_{12}$ and $x_{23}$ indices, and replica~3 into $x_{13}$ and $x_{23}$ indices. In particular,
a nontrivial invariant exists only if the quantities in Eq.~\eqref{eq:k3_bridge operators} are integers in the interval $x_{ij} \in [0, 2n]$.

Since all perfect matchings with the same pairing counts yield the same antisymmetric tensor up
to an overall sign, it is sufficient to choose a canonical pairing $\pi$. Let
$\mathbf{S}=(S^{1},S^{2},S^{3})$, where each $S^{i}$ is an ordered list of $r_i$
indices. We choose $\pi$ so that the first $x_{12}$ entries of replicas~1 and~2 are matched,
the remaining $x_{13}$ entries of replica~1 are matched with the first $x_{13}$ entries of
replica~3, and the remaining $x_{23}$ entries of replica~2 are matched with the remaining
$x_{23}$ entries of replica~3. The corresponding raw pairing tensor is
\begin{equation}
T_{\mathbf{S}}^{\pi}
=
\left( \prod_{a=1}^{x_{12}} \delta_{S^{1}_a,\, S^{2}_a} \right)
\left( \prod_{b=1}^{x_{13}} \delta_{S^{1}_{x_{12}+b},\, S^{3}_b} \right)
\left( \prod_{c=1}^{x_{23}} \delta_{S^{2}_{x_{12}+c},\, S^{3}_{x_{13}+c}} \right).
\label{eq:k3_raw_tensor}
\end{equation}

Projecting to the fermionic subspace with the replica-wise antisymmetrizer
$\mathcal{A}_{r_1,r_2,r_3}=\mathcal{A}^{(1)}_{r_1}\mathcal{A}^{(2)}_{r_2}\mathcal{A}^{(3)}_{r_3}$,
we obtain
\begin{equation}
\begin{split}
\tilde{T}_{\mathbf{S}}^{\pi}
=
&(\mathcal{A}_{r_1,r_2,r_3}T^\pi)_{\mathbf{S}}
=  \frac{1}{r_1!r_2!r_3!} \times \\
&
\sum_{\sigma_1\in\mathrm{S}_{r_1}}
\sum_{\sigma_2\in\mathrm{S}_{r_2}}
\sum_{\sigma_3\in\mathrm{S}_{r_3}}
\sgn(\sigma_1)\sgn(\sigma_2)\sgn(\sigma_3)\,
T^\pi_{\sigma(\mathbf{S})},
\label{eq:k3_antisym_tensor}
\end{split}
\end{equation}
where $\sigma(\mathbf{S})=(\sigma_1(S^{(1)}),\sigma_2(S^{(2)}),\sigma_3(S^{(3)}))$.
The Kronecker-delta structure in Eq.~\eqref{eq:k3_raw_tensor} implies that
$\tilde{T}_{\mathbf{S}}^{\pi}$ is nonzero only if the unordered subsets corresponding to each
bridge coincide across the paired replicas: the $x_{12}$ indices selected in replicas~1 and~2
must define the same subset, the $x_{13}$ indices selected in replicas~1 and~3 must define the
same subset, and the $x_{23}$ indices selected in replicas~2 and~3 must define the same subset.
Equivalently, the support of $\tilde{T}_{\mathbf{S}}^{\pi}$ is labeled by a triple of disjoint
subsets
\begin{equation}
(A_{12},A_{13},A_{23})\subseteq[2n],
\qquad
|A_{ij}|=x_{ij},
\end{equation}
encoding the three bridge types.

Contracting $\tilde{T}_{\mathbf{S}}^{\pi}$ with Majorana strings, one obtains the corresponding
commutant operator in $\Gamma_{r_1,r_2,r_3}$,
\begin{equation}
\tilde{T}^{\pi}
=
\sum_{\mathbf{S}}
\tilde{T}_{\mathbf{S}}^{\pi}\,
\gamma_{S^{(1)}}\otimes\gamma_{S^{(2)}}\otimes\gamma_{S^{(3)}}.
\end{equation}
As in the $k=2$ case, the permutation signs generated by the antisymmetrizer are canceled by the
signs acquired when reordering the Majorana strings into canonical sorted order. Therefore, the
sum collapses from ordered lists to unordered bridge subsets, yielding, up to an overall
normalization constant,
\begin{equation}
\tilde{T}^{(x_{12},x_{13},x_{23})} \!
\propto \!\!
\sum_{\substack{
A_{ij}\subseteq[2n]\\
|A_{ij}|=x_{ij}\\
A_{ij}\ \mathrm{disjoint}
}} \!\!\!
\gamma_{A_{12}\cup A_{13}}
\otimes
\gamma_{A_{12}\cup A_{23}}
\otimes
\gamma_{A_{13}\cup A_{23}}.
\label{eq:k3_final}
\end{equation}
This is the explicit commutant basis element for $k=3$: it is labeled by the pairing numbers $(x_{12},x_{13},x_{23})$, or equivalently by the replica weights $(r_1,r_2,r_3)$ satisfying
Eq.~\eqref{eq:k3_bridge operators}. This reproduces Eq.~\eqref{eq:antiK3} in the Main Text.

\subsubsection{Nonorthogonality and overcompleteness of pairing tensors for $k \geq 4$}
\label{app:pairingTensorsBas4}

We start by giving the simplest explicit example showing that for $k\ge 4$ the
pairing tensors $\tilde T^{(\{x_{ij}\})}$ may be non-orthogonal within a fixed Majorana-weight sector. We consider $k=4$ replicas and a single fermionic mode, $n=1$, so that the two Majoranas are $\gamma_1,\gamma_2$ and the nontrivial Majorana-weight sectors are $\Gamma_0$, $\Gamma_1$, and $\Gamma_2$.

The first sector in which multiple pairing patterns appear is $\Gamma_{1,1,1,1}$.
In this sector, each replica carries a single Majorana index, so any admissible bridge pattern
must define a perfect matching of the four replicas. There are three such matchings:
\begin{equation}
(12)(34),\qquad (13)(24),\qquad (14)(23),
\end{equation}
corresponding to the three choices
\begin{align}
x^{(a)}_{12}=x^{(a)}_{34}=1, &\qquad \text{all other }x^{(a)}_{ij}=0,\\
x^{(b)}_{13}=x^{(b)}_{24}=1, &\qquad \text{all other }x^{(b)}_{ij}=0,\\
x^{(c)}_{14}=x^{(c)}_{23}=1, &\qquad \text{all other }x^{(c)}_{ij}=0.
\end{align}
The corresponding raw pairing tensors are
\begin{align}
T^{(12)(34)}_{j_1j_2j_3j_4} &= \delta_{j_1j_2}\delta_{j_3j_4},\\
T^{(13)(24)}_{j_1j_2j_3j_4} &= \delta_{j_1j_3}\delta_{j_2j_4},\\
T^{(14)(23)}_{j_1j_2j_3j_4} &= \delta_{j_1j_4}\delta_{j_2j_3}.
\end{align}
Since each replica contains only one index, antisymmetrization is trivial in this sector, and the
corresponding commutant operators are
\begin{align}
\tilde T^{(12)(34)}
&=
\sum_{j_1,j_2,j_3,j_4=1}^2
\delta_{j_1j_2}\delta_{j_3j_4}\,
\gamma_{j_1}\otimes\gamma_{j_2}\otimes\gamma_{j_3}\otimes\gamma_{j_4}
\nonumber\\
&=
\left(\sum_{\mu=1}^2 \gamma_\mu\otimes\gamma_\mu\right)
\otimes
\left(\sum_{\nu=1}^2 \gamma_\nu\otimes\gamma_\nu\right),
\\[1mm]
\tilde T^{(13)(24)}
&=
\sum_{j_1,j_2,j_3,j_4=1}^2
\delta_{j_1j_3}\delta_{j_2j_4}\,
\gamma_{j_1}\otimes\gamma_{j_2}\otimes\gamma_{j_3}\otimes\gamma_{j_4}
\nonumber\\
&=
\sum_{\mu,\nu=1}^2
\gamma_\mu\otimes\gamma_\nu\otimes\gamma_\mu\otimes\gamma_\nu,
\\[1mm]
\tilde T^{(14)(23)}
&=
\sum_{j_1,j_2,j_3,j_4=1}^2
\delta_{j_1j_4}\delta_{j_2j_3}\,
\gamma_{j_1}\otimes\gamma_{j_2}\otimes\gamma_{j_3}\otimes\gamma_{j_4}
\nonumber\\
&=
\sum_{\mu,\nu=1}^2
\gamma_\mu\otimes\gamma_\nu\otimes\gamma_\nu\otimes\gamma_\mu.
\end{align}

These three operators belong to the same sector $\Gamma_{1,1,1,1}$, and therefore need not be
orthogonal. Indeed, using the Hilbert--Schmidt orthogonality
\begin{equation}
\Tr(\gamma_\mu\gamma_\nu)=2\,\delta_{\mu\nu},
\end{equation}
one readily finds
\begin{align}
\Tr\!\left[\tilde T^{(12)(34)}\tilde T^{(12)(34)}\right]
=
\Tr\!\left[\tilde T^{(13)(24)}\tilde T^{(13)(24)}\right]
= \nonumber \\
=\Tr\!\left[\tilde T^{(14)(23)}\tilde T^{(14)(23)}\right]
&=16,
\nonumber \\
\Tr\!\left[\tilde T^{(12)(34)}\tilde T^{(13)(24)}\right]
=
\Tr\!\left[\tilde T^{(12)(34)}\tilde T^{(14)(23)}\right]
=\nonumber \\
=\Tr\!\left[\tilde T^{(13)(24)}\tilde T^{(14)(23)}\right]
&=8. \nonumber 
\end{align}
Hence the Gram matrix is
\begin{equation}
G=
\begin{pmatrix}
16 & 8 & 8\\
8 & 16 & 8\\
8 & 8 & 16
\end{pmatrix},
\end{equation}
which is manifestly not diagonal, showing that the three pairing tensors are non-orthogonal.

Now, we continue the example of $k=4$ and $n=1$ to explicitly illustrate the overcompleteness of the pairing operators as a matchgate commutant basis. We focus our attention to the Majorana weight sector $\Gamma_{2,2,2,2}$.

In this sector, each replica carries two indices, so an admissible pairing matrix
$\{x_{ij}\}_{1\le i,j\le 4}$ is a symmetric matrix of nonnegative integers satisfying
\begin{equation}
x_{ii}=0,
\qquad
\sum_{j=1}^4 x_{ij}=2,
\qquad
i=1,2,3,4.
\label{eq:xij_constraints_2222}
\end{equation}
Solving these constraints, one finds six admissible pairing patterns:
\begin{equation}
\begin{gathered}
x_{12}=x_{34}=2,\\
x_{13}=x_{24}=2,\\
x_{14}=x_{23}=2,\\
x_{12}=x_{13}=x_{24}=x_{34}=1,\\
x_{12}=x_{14}=x_{23}=x_{34}=1,\\
x_{13}=x_{14}=x_{23}=x_{24}=1,
\end{gathered}
\label{eq:six_bridge_patterns}
\end{equation}
with all unspecified $x_{ij}$ equal to zero.

Hence, the antisymmetrized pairing construction produces six candidate commutant operators
\begin{equation}
\tilde T^{(\{x_{ij}\})}
=
\sum_{\mathbf S}
\tilde T_{\mathbf S}^{(\{x_{ij}\})}\,\gamma_{\mathbf S},
\label{eq:six_pairing_ops}
\end{equation}
all belonging to the same sector $\Gamma_{2,2,2,2}$.

However, for a single-fermion system ($n=1$), there are only $2n=2$ available Majorana modes, labeled $1$ and $2$. Because each replica in the $\Gamma_{2,2,2,2}$ sector requires an ordered subset of size $r_\ell=2$, the only possible non-vanishing choice of indices for any replica is the fully occupied set $S^\ell = \{1, 2\}$.Consequently, the sum over the $k$-tuples of subsets $\mathbf{S} = (S^1, S^2, S^3, S^4)$ in Eq.~\eqref{eq:six_pairing_ops} collapses to a single term:
\begin{equation}
    \tilde T^{(\{x_{ij}\})} 
= 
\tilde T_{(\{1,2\}, \{1,2\}, \{1,2\}, \{1,2\})}^{(\{x_{ij}\})} \, \gamma_{12} \otimes \gamma_{12} \otimes \gamma_{12} \otimes \gamma_{12},
\label{eq:g2exampl}
\end{equation}where $\gamma_{12} = \gamma_1 \gamma_2$. Because the antisymmetric pairing tensor $\tilde T_{\mathbf S}^{(\{x_{ij}\})}$ evaluates to a scalar coefficient for this single valid mode assignment, all
operators $\tilde T^{(\{x_{ij}\})} $ in Eq.~\eqref{eq:g2exampl}
are strictly proportional to the exact same tensor product of Majorana strings: $(\gamma_{12})^{\otimes 4}$.

Therefore, while there are six admissible pairing patterns in Eq.~\eqref{eq:six_bridge_patterns}, the dimension of the matchgate commutant in the $\Gamma_{2,2,2,2}$ sector for $n=1$ is exactly $1$.

This drastic collapse directly illustrates the fundamental limitation of the pairing tensor basis. The pairing configurations are in one-to-one correspondence with the basis elements of the Brauer algebra $\mathcal{B}_{R/2}(2n)$, which maps to the commutant of $O(2n)$ acting on $(\mathbb{C}^{2n})^{\otimes R/2}$ by conjugation. According to the Second Fundamental Theorem of Invariant Theory of the orthogonal group~\cite{Lehrer2012SFT}, this homomorphism is an isomorphism if and only if $R/2 \le 2n$. 
Consequently, whenever the equivalent number of operator copies $R/2$ strictly exceeds the local space dimension $2n$ (i.e., $R > 4n$), structural linear dependencies (syzygies)
necessarily arise among the pairing configurations. In our counter-example, $R/2 = 4$, which is strictly greater than $2n=2$, mathematically guaranteeing that the antisymmetric tensor pairing basis is \emph{overcomplete}. Finally, we note that this theorem also rigorously guarantees that for any replica weight sector satisfying $R \le 4n$, the homomorphism is faithful, and the pairing operators $\tilde T^{(\{x_{ij}\})}$ corresponding to valid, distinct pairing matrices remain strictly linearly independent.

\section{Bridge operators generate the matchgate commutant}
\label{app:indu}
In this Appendix, we prove that the matchgate commutant is generated by the bridge operators
\begin{equation}
\Lambda_{ab}:=\sum_{\mu=1}^{2n}\gamma^{(a)}_\mu\gamma^{(b)}_\mu,
\qquad
1\le a<b\le k.
\end{equation}
More precisely, we show that every pairing operator
$\tilde T^{(x)}:= \tilde T^{(\{x_{ab}\})}$ belongs to the associative algebra~\eqref{eq:imageU} generated by
the $\Lambda_{ab}$. Since the operators $\tilde T^{(x)}$ span the commutant, this implies
\begin{equation}
\Com_k(\MG_n)=\pi\bigl(U(\mathfrak{so}(k))\bigr),
\label{eq:thesis}
\end{equation}
where $\pi\bigl(U(\mathfrak{so}(k))\bigr)$ denotes the concrete image of the enveloping algebra in
its bridge-operator representation on $\mathcal H_n^{\otimes k}$.

Here $x=\{x_{ab}\}_{1\le a<b\le k}$ denotes admissible bridge data, and we write
\begin{equation}
|x|:=\sum_{a<b}x_{ab}
\end{equation}
for the total bridge number. We prove the claim by strong induction on $|x|$.

\paragraph*{Base case.}
For $|x|=0$, all bridge numbers vanish, and the corresponding pairing operator is
the identity operator,
\begin{equation}
\tilde T^{(0)}=\mathbb 1.
\end{equation}
Thus $\tilde T^{(0)}$ belongs to the algebra generated by the bridge operators, as the empty product of
generators.

For $|x|=1$, exactly one entry of $x$ is nonzero, say $x_{ab}=1$, while all others vanish. By
construction, the corresponding pairing operator is, up to the fixed
normalization convention,
\begin{equation}
\tilde T^{(x)}=\Lambda_{ab}.
\end{equation}
Hence the statement holds for $|x|=1$ as well.

\paragraph*{Induction step.}
Assume that for some $m\ge 1$, every pairing operator $\tilde T^{(y)}$ with
$|y|\le m$ belongs to the algebra generated by the $\Lambda_{ab}$. Let $x$ be admissible with
$|x|=m+1$. Choose any pair $(a,b)$ such that $x_{ab}>0$, and define $x^{-}$ by
\begin{equation}
x^{-}_{ab}=x_{ab}-1,
\qquad
x^{-}_{cd}=x_{cd}\quad\text{for }(c,d)\neq(a,b).
\label{eq:xminus_def}
\end{equation}
Then $|x^-|=m$.

The proof rests on the following triangularity statement.

\paragraph*{Lemma.}
Let $x$ be admissible and let $x^{-}$ be defined by Eq.~\eqref{eq:xminus_def}. Then
\begin{equation}
\tilde T^{(x^-)}\Lambda_{ab}
=
c_x\,\tilde T^{(x)}
+
\sum_{|y|<|x|} c_y\,\tilde T^{(y)},
\qquad
c_x\neq 0,
\label{eq:triangular_bridge_action}
\end{equation}
for some coefficients $c_x,c_y$ depending only on the normalization convention.

\paragraph*{Proof of the lemma.}
Write $\tilde T^{(x^-)}$ as a pairing operator realizing the bridge
configuration $x^-$. Multiplication by
\begin{equation}
\Lambda_{ab}=\sum_{\mu=1}^{2n}\gamma_\mu^{(a)}\gamma_\mu^{(b)}
\end{equation}
adds one Majorana in replica $a$ and one Majorana in replica $b$, with the same mode index
$\mu$. For each monomial in $\tilde T^{(x^-)}$, there are only three possibilities.

(i) Neither $\gamma_\mu^{(a)}$ nor $\gamma_\mu^{(b)}$ contracts with a Majorana already present in
the monomial. Then the product simply adds one new bridge between replicas $a$ and $b$, producing
a monomial of bridge type $x$. After antisymmetrization, these contributions sum to a nonzero
multiple of $\tilde T^{(x)}$.

(ii) Both $\gamma_\mu^{(a)}$ and $\gamma_\mu^{(b)}$ contract with existing Majoranas in the
monomial. Since each contraction removes one Majorana from the monomial, the net number of
Majoranas decreases by two relative to the disjoint-insertion case. Because the total bridge
number is half the total number of Majoranas appearing in the monomial, the resulting bridge data
$y$ satisfy
\begin{equation}
|y|\le |x|-1.
\end{equation}

(iii) Exactly one of the two Majoranas contracts with an existing Majorana. In that case, relative
to the disjoint-insertion case, the net number of Majoranas decreases by one pair after the
product is reordered into antisymmetrized pairing form. Equivalently, one bridge is removed and
another is rerouted. Hence the resulting bridge data $y$ again satisfy
\begin{equation}
|y|\le |x|-1.
\end{equation}

No other terms occur. Therefore the product is triangular with respect to the total bridge number,
and takes the form \eqref{eq:triangular_bridge_action}. It remains to check that the coefficient
$c_x$ is nonzero. This follows because the contributions of type (i) are precisely those in which
the bridge from $\Lambda_{ab}$ is inserted disjointly from the existing bridge operators in
$\tilde T^{(x^-)}$; all such terms contribute to the same pairing operator
$\tilde T^{(x)}$, and hence their sum is nonzero.
\hfill$\square$

\paragraph*{Completion of the induction.}
By the induction hypothesis, $\tilde T^{(x^-)}$ lies in the algebra generated by the bridge operators, and
hence so does $\tilde T^{(x^-)}\Lambda_{ab}$. Using Eq.~\eqref{eq:triangular_bridge_action}, all
terms $\tilde T^{(y)}$ with $|y|<|x|$ also belong to this algebra by the induction hypothesis.
Since $c_x\neq 0$, we may solve for $\tilde T^{(x)}$:
\begin{equation}
\tilde T^{(x)}
=
c_x^{-1}
\left(
\tilde T^{(x^-)}\Lambda_{ab}
-
\sum_{|y|<|x|} c_y\,\tilde T^{(y)}
\right).
\end{equation}
Thus $\tilde T^{(x)}$ lies in the algebra generated by the bridge operators. By strong induction,
this holds for every admissible bridge configuration $x$.

Finally, since the pairing operators $\tilde T^{(x)}$ span the matchgate
commutant, we conclude that the commutant coincides with the algebra generated by the bridge operators, and Eq.~\eqref{eq:thesis} is valid.

\begin{widetext}

\section{The Gelfand--Tsetlin construction illustrated with angular momentum}
\label{app:GT_angular_momentum}

The algebraic construction of the matchgate commutant basis in Sec.~\ref{subsec:gt_basis_elems} relies on a Gelfand--Tsetlin (GT) decomposition along a subgroup chain. While the general formalism involves the orthogonal algebras $\mathfrak{so}(k)$, every essential idea already appears in the textbook theory of angular momentum. In this appendix, we illustrate the construction on the chain
\begin{equation}
\mathrm{SO}(4)\supset\mathrm{SO}(3)\supset\mathrm{SO}(2),
\label{eq:so4_chain_app}
\end{equation}
which is the simplest example that exhibits all the features of the general GT machinery: multiple Casimirs, nontrivial branching rules, multi-label GT patterns, and ladder operators acting at different levels of the chain. This chain mirrors the general GT chain~\eqref{eq:gt_chain} with $k=4$. Before tackling the full $\mathrm{SO}(4)$ example, we first recall the familiar single-step case $\mathrm{SO}(3)\supset\mathrm{SO}(2)$ to establish notation and build intuition.

\subsection{Warm-up: $\mathrm{SO}(3)\supset\mathrm{SO}(2)$ and a single angular momentum}
\label{app:warmup_so3}

Consider a quantum-mechanical system with angular momentum quantum number $\ell$.  The Hilbert space carries the irreducible representation $V_\ell$ of $\mathrm{SO}(3)$, spanned by the $2\ell+1$ states $\ket{\ell,m}$ with $m=-\ell,\dots,\ell$. The Lie algebra $\mathfrak{so}(3)$ is generated by the angular momentum operators $J_x,J_y,J_z$, satisfying the commutation relations
\begin{equation}
[J_x,J_y]=iJ_z,\qquad [J_y,J_z]=iJ_x,\qquad [J_z,J_x]=iJ_y.
\label{eq:su2_comm}
\end{equation}
The subgroup $\mathrm{SO}(2)\subset\mathrm{SO}(3)$ consists of rotations about the $z$-axis and is generated by $J_z$ alone.

\paragraph{Casimirs and quantum numbers.}
For $\mathrm{SO}(3)$, the (unique) Casimir is the total angular momentum squared,
\begin{equation}
\mathbf{J}^2 = J_x^2+J_y^2+J_z^2,
\label{eq:J2_casimir}
\end{equation}
with eigenvalue $\ell(\ell+1)$ on the irrep $V_\ell$. This operator is the analog of the quadratic Casimir~\eqref{eq:quadratic_casimir} in the matchgate setting. For $\mathrm{SO}(2)$, the Casimir is the generator itself, $J_z$, with eigenvalues $m=-\ell,\dots,\ell$.
Together, the pair $(\mathbf{J}^2,J_z)$ forms the commuting family analogous to Eq.~\eqref{eq:gt_commuting_family}. Their joint eigenvalues $(\ell,m)$ uniquely label the basis states. In the language of the main text, the ``GT pattern'' $\mathfrak{m}$ is simply the magnetic quantum number $m$.

\paragraph{Diagonal projectors.}
Once the joint eigenbasis $\{\ket{\ell,m}\}$ is known, the diagonal projectors are
\begin{equation}
P_m = \ket{\ell,m}\bra{\ell,m},\qquad m=-\ell,\dots,\ell,
\label{eq:Pm_proj}
\end{equation}
satisfying $P_m P_{m'} = \delta_{mm'}P_m$ in direct analogy with Eq.~\eqref{eq:GT_projector_relations}. Crucially, these projectors can be built \emph{without knowing the eigenstates} as polynomials in $J_z$ via spectral projection:
\begin{equation}
P_m = \prod_{\substack{m'=-\ell\\m'\neq m}}^{\ell}\frac{J_z - m'}{m-m'},
\label{eq:Pm_spectral}
\end{equation}
completely analogous to Eq.~\eqref{eq:k2_projectors} for the $k=2$ matchgate commutant.

\paragraph{Ladder operators.}
The off-diagonal operators are built from the ladder operators
\begin{equation}
J_+ = J_x + iJ_y,\qquad J_- = J_x - iJ_y,
\label{eq:ladder_ops}
\end{equation}
which shift the magnetic quantum number by one unit:
\begin{equation}
J_+\ket{\ell,m} = \sqrt{\ell(\ell+1)-m(m+1)}\;\ket{\ell,m+1},\qquad
J_-\ket{\ell,m} = \sqrt{\ell(\ell+1)-m(m-1)}\;\ket{\ell,m-1}.
\end{equation}
To connect two states $\ket{\ell,m}$ and $\ket{\ell,m'}$ with $m'>m$, one applies $J_+$ a total of $(m'-m)$ times. Setting $Y_{m',m} = (J_+)^{m'-m}$ and sandwiching between projectors,
\begin{equation}
X_{m,m'} \;\propto\; P_m\,Y_{m,m'}\,P_{m'},
\label{eq:Xmm_angular}
\end{equation}
yields, after normalization, the matrix unit $X_{m,m'} = \ket{\ell,m}\bra{\ell,m'}$. The $(2\ell+1)^2$ operators $\{X_{m,m'}\}$ form a complete orthonormal basis for $\End(V_\ell)$, exactly as in the commutant construction of Eq.~\eqref{eq:Xmn_sandwich}.

The single-step chain $\mathrm{SO}(3)\supset\mathrm{SO}(2)$ is almost too simple: the GT pattern is just one number $m$, and there is only one level of ladder operators. We now turn to the two-step chain where the full structure of the GT construction becomes visible.

\subsection{Two coupled angular momenta and the chain $\mathrm{SO}(4)\supset\mathrm{SO}(3)\supset\mathrm{SO}(2)$}
\label{app:so4_example}

\subsubsection{The Hilbert space and generators}

Consider two independent spin-$\tfrac{1}{2}$ particles, with individual angular momentum operators $\mathbf{J}_1=(J_{1x},J_{1y},J_{1z})$ and $\mathbf{J}_2=(J_{2x},J_{2y},J_{2z})$. The Hilbert space is $\mathcal{H}=\mathbb{C}^2\otimes\mathbb{C}^2\cong\mathbb{C}^4$.

The Lie algebra $\mathfrak{so}(4)$ acts naturally on this four-dimensional space. A key fact from Lie theory is that $\mathfrak{so}(4)\cong\mathfrak{su}(2)\oplus\mathfrak{su}(2)$, so we can describe all six generators of $\mathfrak{so}(4)$ in terms of two sets of angular momentum operators. Concretely, define
\begin{equation}
\mathbf{A} = \tfrac{1}{2}(\mathbf{J}_1+\mathbf{J}_2),\qquad \mathbf{B} = \tfrac{1}{2}(\mathbf{J}_1-\mathbf{J}_2),
\label{eq:AB_def}
\end{equation}
which satisfy $[A_i,A_j]=i\epsilon_{ijk}A_k$, $[B_i,B_j]=i\epsilon_{ijk}B_k$, and $[A_i,B_j]=0$. Together, $\{A_x,A_y,A_z,B_x,B_y,B_z\}$ span the six-dimensional Lie algebra $\mathfrak{so}(4)$. In the matchgate setting, the role of these generators is played by the bridge operators $\Lambda_{ab}$.

The subgroup $\mathrm{SO}(3)\subset\mathrm{SO}(4)$ is embedded as rotations that act identically on both particles, generated by the \emph{total angular momentum}
\begin{equation}
\mathbf{J} = \mathbf{J}_1 + \mathbf{J}_2 = 2\mathbf{A}.
\label{eq:total_J}
\end{equation}
Finally, $\mathrm{SO}(2)\subset\mathrm{SO}(3)$ is generated by $J_z = J_{1z}+J_{2z}$.

\subsubsection{The commuting family of Casimirs}

We now identify the Casimir operators at each level of the chain $\mathrm{SO}(4)\supset\mathrm{SO}(3)\supset\mathrm{SO}(2)$, building up the commuting family~\eqref{eq:gt_commuting_family}.

\paragraph{$\mathrm{SO}(4)$ Casimirs.} Being rank-2, the algebra $\mathfrak{so}(4)$ has two independent Casimirs.
The quadratic Casimir is
\begin{equation}
\mathcal{C}_1^{(4)} = \mathbf{A}^2 + \mathbf{B}^2 = \tfrac{1}{2}(\mathbf{J}_1^2 + \mathbf{J}_2^2),
\label{eq:C1_so4}
\end{equation}
which, for two spin-$\frac{1}{2}$ particles, takes the fixed value $\tfrac{1}{2}\cdot\tfrac{3}{4}+\tfrac{1}{2}\cdot\tfrac{3}{4}=\tfrac{3}{4}$ on the entire Hilbert space. Being rank-2 and even-dimensional, $\mathfrak{so}(4)$ has a second Casimir that is a Pfaffian-type invariant [cf.\ Eq.~\eqref{eq:pfafianCasimir}]:
\begin{equation}
\mathcal{C}_2^{(4)} = \mathbf{A}^2 - \mathbf{B}^2 = \mathbf{J}_1\cdot\mathbf{J}_2.
\label{eq:C2_so4}
\end{equation}
This is precisely the Heisenberg exchange interaction between the two spins. On the four-dimensional Hilbert space, it takes eigenvalues $+\tfrac{1}{4}$ (triplet) and $-\tfrac{3}{4}$ (singlet), so it distinguishes the two $\mathrm{SO}(4)$ sectors.

\paragraph{$\mathrm{SO}(3)$ Casimir.} The total angular momentum squared,
\begin{equation}
\mathcal{C}^{(3)} = \mathbf{J}^2 = (\mathbf{J}_1+\mathbf{J}_2)^2,
\label{eq:C_so3}
\end{equation}
has eigenvalue $j(j+1)$ and distinguishes the $\mathrm{SO}(3)$ irreps inside each $\mathrm{SO}(4)$ sector.

\paragraph{$\mathrm{SO}(2)$ Casimir.} The magnetic quantum number operator
\begin{equation}
\mathcal{C}^{(2)} = J_z = J_{1z}+J_{2z},
\label{eq:C_so2}
\end{equation}
with eigenvalue $m$, completes the chain.

The full commuting family is therefore $\{\mathcal{C}_1^{(4)},\,\mathcal{C}_2^{(4)},\,\mathcal{C}^{(3)},\,\mathcal{C}^{(2)}\}$, in direct correspondence with Eq.~\eqref{eq:gt_commuting_family}.

\subsubsection{Branching rules and GT patterns}
\label{app:branching}

\paragraph{$\mathrm{SO}(4)$ irreps.} For our two spin-$\frac{1}{2}$ particles, the four-dimensional Hilbert space decomposes under $\mathrm{SO}(4)$ as
\begin{equation}
\mathbb{C}^4 = V_{(\frac{1}{2},\frac{1}{2})} \;\oplus\; V_{(\frac{1}{2},-\frac{1}{2})},
\label{eq:so4_decomp}
\end{equation}
where the labels $(a,b)$ refer to the quantum numbers of $(\mathbf{A}^2,\mathbf{B}^2)$, i.e., the $\mathfrak{su}(2)\oplus\mathfrak{su}(2)$ decomposition. In terms of the $\mathrm{SO}(4)$ highest-weight notation $\nu=(\nu_1,\nu_2)$ [cf.\ Eq.~\eqref{eq:highest_weights_Dr}], these correspond to $\nu=(1,0)$ (the triplet sector, spanned by the symmetric spin states) and $\nu=(0,0)$ (the singlet sector).

\paragraph{Branching $\mathrm{SO}(4)\downarrow\mathrm{SO}(3)$.}
The restriction of each $\mathrm{SO}(4)$ irrep to $\mathrm{SO}(3)$ follows the standard addition of angular momenta. The triplet sector $\nu=(1,0)$ restricts to a single $\mathrm{SO}(3)$ irrep with $j=1$:
\begin{equation}
V_{(1,0)}\big\downarrow_{\mathrm{SO}(3)} = V_{j=1}.
\end{equation}
The singlet sector $\nu=(0,0)$ restricts to $j=0$:
\begin{equation}
V_{(0,0)}\big\downarrow_{\mathrm{SO}(3)} = V_{j=0}.
\end{equation}
In this example, the branching is multiplicity-free: each $\mathrm{SO}(3)$ irrep appears at most once.

\paragraph{Branching $\mathrm{SO}(3)\downarrow\mathrm{SO}(2)$.}
Within the $j=1$ sector, the further restriction to $\mathrm{SO}(2)$ gives
\begin{equation}
V_{j=1}\big\downarrow_{\mathrm{SO}(2)} = V_{m=+1}\oplus V_{m=0}\oplus V_{m=-1},
\end{equation}
while the $j=0$ sector yields only $V_{m=0}$.

\paragraph{GT patterns.}
A GT pattern records the full chain of labels. For the chain $\mathrm{SO}(4)\supset\mathrm{SO}(3)\supset\mathrm{SO}(2)$, a GT pattern is a triple $(\nu;\,j;\,m)$ specifying the irrep at each level. The allowed patterns for our Hilbert space are:
\begin{equation}
\begin{array}{c|c|c}
\mathrm{SO}(4)\text{ label }\nu & \mathrm{SO}(3)\text{ label }j & \mathrm{SO}(2)\text{ label }m \\
\hline
(1,0) & 1 & +1 \\
(1,0) & 1 & 0 \\
(1,0) & 1 & -1 \\
(0,0) & 0 & 0
\end{array}
\label{eq:GT_patterns_so4}
\end{equation}
These four patterns are in one-to-one correspondence with the four basis states $\ket{j,m}\in\{\ket{1,1},\ket{1,0},\ket{1,-1},\ket{0,0}\}$. In the general matchgate construction, the GT patterns play exactly the same role, but with more intermediate levels.

\subsubsection{Diagonal projectors}
\label{app:projectors_so4}

The diagonal projectors are the joint eigenspace projectors of the commuting Casimir family $\{\mathcal{C}_2^{(4)},\,\mathbf{J}^2,\,J_z\}$. (The first Casimir $\mathcal{C}_1^{(4)}$ is constant on our Hilbert space and does not contribute to the resolution.) In the coupled basis $\ket{j,m}$, they are simply
\begin{equation}
P_{j,m} = \ket{j,m}\bra{j,m}.
\label{eq:Pjm}
\end{equation}
The four projectors $\{P_{1,1},\,P_{1,0},\,P_{1,-1},\,P_{0,0}\}$ are mutually orthogonal and sum to the identity on $\mathbb{C}^4$.

As in the warm-up, these can be constructed explicitly as polynomials in the Casimir operators without knowing the eigenstates. For example, the projector onto the triplet sector ($j=1$) is
\begin{equation}
\Pi_{j=1} = \frac{\mathbf{J}^2 - 0\cdot(0+1)}{1\cdot(1+1)-0\cdot(0+1)} = \frac{\mathbf{J}^2}{2},
\label{eq:proj_triplet}
\end{equation}
and onto the singlet ($j=0$):
\begin{equation}
\Pi_{j=0} = \frac{\mathbf{J}^2 - 1\cdot(1+1)}{0\cdot(0+1)-1\cdot(1+1)} = \frac{2-\mathbf{J}^2}{2} = \mathbf{1} - \Pi_{j=1}.
\label{eq:proj_singlet}
\end{equation}
Within the triplet sector, the projectors onto individual $m$ values are obtained by further applying the spectral projection of $J_z$ restricted to the $j=1$ subspace:
\begin{equation}
P_{1,m} = \Pi_{j=1}\prod_{\substack{m'=-1\\m'\neq m}}^{1}\frac{J_z - m'}{m-m'}.
\label{eq:P1m_spectral}
\end{equation}
This is the two-step analog of Eq.~\eqref{eq:Pm_spectral}: one first projects onto the $\mathrm{SO}(3)$ sector using $\mathbf{J}^2$, then resolves within it using $J_z$. In the matchgate construction, the same hierarchical projection is carried out level by level along the full chain~\eqref{eq:gt_chain}.

\subsubsection{Ladder operators at each level of the chain}
\label{app:ladders_so4}

This is where the two-step chain reveals structure absent from the single-step warm-up. There are now ladder operators acting at \emph{two different levels}: the $\mathrm{SO}(3)$ level (changing $m$ within a fixed $j$) and the $\mathrm{SO}(4)$ level (changing $j$ itself).

\paragraph{Level 1: $\mathrm{SO}(3)$ ladders (changing $m$).}
Within a fixed $j$ sector, the operators $J_\pm = J_x\pm iJ_y$ raise or lower the magnetic quantum number exactly as in the warm-up:
\begin{equation}
J_+\ket{j,m} = \sqrt{j(j+1)-m(m+1)}\;\ket{j,m+1}.
\label{eq:Jpm_action_so4}
\end{equation}
For instance, inside the triplet:
\begin{equation}
J_+\ket{1,0} = \sqrt{2}\,\ket{1,1},\qquad J_+\ket{1,-1} = \sqrt{2}\,\ket{1,0}.
\end{equation}
These are the ``lowest-level'' ladder operators; they move along the bottom row of the GT pattern (changing $m$) while keeping everything above it fixed.

\paragraph{Level 2: $\mathrm{SO}(4)$ ladders (changing $j$).}
To connect different $\mathrm{SO}(3)$ sectors---i.e., to move between the triplet ($j=1$) and the singlet ($j=0$)---one needs generators of $\mathfrak{so}(4)$ that do \emph{not} belong to the $\mathfrak{so}(3)$ subalgebra. These are the components of the ``relative angular momentum'' $\mathbf{B}=\tfrac{1}{2}(\mathbf{J}_1-\mathbf{J}_2)$. Since $\mathbf{B}$ does not commute with $\mathbf{J}$, it mixes different $j$ sectors. Explicitly, defining
\begin{equation}
B_\pm = B_x \pm iB_y = \tfrac{1}{2}(J_{1\pm}-J_{2\pm}),\qquad B_z = \tfrac{1}{2}(J_{1z}-J_{2z}),
\label{eq:Bpm_def}
\end{equation}
we find that these operators connect the singlet and triplet. For example:
\begin{equation}
B_+\ket{0,0} = \tfrac{1}{\sqrt{2}}\ket{1,1}-\text{(components already in triplet)},
\end{equation}
and more concretely, writing $\ket{0,0}=\tfrac{1}{\sqrt{2}}(\ket{\!\uparrow\downarrow}-\ket{\!\downarrow\uparrow})$, one can verify:
\begin{equation}
B_z\ket{0,0} = \tfrac{1}{2}(J_{1z}-J_{2z})\tfrac{1}{\sqrt{2}}(\ket{\!\uparrow\downarrow}-\ket{\!\downarrow\uparrow}) = \tfrac{1}{\sqrt{2}}\cdot\tfrac{1}{2}(\ket{\!\uparrow\downarrow}+\ket{\!\downarrow\uparrow}) = \tfrac{1}{\sqrt{2}}\ket{1,0}.
\label{eq:Bz_singlet}
\end{equation}
The operator $B_z$ has taken us from the singlet ($j=0,\,m=0$) to the triplet ($j=1,\,m=0$), changing the $\mathrm{SO}(3)$ label while preserving the $\mathrm{SO}(2)$ label. This is the essential mechanism by which off-diagonal operators are built in the matchgate construction: generators at a \emph{higher} level of the chain act as ladder operators that change the intermediate subgroup labels.

\subsubsection{Building the matrix units}
\label{app:matrix_units_so4}

We can now assemble the complete set of matrix units for $\End(\mathbb{C}^4)$ using the projectors and ladder operators identified above. The Hilbert space decomposes into two $\mathrm{SO}(4)$ irreps: the triplet $V_{(1,0)}$ (dimension 3) and the singlet $V_{(0,0)}$ (dimension 1). The endomorphism algebra correspondingly decomposes as
\begin{equation}
\End(\mathbb{C}^4) = \End(V_{(1,0)})\oplus \End(V_{(0,0)}) \oplus \text{off-diagonal blocks},
\end{equation}
but for building the \emph{commutant basis}, one works within each irreducible block separately [cf.\ Eq.~\eqref{eq:block_decomp}].

\paragraph{Singlet block: $\End(V_{(0,0)})$.}
This block is one-dimensional, spanned by the single projector
\begin{equation}
X^{(0,0)}_{0,0} = P_{0,0} = \ket{0,0}\bra{0,0}.
\end{equation}

\paragraph{Triplet block: $\End(V_{(1,0)})$.}
This block is $3\times 3 = 9$-dimensional. The three diagonal matrix units are simply the projectors $P_{1,m}$ for $m\in\{-1,0,1\}$.

The six off-diagonal matrix units are constructed by the sandwiching procedure of Eq.~\eqref{eq:Xmn_sandwich}. For example, to build $X^{(1,0)}_{1,0}=\ket{1,1}\bra{1,0}$, we use the $\mathrm{SO}(3)$ ladder operator $J_+$ and sandwich:
\begin{equation}
X^{(1,0)}_{1,0} \;\propto\; P_{1,1}\,J_+\,P_{1,0}.
\label{eq:X10_example}
\end{equation}
To verify: $P_{1,0}$ selects the state $\ket{1,0}$, then $J_+$ raises it to $\sqrt{2}\ket{1,1}$, and finally $P_{1,1}$ projects onto $\ket{1,1}$. After dividing by $\sqrt{2}$, we obtain $\ket{1,1}\bra{1,0}$ as desired.

Similarly, to jump two steps, e.g., from $m=-1$ to $m=+1$:
\begin{equation}
X^{(1,0)}_{1,-1} \;\propto\; P_{1,1}\,(J_+)^2\,P_{1,-1} = P_{1,1}\,Y_{1,-1}\,P_{1,-1},
\end{equation}
where $Y_{1,-1}=(J_+)^2$ is the ordered ladder product. This is the direct analog of $Y^{(\nu)}_{\mathfrak{m},\mathfrak{n}}$ in Eq.~\eqref{eq:Xmn_sandwich}.

After normalization, the nine operators $\{X^{(1,0)}_{m,m'}\}_{m,m'=-1}^{1}$ satisfy the matrix-unit algebra
\begin{equation}
X^{(1,0)}_{m_1,m_2}\,X^{(1,0)}_{m_3,m_4} = \delta_{m_2 m_3}\,X^{(1,0)}_{m_1,m_4},
\end{equation}
and are mutually orthonormal under the Hilbert--Schmidt inner product.

\begin{table*}[t!]
\centering
\renewcommand{\arraystretch}{1.3}
\begin{tabular}{l|l|l|l}
\hline\hline
\textbf{Concept} & \textbf{$\quad\mathrm{SO}(3)\supset\mathrm{SO}(2)\quad $} & \textbf{$\quad\mathrm{SO}(4)\supset\mathrm{SO}(3)\supset\mathrm{SO}(2)\quad $} & $\quad$\textbf{Matchgate commutant} $\quad$\\
\hline
Subgroup chain & $\mathrm{SO}(3)\supset\mathrm{SO}(2)$ & $\mathrm{SO}(4)\supset\mathrm{SO}(3)\supset\mathrm{SO}(2)$ & $\mathrm{SO}(k)\supset\cdots\supset\mathrm{SO}(2)$ \\
Full Casimirs & $\mathbf{J}^2$ & $\mathbf{A}^2+\mathbf{B}^2$, $\mathbf{A}^2-\mathbf{B}^2$ & $\mathcal{C}^{(k)}_j$ \\
Intermediate Casimir & --- & $\mathbf{J}^2$ & $\mathcal{C}^{(k-1)}_j,\dots$ \\
Bottom Casimir & $J_z$ & $J_z$ & $H = \tfrac{i}{2}\Lambda_{12}$ \\
Irrep label & $\ell$ & $\nu=(\nu_1,\nu_2)$ & highest weight $\nu$ \\
GT pattern & $m$ & $(j,\,m)$ & $\mathfrak{m}\in\mathrm{GT}(\nu)$ \\
Diagonal projector & $P_m$ & $P_{j,m}$ & $P_{\nu,\mathfrak{m}}$ \\
Low-level ladders & $J_\pm$ & $J_\pm$ (change $m$) & ladders in $\mathfrak{so}(3)$ \\
High-level ladders & --- & $B_\pm,B_z$ (change $j$) & ladders in $\mathfrak{so}(m)\setminus\mathfrak{so}(m\!-\!1)$ \\
Matrix unit & $\ket{\ell,m}\!\bra{\ell,m'}$ & $\ket{j,m}\!\bra{j,m'}$ & $X^{(\nu)}_{\mathfrak{m},\mathfrak{n}}$ \\
\hline\hline
\end{tabular}
\caption{Dictionary between the angular-momentum GT constructions of this appendix and the general matchgate commutant construction (Sec.~\ref{subsec:gt_basis_elems}). The middle column shows the new features that arise with a two-step chain: multiple Casimirs, intermediate labels, and ladder operators at different levels.}
\label{tab:GT_dictionary}
\end{table*}

\paragraph{Counting.}
The total number of independent operators in the two blocks is
\begin{equation}
\dim\End(V_{(1,0)}) + \dim\End(V_{(0,0)}) = 3^2 + 1^2 = 10.
\end{equation}
This is less than $\dim\End(\mathbb{C}^4)=16$ because the GT construction produces only those operators that commute with the symmetry group---here, the six off-diagonal operators mixing the triplet and singlet blocks are excluded. This counting mirrors the dimension formula~\eqref{eq:dim_sum_squares} of the matchgate commutant: $\dim\Com_k = \sum_\nu (\dim V_\nu)^2$.

\subsubsection{Why the two-step chain matters}
\label{app:why_two_step}

The key conceptual point that the two-step chain illustrates, and that the warm-up with $\mathrm{SO}(3)\supset\mathrm{SO}(2)$ alone does not, is the following. When the subgroup chain has multiple levels, the GT pattern carries \emph{several} intermediate labels---here $(j,m)$ rather than just $m$---and the ladder operators act at \emph{different} levels of the hierarchy:

\begin{itemize}
\item \textbf{$\mathrm{SO}(3)$ ladders} ($J_\pm$): change $m$ within a fixed $j$ sector. These move along the \emph{bottom} of the GT pattern.
\item \textbf{$\mathrm{SO}(4)$ ladders} ($B_\pm, B_z$): change $j$ itself by connecting different $\mathrm{SO}(3)$ sectors. These move along a \emph{higher} level of the GT pattern.
\end{itemize}
In the general matchgate commutant construction with the chain $\mathrm{SO}(k)\supset\mathrm{SO}(k-1)\supset\cdots\supset\mathrm{SO}(2)$, this structure repeats at every level: the generators of $\mathfrak{so}(m)$ that do not belong to $\mathfrak{so}(m-1)$ serve as ladder operators connecting different $\mathrm{SO}(m-1)$ sectors, and the full off-diagonal operator $Y^{(\nu)}_{\mathfrak{m},\mathfrak{n}}$ is built by composing ladders from all the levels needed to connect the GT patterns $\mathfrak{m}$ and $\mathfrak{n}$.

\subsection{Dictionary}
\label{app:GT_dictionary}

Table~\ref{tab:GT_dictionary} summarizes the correspondence between the angular-momentum constructions of this appendix and the general matchgate commutant framework.

The essential lesson is that the abstract procedure of Sec.~\ref{subsec:gt_basis_elems}---simultaneously diagonalize a commuting family of Casimirs to obtain projectors, then use ladder operators at each level of the chain and sandwiching to build off-diagonal matrix units---is a systematic generalization of the textbook construction of angular-momentum eigenstates and transition operators. The two-step chain $\mathrm{SO}(4)\supset\mathrm{SO}(3)\supset\mathrm{SO}(2)$ already exhibits the essential complication: GT patterns carry multiple labels, and ladder operators at different levels of the hierarchy play distinct roles. In the matchgate setting, the chain is longer but the logic is identical.

\end{widetext}

\section{Details of algebraic commutant matchgate basis construction}

\subsection{Off-diagonal GT operators at $k=4$}
\label{app:k4_detailsT}

In this Appendix, we make explicit the construction of the off-diagonal GT operators for
$k=4$. For fixed highest weight $\nu=(\nu_1,\nu_2)\in\mathcal I_{n,4}$, the GT labels
$\mathfrak m=(s,m)$ are given by Eq.~\eqref{eq:k4_GT_set}. The corresponding one-dimensional GT
projectors, denoted by
$P_{\nu,\mathfrak m}:= P_{\nu,s,m}$,
are given by Eq.~\eqref{eq:k4_GT_projectors}.
We now show how to generate the transition operators
$X^{(\nu)}_{\mathfrak m,\mathfrak n}$ from the bridge algebra.

\subsubsection{Transitions at fixed $s$}

We first consider the $\mathrm{SO}(3)\subset \mathrm{SO}(4)$ subgroup generated by the bridge
operators acting on the first three replicas, with generators $M_1$, $M_2$, $M_3$ defined
in Eq.~\eqref{eq:k4_Mi}. Its ladder operators are
\begin{equation}
M_\pm:=M_1\pm iM_2
=
\tfrac12\bigl(\Lambda_{23}\mp i\Lambda_{13}\bigr).
\label{eq:k4_appendix_Mpm}
\end{equation}
Inside a fixed $\nu$-block, the projected operators
$P_{\nu,s,m}\,M_\pm\,P_{\nu,s,m'}$ preserve $\nu$ and $s$, while shifting the
$\mathrm{SO}(2)$ weight $m$ by $\pm1$. Hence, for fixed $s$, the transition operators are
obtained by sandwiching suitable powers of $M_\pm$ between the one-dimensional GT projectors:
\begin{equation}
X^{(\nu)}_{(s,m),(s,m')}
=
\mathcal N^{(\nu)}_{s;m,m'}\,
P_{\nu,s,m}\,
 M_{m,m'}\,
P_{\nu,s,m'}.
\label{eq:k4_fixed_s_transition}
\end{equation}
Here
\begin{equation}
 M_{m,m'}
:=
\begin{cases}
M_+^{\,m-m'}, & m\ge m',\\[1mm]
M_-^{\,m'-m}, & m<m'.
\end{cases}
\label{eq:k4_fixed_s_operator}
\end{equation}
As in the $k=3$ case, the normalization factor may be chosen so that
$X^{(\nu)}_{(s,m),(s,m')}$ has unit Hilbert--Schmidt norm. Explicitly,
\begin{equation}
\mathcal N^{(\nu)}_{s;m,m'}
=
\sqrt{
\frac{(s-m_>)!\,(s+m_<)!}
     {(s-m_<)!\,(s+m_>)!}
},
\label{eq:k4_fixed_s_norm}
\end{equation}
where
\begin{equation}
m_>:=\max(m,m'),
\qquad
m_<:=\min(m,m').
\end{equation}
This is the standard $\mathfrak{so}(3)$ matrix-unit normalization inside a fixed spin-$s$
multiplet.

\subsubsection{Changing the $\mathrm{SO}(3)$ label $s$}

To connect different values of $s$, one uses operators transforming as a vector under the
$\mathrm{SO}(3)$ generated by the $M_i$. A convenient choice is provided by the operators
$T_{+1}$, $T_0$, $T_{-1}$ defined in Eq.~\eqref{eq:Tops}. These preserve the highest-weight
label $\nu$ and connect adjacent $s$-sectors inside the fixed $\nu$-block. Equivalently, they
satisfy
\begin{equation}
[M_3,T_q]=i\,q\,T_q,
\qquad
q\in\{-1,0,1\},
\label{eq:k4_appendix_tensor_M3}
\end{equation}
together with the usual raising and lowering relations under $M_\pm$. Therefore the standard
selection rules apply:
\begin{equation}
s'\in\{s-1,s,s+1\},
\qquad
m'=m+q,
\qquad
q\in\{-1,0,1\},
\label{eq:k4_appendix_selection}
\end{equation}
subject to the finite range
\begin{equation}
|\nu_2|\le s,s'\le \nu_1.
\end{equation}
Whenever these conditions are satisfied and the projected operator is nonzero, a nearest-neighbor
transition in the $s$ direction may be written as
\begin{equation}
X^{(\nu)}_{(s',m'),(s,m)}
=
\mathcal N^{(\nu)}_{s',m';\,s,m}\,
P_{\nu,s',m'}\,T_q\,P_{\nu,s,m},
\label{eq:k4_appendix_s_transition}
\end{equation}
with $q=m'-m$. The overall coefficient
$\mathcal N^{(\nu)}_{s',m';\,s,m}$ may again be fixed by Hilbert--Schmidt normalization.

\subsubsection{Generation of all off-diagonal operators}

The operators in Eqs.~\eqref{eq:k4_fixed_s_transition} and
\eqref{eq:k4_appendix_s_transition} generate all off-diagonal GT transitions inside a fixed irrep
$V_\nu$. At fixed $s$, arbitrary changes of $m$ are generated by powers of $M_\pm$. The
operators $T_q$ connect adjacent $s$-sectors with $\Delta s=\pm1,0$. By composing these
elementary moves, one may connect any two GT labels $(s,m)$ and $(s',m')$ in $\GT(\nu)$.

Concretely, if
\begin{equation}
\mathfrak m=(s,m),
\qquad
\mathfrak n=(s',m'),
\end{equation}
one may choose a sequence of elementary steps
\begin{equation}
(s,m)\to (s_1,m_1)\to \cdots \to (s_r,m_r)\to (s',m').
\end{equation}
Provided each elementary transition in the sequence is nonzero, the corresponding product of
elementary transitions produces an operator proportional to
$X^{(\nu)}_{\mathfrak m,\mathfrak n}$:
\begin{equation}
X^{(\nu)}_{\mathfrak m,\mathfrak n}
\propto
X^{(\nu)}_{\mathfrak m,\mathfrak m_1}
X^{(\nu)}_{\mathfrak m_1,\mathfrak m_2}
\cdots
X^{(\nu)}_{\mathfrak m_r,\mathfrak n}.
\label{eq:k4_appendix_composition}
\end{equation}
Since each GT sector is one-dimensional, any nonzero operator with the correct source and target
sectors is necessarily proportional to the corresponding transition operator.

The off-diagonal GT operators for $k=4$ are therefore generated explicitly from the bridge
algebra in two steps. Powers of the ladder operators $M_\pm$ produce transitions between
different $\mathrm{SO}(2)$ weights at fixed $s$, while the tensor operators $T_q$, built from
the bridges involving the fourth replica, connect neighboring $\mathrm{SO}(3)$ sectors.
Together with the one-dimensional GT projectors $P_{\nu,s,m}$, these operators generate the full
block $\End(V_\nu)$ for fixed $\nu$.

\subsection{Off-diagonal GT operators for $k=5$.}
\label{app:offk5}

Once the one-dimensional GT projectors $P_{\nu,\mathfrak m}$ are known, the off-diagonal
transition operators may be constructed by sandwiching suitable bridge operators between the source
and target sectors. Writing
\begin{equation}
\mathfrak m=(\mu;s,m),
\qquad
\mathfrak n=(\mu';s',m'),
\end{equation}
with $\mu=(\mu_1,\mu_2),\ \mu'=(\mu'_1,\mu'_2)$,
a general transition operator has the form
\begin{equation}
X^{(\nu)}_{\mathfrak m,\mathfrak n}
\propto
P_{\nu,\mathfrak m}\,
\mathcal O_{\mathfrak m,\mathfrak n}\,
P_{\nu,\mathfrak n},
\label{eq:k5_transition_general}
\end{equation}
where $\mathcal O_{\mathfrak m,\mathfrak n}$ is chosen so that it has a nonvanishing matrix
element between the two GT sectors.

The construction follows the GT chain
\begin{equation}
\mathrm{SO}(5)\downarrow \mathrm{SO}(4)\downarrow \mathrm{SO}(3)\downarrow \mathrm{SO}(2),
\end{equation}
and therefore proceeds by changing, in turn, the final $\mathrm{SO}(2)$ weight $m$, the
intermediate $\mathrm{SO}(3)$ spin $s$, and the $\mathrm{SO}(4)$ label $\mu$, while keeping the
highest-weight label $\nu$ fixed.

\paragraph*{1. Changing the $\mathrm{SO}(2)$ weight $m$ at fixed $(\mu,s)$.}
Inside the $\mathrm{SO}(3)$ subgroup generated by
$\Lambda_{12},\Lambda_{13},\Lambda_{23}$, we consider the operators $M_1$, $M_2$, $M_3$
introduced in the $k=4$ discussion, cf. Eq.~\eqref{eq:k4_Mi}, together with the ladder operators
\begin{equation}
M_\pm:=M_1\pm iM_2
=
\tfrac12\bigl(\Lambda_{23}\mp i\Lambda_{13}\bigr).
\label{eq:k5_Mpm}
\end{equation}
Inside a fixed $\nu$-block, the projected operators
$P_{\nu,\mu,s,m}\,M_\pm\,P_{\nu,\mu,s,m'}$ preserve $\nu$, $\mu$, and $s$, while changing
$m$ by $\pm1$. Hence, for fixed $(\nu,\mu,s)$, one may take
\begin{equation}
X^{(\nu)}_{(\mu;s,m),(\mu;s,m')}
=
\mathcal N^{(\nu)}_{\mu;s;m,m'}\,
P_{\nu,\mu,s,m}\,
 M_{m,m'}\,
P_{\nu,\mu,s,m'},
\label{eq:k5_transition_m}
\end{equation}
with
\begin{equation}
 M_{m,m'}
:=
\begin{cases}
M_+^{\,m-m'}, & m\ge m',\\[1mm]
M_-^{\,m'-m}, & m<m'.
\end{cases}
\label{eq:k5_Mmmprime}
\end{equation}
The normalization factor $\mathcal N^{(\nu)}_{\mu;s;m,m'}$ may be chosen so that
$X^{(\nu)}_{(\mu;s,m),(\mu;s,m')}$ has unit Hilbert--Schmidt norm.

\paragraph*{2. Changing the $\mathrm{SO}(3)$ spin $s$ at fixed $\mu$.}
To connect different $\mathrm{SO}(3)$ sectors inside a fixed $\mathrm{SO}(4)$ irrep, one uses
operators transforming as vectors under the $\mathrm{SO}(3)$ generated by the $M_i$. A convenient
choice is provided by the bridges involving the fourth replica,
$T_{+1}$, $T_0$, $T_{-1}$, defined in Eq.~\eqref{eq:Tops}. Inside a fixed $\nu$-block, these
operators preserve $\nu$ and $\mu$, while connecting adjacent $s$-sectors. They satisfy the
standard rank-one tensor selection rules
\begin{equation}
s'\in\{s-1,s,s+1\},
\qquad
m'=m+q,
\qquad
q\in\{-1,0,1\},
\label{eq:k5_s_selection}
\end{equation}
subject to the admissible range
\begin{equation}
|\mu_2|\le s,s'\le \mu_1.
\end{equation}
Thus the nearest-neighbor transitions in the $s$-direction may be written as
\begin{equation}
X^{(\nu)}_{(\mu;s',m'),(\mu;s,m)}
=
\mathcal N^{(\nu)}_{\mu;s',m';\,s,m}\,
P_{\nu,\mu,s',m'}\,T_q\,P_{\nu,\mu,s,m},
\label{eq:k5_transition_s}
\end{equation}
with $q=m'-m$, whenever the projected operator is nonzero. More general transitions with fixed
$\mu$ are obtained by composing these elementary steps with the weight-shifting operators in
Eq.~\eqref{eq:k5_transition_m}.

\paragraph*{3. Changing the $\mathrm{SO}(4)$ label $\mu$ inside a fixed $\mathrm{SO}(5)$ irrep.}
To connect different $\mathrm{SO}(4)$ sectors inside a fixed $\mathrm{SO}(5)$ irrep $V_\nu$, one
uses bridge operators outside the $\mathrm{SO}(4)$ subalgebra, namely those involving the fifth
replica,
\begin{equation}
Y_a:=\Lambda_{a5},
\qquad
a=1,2,3,4.
\label{eq:k5_Ya}
\end{equation}
These operators transform in the vector representation of $\mathrm{SO}(4)$. Accordingly, they
connect $\mathrm{SO}(4)$ sectors $\mu$ and $\mu'$ whenever $V_{\mu'}$ appears in the tensor
product
\begin{equation}
V_\mu\otimes V_{(1,0)},
\end{equation}
subject to the constraint that both $\mu$ and $\mu'$ occur in the restriction of the fixed
$\mathrm{SO}(5)$ irrep $V_\nu$. Thus the elementary transitions in the $\mu$-direction may be
written schematically as
\begin{equation}
X^{(\nu)}_{(\mu';s',m'),(\mu;s,m)}
=
\mathcal N^{(\nu)}_{\mu',s',m';\,\mu,s,m}\,
P_{\nu,\mu',s',m'}\,Y_a\,P_{\nu,\mu,s,m},
\label{eq:k5_transition_mu}
\end{equation}
whenever the corresponding projected operator is nonzero. In general, the action of $Y_a$
changes both the intermediate $\mathrm{SO}(4)$ label $\mu$ and the lower GT labels $(s,m)$, so
projection onto the target sector must be carried out explicitly.

\paragraph*{Generation of all off-diagonal GT operators.}
The operators in Eqs.~\eqref{eq:k5_transition_m}, \eqref{eq:k5_transition_s}, and
\eqref{eq:k5_transition_mu} generate the full off-diagonal GT basis. Indeed, powers of $M_\pm$
generate arbitrary changes of the final $\mathrm{SO}(2)$ weight $m$ at fixed $(\mu,s)$; the
tensor operators $T_q$ connect neighboring $\mathrm{SO}(3)$ sectors inside a fixed
$\mathrm{SO}(4)$ irrep; and the vector operators $Y_a=\Lambda_{a5}$ connect different
$\mathrm{SO}(4)$ sectors inside the fixed $\mathrm{SO}(5)$ irrep $V_\nu$.

By composing these elementary moves and resolving with the GT projectors after each step, one
obtains all transition operators $X^{(\nu)}_{\mathfrak m,\mathfrak n}$. Since each GT sector is
one-dimensional, any nonzero operator with the correct source and target sectors is necessarily
proportional to the corresponding transition operator. In this way, the full off-diagonal GT basis
is generated explicitly from the bridge algebra itself.

\section{Details on the matchgate commutant dimension}

\subsection{Weyl dimension formula}
\label{app:Weylfromulas}
Here, we fix provide the formulas for the roots of the orthogonal algebras 

When $k=2r$, the positive roots are
\begin{equation}
\Phi^+(D_r)=\{\,e_i-e_j,\ e_i+e_j\ :\ 1\le i<j\le r\,\},
\end{equation}
and the Weyl vector is
\begin{equation}
\rho=\sum_{i=1}^r (r-i)\,e_i.
\end{equation}
Evaluating Eq.~\eqref{eq:weyl_dim_general} gives
\begin{equation}
\dim V_\nu^{D_r}
=
\prod_{1\le i<j\le r}
\frac{
(\nu_i-\nu_j+j-i)\,(\nu_i+\nu_j+2r-i-j)
}{
(j-i)\,(2r-i-j)
}.
\label{eq:weyl_Dr_final}
\end{equation}

If $k=2r+1$,  the positive roots are
\begin{equation}
\Phi^+(B_r)=\{\,e_i-e_j,\ e_i+e_j\ :\ 1\le i<j\le r\,\}\cup\{\,e_i:\ 1\le i\le r\,\},
\end{equation}
and the Weyl vector becomes
\begin{equation}
\rho=\sum_{i=1}^r \Bigl(r-i+\tfrac12\Bigr)e_i.
\end{equation}
The resulting dimension formula is
\begin{equation}
\begin{split}
\dim V_\lambda^{B_r}
&=
\prod_{i=1}^r
\frac{\lambda_i+r-i+\tfrac12}{r-i+\tfrac12}
\\
&\times
\prod_{1\le i<j\le r}
\frac{
(\lambda_i-\lambda_j+j-i)
(\lambda_i+\lambda_j+2r-i-j+1)
}{
(j-i)(2r-i-j+1)
}.
\end{split}
\label{eq:weyl_Br}
\end{equation}

\subsection{Matchgate commutant dimension derivation} 
\label{app:matchgateDIM}

The dimension of the matchgate commutant is determined by the allowed highest weights
$\mathcal I_{n,k}$, cf. Eqs.~\eqref{eq:highest_weights_Br}
and \eqref{eq:highest_weights_Dr}, together with the corresponding Weyl dimensions:
\begin{equation}
\dim \Com_k(\MG_n)
=
\sum_{\nu\in\mathcal I_{n,k}} \bigl(\dim V_\nu\bigr)^2.
\label{eq:dim_sum_squares_recalled_app}
\end{equation}
Below, we show that this sum can be evaluated by rewriting the Weyl dimensions as Vandermonde-type
determinants and then applying the discrete Cauchy--Binet identity
\cite{Krattenthaler1999,deBruijn1955}.

\paragraph{Odd case: $k=2r+1$.}
For the algebra $B_r$, introduce shifted highest weights
\begin{equation}
l_i=\nu_i+r-i+\tfrac12.
\end{equation}
The dominance conditions and the truncation $\nu_1\le n$ become
\begin{equation}
n+r-\tfrac12 \ge l_1>l_2>\cdots>l_r\ge \tfrac12,
\qquad
l_i\in\mathbb Z+\tfrac12.
\label{eq:Br_shifted_domain_app}
\end{equation}
In these variables, the Weyl dimension formula takes the form
\begin{equation}
\dim V_\nu^{B_r}
=
C_{B_r}
\prod_{i=1}^r l_i
\prod_{1\le i<j\le r}(l_i^2-l_j^2),
\label{eq:Br_dim_shifted_app}
\end{equation}
where
\begin{equation}
C_{B_r}
=
\prod_{i=1}^r \Bigl(r-i+\tfrac12\Bigr)^{-1}
\prod_{1\le i<j\le r}(j-i)^{-1}(2r-i-j+1)^{-1}
\end{equation}
is independent of $\nu$. Using
\begin{equation}
\prod_{i=1}^r l_i\prod_{1\le i<j\le r}(l_i^2-l_j^2)
=
\det\!\bigl(l_i^{\,2(r-j)+1}\bigr)_{i,j=1}^r,
\end{equation}
we obtain
\begin{equation}
\dim V_\nu^{B_r}
=
C_{B_r}\det\!\bigl(l_i^{\,2(r-j)+1}\bigr)_{i,j=1}^r.
\label{eq:Br_dim_det_app}
\end{equation}
Substituting this into Eq.~\eqref{eq:dim_sum_squares_recalled_app} gives
\begin{equation}
\dim \Com_{2r+1}(\MG_n)
=
C_{B_r}^2
\sum_{l_1>\cdots>l_r\in\mathcal L_{B_r}}
\det\!\bigl(l_i^{\,2(r-j)+1}\bigr)^2,
\end{equation}
with
\begin{equation}
\mathcal L_{B_r}
=
\Bigl\{\tfrac12,\tfrac32,\dots,n+r-\tfrac12\Bigr\}.
\end{equation}
Applying the discrete Cauchy--Binet identity,
\begin{equation}
\sum_{x_1>\cdots>x_r\in\mathcal L}
\det\!\bigl(f_a(x_b)\bigr)\det\!\bigl(g_a(x_b)\bigr)
=
\det\!\left(\sum_{x\in\mathcal L} f_a(x)g_b(x)\right),
\label{eq:CB_general_app}
\end{equation}
with $f_a(x)=g_a(x)=x^{2(r-a)+1}$, we arrive at the Hankel determinant
\begin{equation}
\dim \Com_{2r+1}(\MG_n)
=
C_{B_r}^2
\det\!\left(
\sum_{l\in\mathcal L_{B_r}} l^{\,2(2r-a-b+1)}
\right)_{a,b=1}^r.
\label{eq:Br_Hankel_app}
\end{equation}

\paragraph{Even case: $k=2r$.}
The construction for $D_r$ is entirely analogous. After introducing shifted
weights $l_i=\nu_i+r-i$, the Weyl dimension formula again reduces to a
Vandermonde determinant in the squared variables $l_i^2$. The only additional
subtlety is the $D_r$ degeneracy associated with $\nu_r\leftrightarrow -\nu_r$,
which is absorbed into the discrete measure. One therefore obtains, in complete
analogy with Eq.~\eqref{eq:Br_Hankel_app}, a second Hankel determinant for
$\dim \Com_{2r}(\MG_n)$.

\paragraph{Orthogonal-polynomial evaluation.}
The determinants obtained in the odd and even sectors are standard discrete Hankel
determinants. By the Heine--Christoffel formula, they can be written as
Vandermonde-squared sums over a finite lattice, or equivalently as products of norms
of the corresponding monic orthogonal polynomials
\cite{Krattenthaler1999,ForresterWarnaar2008,Ismail2005}. In the present case, after
symmetrizing the lattice, the relevant orthogonal polynomials are Hahn polynomials
with $\alpha=\beta=0$, i.e. discrete Chebyshev (Gram) polynomials
\cite{KoekoekSwarttouw1998}. Evaluating the associated norms and
restoring the prefactors yields the same final expression in both parity sectors:
\begin{equation}
\dim \Com_k(\MG_n)
=
\prod_{1\le i\le j\le k-1}
\frac{2n+i+j-1}{i+j-1}.
\label{eq:rmt_tableau_prod2_app}
\end{equation}
Thus the parity dependence enters only at the intermediate level of the discrete
measure, while the final commutant dimension is given by the universal product
formula \eqref{eq:rmt_tableau_prod2_app}.

\section{Twirled vacuum normalization}
\label{app:P0norm}

To determine the normalization \(\Tr P_{\mathbf 0}^{(k)}\), observe first that
\(P_{\mathbf 0}^{(k)}\) projects onto the linear span
\begin{equation}
\mathcal K_{n,k}
:=
\Span\Bigl\{
\bigl(U\ket{\mathbf 0}\bigr)^{\otimes k}
:\;
U\in \MG_n
\Bigr\}.
\end{equation}
Hence,
\begin{equation}
\Tr P_{\mathbf 0}^{(k)}=\dim \mathcal K_{n,k}.
\end{equation}

Now, the even component of the matchgate group realizes the spin representation of
\(\Spin(2n)\), and the vacuum \(\ket{\mathbf 0}\) is a highest-weight vector of one of the chiral
spinor representations, with highest weight
\(\omega_n=(\tfrac12,\dots,\tfrac12)\) in orthogonal coordinates (up to chirality convention).
Therefore, under the diagonal action on the \(k\)-fold tensor product,
\(\ket{\mathbf 0}^{\otimes k}\) is a highest-weight vector of weight
\begin{equation}
k\omega_n=\left(\frac{k}{2},\dots,\frac{k}{2}\right),
\end{equation}
and the \(\mathrm{SO}(2n)\)-orbit span is the irreducible \(\Spin(2n)\) module \(V_{k\omega_n}\).

To account for \(\MG_n\) corresponding to the full orthogonal group \(\mathrm O(2n)\), the odd component maps
\(\ket{\mathbf 0}\) to the opposite chiral spinor orbit. Hence the full orbit span is the direct sum
of the two chiral sectors,
\begin{equation}
\mathcal K_{n,k}\cong V_{k\omega_n}\oplus V_{k\omega_{n-1}},
\end{equation}
where \(V_{k\omega_{n-1}}\) is the module generated from the opposite-chirality highest-weight
vector. Since these two chiral modules have the same dimension,
\begin{equation}
\Tr P_{\mathbf 0}^{(k)}
=
2\,\dim V_{k\omega_n}.
\end{equation}

Using the Weyl dimension formula for \(D_n=\mathfrak{so}(2n)\),
\begin{equation}
\dim V_\lambda
=
\prod_{1\le i<j\le n}
\frac{(\lambda_i-\lambda_j+j-i)(\lambda_i+\lambda_j+2n-i-j)}
{(j-i)(2n-i-j)},
\end{equation}
and substituting \(\lambda_i=k/2\) for all \(i\), the factors
\((\lambda_i-\lambda_j+j-i)\) reduce to \(j-i\) and cancel against the denominator. One obtains
\begin{equation} \label{eq:dim_proj_gsym}
\Tr P_{\mathbf 0}^{(k)}
=
2\,
\prod_{1\le i<j\le n}
\frac{k+2n-i-j}{2n-i-j}.
\end{equation}

\section{Additional details on nonstabilizerness of fermionic Gaussian states}
\label{app:sre}

\subsection{The $Q_4$ operator in terms of Majorana strings}
Here, we show how to rewrite $Q_4$ in terms of Majorana operators.
Under the Jordan--Wigner transformation, the Pauli operators at site $j$ map to the $2n$ Majorana operators $\gamma_\mu$. When raised to the fourth power across the replicas, the complex phases vanish, but the non-local Jordan--Wigner strings`\cite{Jordan1928} remain. 
Introducing the local chiral parity operator $\Xi_j := \gamma_{2j-1}^{\otimes 4}\gamma_{2j}^{\otimes 4}$, the four-copy Pauli operators at site $j$ become
\begin{align}
Z_j^{\otimes 4} &= \Xi_j, \nonumber \\
X_j^{\otimes 4} &= \Bigl(\prod_{k<j} \Xi_k\Bigr) \gamma_{2j-1}^{\otimes 4}, \nonumber \\
Y_j^{\otimes 4} &= \Bigl(\prod_{k<j} \Xi_k\Bigr) \gamma_{2j}^{\otimes 4}.
\end{align}
Summing these gives the local factor $Q_{4,j}$ at site $j$:
\begin{equation}
Q_{4,j} = I^{\otimes 4} + \Bigl(\prod_{k<j} \Xi_k\Bigr) \bigl(\gamma_{2j-1}^{\otimes 4} + \gamma_{2j}^{\otimes 4}\bigr) + \Xi_j.
\label{eq:single_site_Q4_with_string}
\end{equation}
To resolve the non-local strings, we define the string-free local factor $h_j$:
\begin{equation}
h_j := I^{\otimes 4} + \gamma_{2j-1}^{\otimes 4} + \gamma_{2j}^{\otimes 4} + \Xi_j = \bigl(I^{\otimes 4} + \gamma_{2j-1}^{\otimes 4}\bigr)\bigl(I^{\otimes 4} + \gamma_{2j}^{\otimes 4}\bigr).
\label{eq:hj_factor}
\end{equation}
Because $(\gamma_\mu^{\otimes 4})^2 = I^{\otimes 4}$, multiplying $h_j$ by its own chiral parity simply permutes its terms, leaving the operator strictly invariant: $h_j \Xi_j = h_j$.
Consequently, $h_j$ perfectly absorbs any $\Xi_j$ string acting on it from the right. When we expand the global product $Q_4 = Q_{4,1} Q_{4,2} \dots Q_{4,n}$ sequentially from left to right, the preceding factors $h_k$ ($k < j$) completely absorb the Jordan--Wigner string $\prod_{k<j} \Xi_k$ present in $Q_{4,j}$. The strings cancel out of the global product, yielding Eq.~\eqref{eq:Q4_product_gamma_final}.

\section{Random Matrix Theory Derivations}
\label{app:RMT}

Random-matrix theory (RMT) provides a complementary route to both the state-overlap moments and the unitary trace moments entering commutant counting. The key point is that many quantities of interest depend only on eigenvalues of a $2n\times 2n$ matrix, such as the covariance-matrix combination $\Gamma_1\Gamma_2$ or the relative orthogonal matrix generating the Gaussian orbit. Once the dependence is purely spectral, orthogonal invariance lets us diagonalize the matrix and replace the Haar matrix integral by an integral over eigenangles with a known Jacobian. This step converts the problem into a Jacobi--Selberg integral, whose value is known in closed form as a product of Euler Gamma functions. In practice, the RMT derivation is therefore a systematic change of variables: rewrite the average in terms of eigenvalues, match to a Selberg class integral, and read off the closed formula.

\subsection{Frame Potential for Random Gaussian States}
\label{app:RMTfp}
We now compute the state frame potential~\eqref{eq:state_fp_def} for random pure fermionic Gaussian states,
\begin{equation}
\mathcal{F}^{(k)}_{\MG_n}
:=
\mathbb E_{O_1,O_2\sim \mathrm{O}(2n)}
\!\left[
|\langle\psi_{O_1}|\psi_{O_2}\rangle|^{2k}
\right],
\label{eq:FGS-frame-potential-def}
\end{equation}
where $\ket{\psi_O}=U_O\ket{\mathbf 0}$ is the pure Gaussian state generated by the orthogonal matrix $O\in \mathrm{O}(2n)$ and $U_O$ is the corresponding Gaussian unitary.
To fix conventions, let $\ket{\mathbf 0}=\ket{0}^{\otimes n}$ be the fermionic vacuum state with covariance matrix
\begin{equation}
\CorrMat_0
=
\bigoplus_{j=1}^{n}
\begin{pmatrix}
0 & 1\\
-1 & 0
\end{pmatrix}.
\end{equation}
Any $O\in \mathrm{O}(2n)$ can be written as $O=Q R_1^{(1-\det O)/2}$, where
$Q\in \mathrm{SO}(2n)$ and $R_1=\mathrm{diag}(-1,1,\dots,1)$. The associated state is
$\ket{\psi_O} = U_Q\,\gamma_1^{(1-\det(O))/2}\ket{\mathbf 0}$, with covariance matrix
$\CorrMat = O\CorrMat_0 O^T$.

It is convenient to first average within a fixed parity sector:
\begin{equation}
\mathbb E_{Q_1,Q_2\sim \mathrm{SO}(2n)}
\!\left[
|\langle\psi_{Q_1}|\psi_{Q_2}\rangle|^{2k}
\right]
=
\mathbb E_{Q\sim \mathrm{SO}(2n)}
\!\left[
|\langle\mathbf 0|\psi_Q\rangle|^{2k}
\right],
\label{eq:SO-reduction}
\end{equation}
where we used Haar invariance and $U_{Q_1}^\dagger U_{Q_2}=U_Q$ with $Q=Q_1^{-1}Q_2$.

For any two pure fermionic Gaussian states $\ket{\psi_{O_1}}$, $\ket{\psi_{O_2}}$ with covariance matrices $\CorrMat_1$ and $\CorrMat_2$,
their squared overlap is~\cite{Bravyi05flo}
\begin{equation}
|\langle\psi_{O_1}|\psi_{O_2}\rangle|^2
=
\sqrt{\det\!\left(\frac{I-\CorrMat_1\CorrMat_2}{2}\right)}.
\label{eq:RMT_overlap_general}
\end{equation}
In Eq.~\eqref{eq:SO-reduction}, we set $\CorrMat_1=\CorrMat_0$ and $\CorrMat_2=Q\CorrMat_0 Q^T$, and define
$\Upsilon:=\CorrMat_0 Q\CorrMat_0 Q^T$. The matrix $\Upsilon$ is orthogonal, with eigenvalues
$e^{\pm 2i\theta_j}$ for $j\in[n]$. Using Eq.~\eqref{eq:RMT_overlap_general}, one finds
\begin{equation}
\det\left(\frac{I-\Upsilon}{2}\right)=\prod_{j=1}^n \left(\frac{1-e^{2i\theta_j}}{2}\right)\left(\frac{1-e^{-2i\theta_j}}{2}\right)
= \prod_{j=1}^n\sin^2\theta_j,
\end{equation}
so $|\langle\mathbf 0|\psi_Q\rangle|^{2k}=\prod_{j=1}^n\sin^k\theta_j$.

This quantity is integrated against the Weyl measure,
\begin{equation}
d\Upsilon\propto
\prod_{1\le i<j\le n}
\bigl(\cos (2\theta_i)-\cos (2\theta_j)\bigr)^2
\prod_{j=1}^n d\theta_j .
\label{eq:weyl-so2n-u_n}
\end{equation}
where the proportionality constant is fixed by $\int d\Upsilon=1$. To perform the integral explicitly, we introduce the radial variables
\begin{equation}
x_j := \sin^2\theta_j \in [0,1]\;.
\end{equation}
The Haar measure of $\mathrm{SO}(2n)$ induces a Jacobi ensemble measure on $\{x_j\}$. In terms of these variables, we have
\begin{equation}
\begin{split}
\cos(2\theta_j) &= 1-2x_j,\\
\sin^k(\theta_j) &= x_j^{k/2},\\
d\theta_j &= \frac{dx_j}{2\sqrt{x_j(1-x_j)}}.
\end{split}
\end{equation}
so the moment factor is $|\langle\psi_{O_1}|\psi_{O_2}\rangle|^{2k}=\prod_{j=1}^n x_j^{k/2}$.
Finally, since
\begin{equation}
\cos 2\theta_i-\cos 2\theta_j = -2(x_i-x_j),
\end{equation}
we obtain the measure
\begin{equation}
d\Upsilon\propto
\prod_{1\le i<j\le n}(x_i-x_j)^2
\prod_{j=1}^n x_j^{-1/2}(1-x_j)^{-1/2}\,dx_j .
\label{eq:jacobi-density-overlap}
\end{equation}
Hence the frame-potential moment is
\begin{equation}
\begin{split}
    \mathbb E_{Q\sim\mathrm{SO}(2n)}
\!\left[
|\langle\mathbf 0|\psi_Q\rangle|^{2k}
\right]
&=
\frac{1}{Z_n}
\int_{[0,1]^n}
\prod_{1\le i<j\le n}(x_i-x_j)^2\\
&\quad\times
\prod_{j=1}^n \left[x_j^{\frac{k-1}{2}}(1-x_j)^{-1/2}\right]\,dx_j ,
\end{split}
\label{eq:overlap-selberg-ratio}
\end{equation}
where $Z_n$ is the same integral at $k=0$. 
Equation~\eqref{eq:overlap-selberg-ratio} has Selberg form:
\begin{equation}
\begin{split}
\Sigma_n(\alpha,\beta,\gamma)
&:=
\int_{[0,1]^n}
\prod_{1\le i<j\le n}|x_i-x_j|^{2\gamma}\\
&\qquad\times
\prod_{j=1}^n \left[x_j^{\alpha-1}(1-x_j)^{\beta-1}\right]\,dx_j,\\
&=
\prod_{j=0}^{n-1}
\frac{
\Gamma(\alpha+j\gamma)\,
\Gamma(\beta+j\gamma)\,
\Gamma(1+(j+1)\gamma)
}{
\Gamma(\alpha+\beta+(n+j-1)\gamma)\,
\Gamma(1+\gamma)
}
\end{split}\label{eq:selmonta}
\end{equation}
with closed form given by the second line. In our case, the denominator uses
$\alpha=\beta=1/2$, $\gamma=1$, while the numerator shifts to $\alpha=(k+1)/2$.
Therefore,
\begin{equation}
\mathbb E_{Q\sim\mathrm{SO}(2n)}
\!\left[
|\langle\mathbf 0|\psi_Q\rangle|^{2k}
\right]
=
\frac{
\Sigma_n\!\left(\frac{k+1}{2},\frac12,1\right)
}{
\Sigma_n\!\left(\frac12,\frac12,1\right)
}.
\label{eq:overlap-selberg-quotient}
\end{equation}
Specializing the Selberg product gives
\begin{equation}
\mathbb E_{Q\sim\mathrm{SO}(2n)}
\!\left[
|\langle\mathbf 0|\psi_Q\rangle|^{2k}
\right]
=
\prod_{j=0}^{n-1}
\frac{
\Gamma\!\left(\frac{n+j}{2}\right)
\Gamma\!\left(\frac{k+j+1}{2}\right)
}{
\Gamma\!\left(\frac{j+1}{2}\right)
\Gamma\!\left(\frac{n+k+j}{2}\right)
}.
\label{eq:RMT_overlap_SO}
\end{equation}
Sampling from the full orthogonal group $\mathrm{O}(2n)$, the two states have independent fermion parities. With probability $1/2$ they lie in opposite parity sectors, in which case the overlap vanishes identically. Therefore, for every $k>0$,
\begin{equation}
\mathcal{F}^{(k)}_{\MG_n}
:=
\mathbb E_{O_1,O_2\sim\mathrm{O}(2n)}
\!\left[
|\langle\psi_{O_1}|\psi_{O_2}\rangle|^{2k}
\right]
=
\frac{1}{2}\,
\mathbb E_{Q\sim\mathrm{SO}(2n)}
\!\left[
|\langle\mathbf 0|\psi_Q\rangle|^{2k}
\right].
\label{eq:RMT_overlap_O}
\end{equation}
Combining \eqref{eq:RMT_overlap_SO} and \eqref{eq:RMT_overlap_O}, we obtain the state frame potential of the matchgate group
\begin{equation}
\mathcal{F}^{(k)}_{\MG_n}
=
\frac12
\prod_{j=0}^{n-1}
\frac{
\Gamma\!\left(\frac{n+j}{2}\right)
\Gamma\!\left(\frac{k+j+1}{2}\right)
}{
\Gamma\!\left(\frac{j+1}{2}\right)
\Gamma\!\left(\frac{n+k+j}{2}\right)
}.
\end{equation}

\subsection{Frame Potential for Random Gaussian Unitaries}

We now compute the matchgate unitary frame potential~\eqref{eq:unitary_fp_def}. The target quantity is the full orthogonal average
\begin{equation}
\mathfrak{F}^{(k)}_{\MG_n}
:=
\mathbb E_{O_1,O_2\sim \mathrm{O}(2n)}
\!\left[
\bigl|\Tr(U_{O_1}U_{O_2}^\dagger)\bigr|^{2k}
\right],
\end{equation}
where $U_O$ is the fermionic Gaussian unitary implementing
\begin{equation}
U_O\,\gamma\,U_O^\dagger = O\,\gamma,
\qquad O\in \mathrm{O}(2n).
\end{equation}
By Haar invariance, with $O=O_1^{-1}O_2$, we have
\begin{equation}
\mathfrak{F}^{(k)}_{\MG_n}
=
\mathbb E_{O\sim\mathrm{O}(2n)}
\!\left[
|\Tr(U_O)|^{2k}
\right].
\end{equation}
As in the previous subsection, we first evaluate the simplified $\mathrm{SO}(2n)$ average and then lift the result to $\mathrm{O}(2n)$.
For $O\in \mathrm{SO}(2n)$ with eigenvalues $e^{\pm 2i\theta_j}$ with $j\in [n]$, define
\begin{equation}
\Tr(U_O)
=
\prod_{j=1}^n \left(e^{i\theta_j}+e^{-i\theta_j}\right)
=
2^n \prod_{j=1}^n \cos\!\left(\theta_j\right),
\end{equation}
so $|\Tr(U_O)|^{2k} = 2^{2kn}\prod_{j=1}^n \cos^{2k}\!\left(\theta_j\right)$.
We introduce the variables
\begin{equation}
x_j := \cos^2\!\left(\theta_j\right)
= \frac{1+\cos 2\theta_j}{2}\in[0,1],
\end{equation}
for which we have 
\begin{equation}
\begin{split}
  d\theta_j = \frac{dx_j}{2\sqrt{x_j(1-x_j)}}, &\qquad \cos^{2k}\!\left(\theta_j\right)=x_j^k.  
\\
\cos (2\theta_i)-\cos(2\theta_j) &= 2(x_i-x_j),
\end{split}
\end{equation}
Combining these ingredients we obtain
\begin{equation}
\begin{split}
    \mathbb E_{\mathrm{SO}(2n)}
[|\Tr(U_O)|^{2k}]
&=
\frac{2^{2kn}}{Z_n^{\mathrm{u}}}
\int_{[0,1]^n}
\prod_{1\le i<j\le n}|x_i-x_j|^2\\ &
\qquad \qquad \prod_{j=1}^n
x_j^{k-\frac12}(1-x_j)^{-\frac12}\,dx_j,
\end{split}
\end{equation}
which is a Selberg integral with 
\begin{equation}
\alpha = k+\frac12,
\qquad
\beta = \frac12,
\qquad
\gamma = 1.
\end{equation}
(Instead, $Z_n^{\mathrm{u}}$ is the same integral at $k=0$.)
Evaluating the Selberg ratio gives
\begin{equation}
\mathbb E_{\mathrm{SO}(2n)}
\!\left[
|\Tr(U_O)|^{2k}
\right]
=
2^{2kn}
\prod_{j=0}^{n-1}
\frac{
\Gamma\!\left(k+\frac12+j\right)
}{
\Gamma\!\left(\frac12+j\right)
}
\frac{
\Gamma(n+j)
}{
\Gamma(n+k+j)
}.
\label{eq:RMT_trace_SO}
\end{equation}
Returning to the $\mathrm{O}(2n)$ case amounts to consider an additional factor $2$ coming from the parity, which therefore leads to 
\begin{equation}
\mathfrak{F}^{(k)}_{\MG_n}
=
\frac12\,2^{2kn}
\prod_{j=0}^{n-1}
\frac{
\Gamma\!\left(k+\frac12+j\right)
}{
\Gamma\!\left(\frac12+j\right)
}
\frac{
\Gamma(n+j)
}{
\Gamma(n+k+j)
}.
\label{eq:RMT_trace_O_final}
\end{equation}
By definition, 
\begin{equation}
\mathfrak{F}^{(k)}_{\MG_n}
=
\dim\!\bigl(\Com_k(\MG_n)\bigr),
\end{equation}
thus giving a closed formula for arbitrary $k$ and $n$.
We conclude with a remark. 
Using the Gamma duplication formula, Eq.~\eqref{eq:RMT_trace_O_final} can be rewritten as the rational product
\begin{equation}\label{eq:rmt_tableau_prod}
\mathfrak{F}^{(k)}_{\MG_n}
=
\dim \Com_k(\MG_n)
=
T(2n,k-1),
\end{equation}
where
\begin{equation}
T(n,m)
=
\prod_{1\le i\le j\le m}\frac{n+i+j-1}{i+j-1}.
\end{equation}
This is precisely the quantity appearing in OEIS sequence A102539. The same expression also admits a direct combinatorial derivation: the pairing basis can be put in bijection with semistandard Young tableaux with entries in $1,2,\dots,k-1$ and at most $2n$ columns.

%

\end{document}